\documentclass[aps,10pt,superscriptaddress,notitlepage,nofootinbib,pra,twocolumn]{revtex4-2}
\usepackage{graphicx} 
\usepackage{amsmath,amssymb,amsfonts,siunitx,xcolor,xfrac,bbold,hyperref,amsthm,physics}
\newcommand{\eg}{, e.g.~}
\newcommand{\ie}{, i.e.~}

\newtheorem{conjecture}{Conjecture}

\makeatother

\begin{document}

\date{\today}
\author{David D. Noachtar}
\email{dnoachtar@uchicago.edu}

\author{Aashish A. Clerk}
\email{aaclerk@uchicago.edu}
\affiliation{Pritzker School of Molecular Engineering and Chicago Quantum Institute, The University of Chicago, 60637 Chicago, Illinois, USA}

\title{Exact metastability in a class of driven-dissipative quantum many-body systems}

\begin{abstract}
    Metastability in many-body quantum systems and its associated exponentially-long timescales have been the subject of considerable recent interest.  Here, we focus on a class of driven-dissipative many-body open quantum systems described by a Lindbladian having hidden time-reversal symmetry (a form of quantum detailed balance). Examples include boundary-driven interacting spin chains, bosonic lattice models and driven-dissipative collective spin models.  
    We suggest that for such systems, slow timescales in the vicinity of a dissipative first-order phase transition can be {\it analytically} predicted using a special purification of the non-equilibrium steady state.  We show the accuracy of our conjecture through detailed studies of a dissipative transverse-field Ising model with collective and local decay, and a driven-dissipative nonlinear cavity model.  Our results allow quantitative insights into metastability and slow dynamics for a range of systems, including cases where semiclassical or path-integral instanton approaches are intractable.  
\end{abstract}

\maketitle

\section{Introduction}

Metastability is a ubiquitous non-equilibrium phenomenon in both classical and quantum systems, whereby a system gets ``stuck" in an extremely long-lived intermediate state before relaxing to a final stationary state.  Examples range from
the classical hysteresis of ferromagnets, diamonds and supercooled water~\cite{langer_statistical_1969}, to quantum examples spanning the radioactive decay of nuclei to cosmological scenarios such as false vacuum decay.  
Perhaps the most surprising and theoretically challenging feature of metastability is the emergence of an exponentially-long timescale (i.e.~the lifetime of the metastable state), a quantity that is generally non-perturbative in system parameters.  In classical contexts there are well established methods for understanding these times (e.g.~Kramers' theory for thermal activation~\cite{hanggi_reaction-rate_1990}).  Recent work has also sought to establish a general theory for such long timescales in closed (non-dissipative) quantum systems~\cite{yin_theory_2025}.

\begin{figure}
    \centering
    \includegraphics[width=\linewidth]{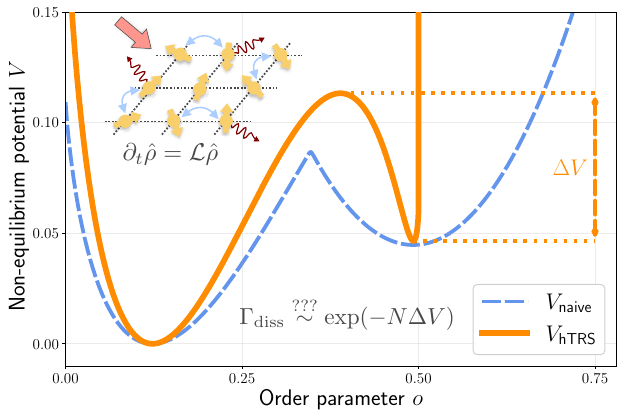}
    \caption{
    Hidden time-reversal symmetry allows one to directly predict the dissipative gap $\Gamma_\text{diss}$ (i.e.~slowest relaxation rate) of open many-body Lindbladians (see inset). A special steady-state purification defines a potential function $V_{\rm hTRS}(o)$ (solid, orange) whose barrier height $\Delta V$ controls the dissipative gap; here $o$ is an effective order parameter.  This potential does not match a more naively-defined potential $V_{\rm naive}(o)$ obtained directly from the steady-state density matrix (dashed, blue).  Results are for a dissipative transverse-field Ising model (with $o$ related to the magnetization), tuned near a first-order phase transition; see Appendix~\ref{app:dtfim-pot-barriers} and Fig.~\ref{fig:fig12}).
    }
    \label{fig:fig1}
\end{figure}

This leaves the case of open (dissipative) quantum systems, especially those that do not simply admit an effective classical description. 
The emergence of slow timescales is generic in dissipative quantum systems exhibiting a first-order phase transition~\cite{ptaszynski_dynamical_2024}, a phenomenon occurring in a wide variety of systems~\cite{rose_metastability_2016,leppenen_quantum_2024,minganti_spectral_2018,xiang_switching_2025,gelhausen_dissipative_2018,chen_quantum_2023,beaulieu_observation_2025}. 
Characterizing emergent slow timescales in many-body open systems can also be of practical interest: often the long-lived metastable states can possess useful entanglement structures that are absent in the long-time steady state (see e.g. Ref.~\cite{groszkowski_reservoir-engineered_2022}) or they can encode logical information~\cite{chamberland_building_2022}.   
While there exists understanding for such long timescales in certain paradigmatic single-body systems (e.g.~slow switching rates in driven-dissipative nonlinear quantum oscillators~\cite{marthaler_switching_2006,dykman_critical_2007,lee_real-time_2025,carde_nonperturbative_2026}), for truly many-body open quantum systems this question has for the most part only been treated numerically (something that is challenging given that one needs large system sizes and the ability to probe extremely long timescales). In addition to numerics, recent work has established a quantum bottleneck theorem~\cite{rakovszky_bottlenecks_2024} for quantum channels with a specific local structure that allows a lower bound on the relaxation timescale. There has also been an effort to obtain effective theories of the slow dynamics in open quantum metastable systems~\cite{macieszczak_theory_2021,macieszczak_towards_2016}; while yielding many insights, these approaches use the slow timescale as an input rather than providing a way to calculate them. Path-integral methods and instanton calculations offer a route to calculating these slow timescales (see Refs.~\cite{carde_nonperturbative_2026,lee_real-time_2025,thompson_qubit_2022,mylnikov_qubit_2025,ptaszynski_quantum_2026,dutta_quantum_2025}), but
can be difficult to implement.  In many cases one cannot formulate a tractable many-body path-integral description, and even if one can, solving for instanton trajectories might be intractable.  
    \begin{figure}
        \centering
        \includegraphics[width=\linewidth]{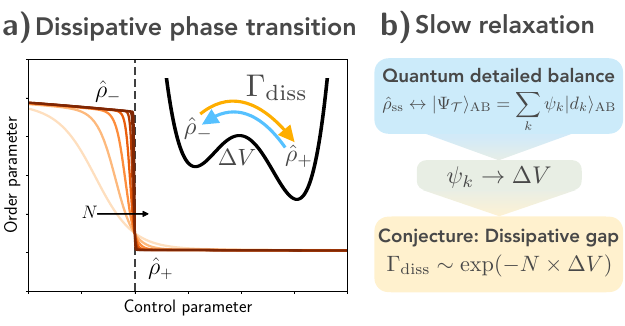}
        \caption{
        \textbf{a)} Generic plot of the steady-state average order parameter versus the relevant system parameter for a system exhibiting a first-order dissipative phase transition (DPT), for various system sizes $N$.
        At the DPT (dashed line) the steady state $\hat \rho_\text{ss}$ abruptly changes from $\hat \rho_-$ to $\hat \rho_+$.  Slow dynamics are governed by the dissipative gap $\Gamma_\text{diss}$, which close to the transition is akin to the slow switching rate between minima of a double-well potential.  
        \textbf{c)} For systems having hTRS, despite not being in thermal equilibrium, we conjecture (and show for several examples) that
        $\hat \rho_\text{ss}$ directly determines  $\Gamma_\text{diss}$.  
        A special purification of $\hat \rho_\text{ss}$ sets
        an effective potential barrier $\Delta V$, which then sets 
        $\Gamma_\text{diss}$ in the vicinity of a first-order DPT. 
        }
        \label{fig:fig2}
    \end{figure}

In this work, we consider metastability and slow timescales in a class of open quantum systems that are described by a Lindblad master equation, that exhibit steady-state phase transitions, and which satisfy a subtle version of quantum detailed balance, so-called hidden time-reversal symmetry (hTRS)~\cite{roberts_hidden_2021}.  Such systems can have complex, non-thermal steady states that do not commute with the system Hamiltonian.  They span a range of many-body systems including boundary-driven spin chains~\cite{yao_hidden_2024,lingenfelter_exact_2024}, bosonic~\cite{roberts_competition_2023} and fermionic~\cite{lingenfelter_exact_2026} lattice models and disordered interacting spin ensembles with collective decay~\cite{roberts_exact_2023}.  While hTRS has applications in quantum metrology~\cite{lee_timescales_2025,yang_efficient_2023,godley_adaptive_2023,girotti_estimating_2025}, it has mostly been applied as a tool for finding quantum steady states. hTRS imposes a particular structure on the non-equilibrium steady state (NESS). While previous literature has largely leveraged this structure in a model-dependent way, we formalize a systematic, model-agnostic ``recipe" for deriving the exact steady-state solution. Furthermore, we conjecture that one can do even more with hTRS: {\it it enables one to predict slow timescales associated with metastability directly from the non-equilibrium steady state}. 

We argue that this can be done without any need to construct an effective action and perform an instanton calculation, and goes beyond simply assuming that the steady state is thermal. While there are many naive ways of trying to associate an \emph{effective, non-equilibrium potential} with a non-equilibrium steady state, our ansatz $V_\text{hTRS}$ has direct utility.  It defines a potential barrier $\Delta V$ (see Fig.~\ref{fig:fig1}) that determines (up to a prefactor) the slow switching rate at a phase transition: $\tau_{\rm slow} \sim \exp{N \Delta V}$, where $N$ is the effective system size (e.g.~number of unit cells, qubits, etc.). We stress that this potential is not simply obtained from the modular Hamiltonian, i.e. the log of the steady-state density matrix. The basic idea is sketched in Fig.~\ref{fig:fig2}.  
Our potential $V_\text{hTRS}$ is a function of the system's order parameter (generically a charge), and is directly obtained from a special purification of the steady state in terms of definite-charge states (denoted as $\ket{d_k}_\text{AB}$ in Fig.~\ref{fig:fig2}b.

We verify the accuracy of this prediction by studying two very different models that exhibit first-order dissipative phase transitions and that have hTRS: a driven-dissipative nonlinear cavity model and a dissipative transverse-field Ising model (DTFIM) with all-to-all interactions, which has a true many-body character. Besides providing fundamental insights into the nature of metastability, our investigation also marks the first application of hTRS to constrain dynamics of open quantum systems.  Our work also provides a new way to understand the structure of exactly-solvable hTRS many-body open systems, showing that in many physically-relevant cases, hTRS lets one understand the NESS in terms of eigenstates of an effective Hermitian ladder model, which only features inter-leg couplings.  Within this mapping the purification of the NESS corresponds to an eigenstate of the ladder model that is localized completely on a single ladder leg. We exemplify this new approach in the case of the DTFIM in a way that complements previous derivations of the steady state~\cite{roberts_exact_2023} and provides new intuition into the physics (see Appendix~\ref{app:DTFIM-purified-steady-state}).

The rest of this manuscript is organized as follows. In Sec.~\ref{sec:sec2} we provide an overview over our problem of interest: the slow dynamics close to first-order phase transitions in Markovian open quantum systems. Then, Sec.~\ref{sec:sec3} highlights the ingredients of our conjecture: we first discuss slow relaxation in a classical random walk and then introduce the hTRS framework. Sec.~\ref{sec:sec4} first establishes a ``recipe" for obtaining steady states using hTRS (Secs.~\ref{subsec:sec4A} and~\ref{subsec:sec4B}) and then establishes a connection between these two ingredients which we use to define a non-equilibrium potential in Sec.~\ref{subsec:4C}. This potential plays the central role of defining a potential barrier in our conjecture in Sec.~\ref{subsec:sec4D}. Sec.~\ref{sec:sec5} applies this conjecture to two paradigmatic models with hTRS. Finally, Sec.~\ref{sec:sec6} discusses our results and highlights its significance in developing an analytical understanding for slow relaxation in open quantum systems.

\section{Statement of the problem: slow dynamics in Markovian open quantum systems}\label{sec:slow-dynamics-overview}\label{sec:sec2}

The basic problem of interest is sketched in Fig.~\ref{fig:fig2}:  we consider an open (many-body) quantum system whose parameters are tuned to be close to a first-order phase transition.  One expects slow dynamics and relaxation in such a situation, corresponding to slow switching between long-lived states.  The goal is to understand how the slow rate scales with increasing system size $N$ (where $N$ is the number of qubits, cavities, photons, etc.~in the many-body model).  

We focus throughout on Markovian open quantum systems described by a GKSL (or Lindblad) type quantum master equation.  Letting $\hat{\rho}$ describe the reduced density matrix of the system, the dynamics follows~\cite{lindblad_generators_1976,gorini_completely_1976}
        \begin{align}
            \mathcal L \hat \rho \equiv \partial_t \hat \rho &= -i [\hat H, \hat \rho] + \sum_k \mathcal D[\hat L_k] \hat \rho.
            \label{eq:lindbladian}
        \end{align}
Here, $\hat H$ is the (time-independent) system Hamiltonian and the jump operators $\{\hat L_k\}$ parameterize the effect of dissipation; the Lindblad superoperator is defined as $\mathcal D[\hat A] \hat \rho = \hat A \hat \rho \hat A^\dagger - \{\hat A^\dagger \hat A, \hat \rho\}/2$. 
 We are primarily interested in the steady-state density matrix $\hat \rho_\text{ss}$ which is defined by $\mathcal L \hat \rho_\text{ss}=0$; we will focus on situations where for any finite system size $N$, the steady state is unique.  

Our focus is on systems where Eq.~\eqref{eq:lindbladian} describes a many-body system tuned close to a first-order (or discontinuous) dissipative phase transition. Such transitions are defined as nonanalytic behavior of the steady state $\hat \rho_\text{ss}$~\cite{fazio_many-body_2025} in
the limit $N \rightarrow \infty$.  
More precisely, given a Lindbladian that is a function of a control parameter $\delta$, $\mathcal{L} = \mathcal{L}(\delta)$, there is a first-order dissipative phase transition (DPT) at $\delta = 0$ if
        \begin{align}
            \lim_{\delta\to0^+}\lim_{N\to\infty}\hat \rho_{\text{ss},\delta}\neq \lim_{\delta\to0^-}\lim_{N\to\infty}\hat \rho_{\text{ss},\delta}.
        \end{align}
For such phase transitions, the nonanalyticity of $\hat \rho_\text{ss}$ directly results in discontinuous behavior of one or more steady-state expectation values (which can then serve as an effective order parameter for the transition).  


Like their classical counterparts, dissipative first-order phase transitions in our Lindblad setting are associated with a slow relaxation rate that vanishes in the thermodynamic limit.  The relevant quantity for the Lindblad master equation in Eq.~\eqref{eq:lindbladian} is the slowest system relaxation rate, the dissipative gap $\Gamma_\text{diss}$.  Noting that all eigenvalues $\lambda$ of $\mathcal L$ have non-positive real part, the dissipative gap is given by:
        \begin{align}
            \Gamma_\text{diss} &= -\max_{\text{Re}(\lambda) < 0} \text{Re}(\lambda).
        \end{align}
The dissipative gap generally determines the long-time relaxation rate of an arbitrary initial state to the final steady state~\cite{fazio_many-body_2025}.  In this work we do not consider anomalous cases where this is not true, e.g. Refs.~\cite{mori_resolving_2020,wang_accelerating_2023,lee_anomalously_2023}.  



        
For a dissipative phase transition, $\Gamma_\text{diss}$ will vanish close to the transition $\delta = 0$ in the large-$N$ limit, corresponding to a steady-state degeneracy\footnote{Since we are interested in how the (long, but finite) metastable timescales diverge as $N\to\infty$, we will formally always take $t\to\infty$ \emph{before} taking the large-$N$ limit. This ensures a unique steady state and avoids multistability.}. 
At finite $N$, general arguments suggest that the dissipative gap will be exponentially suppressed with increasing system size $N$ near a first-order DPT~\cite{casteels_critical_2017,ptaszynski_dynamical_2024}, allowing us to write:
        \begin{align}
            \Gamma_\text{diss} \sim \exp[-N \Delta V(\delta)].
            \label{eq:slowrate}
        \end{align}
This equation defines an effective potential barrier $\Delta V$, which will depend (among other things) on the control parameter $\delta$, and whose form is expected to be independent of $N$ in the large-$N$ limit.   

The key question we seek to understand is: is there a meaningful potential $V$ that gives the effective potential barrier $\Delta V$ (and, if so, can we find $V$)? While several studies have investigated slow relaxation close to a first-order DPT~\cite{ptaszynski_dynamical_2024,xiang_switching_2025,casteels_critical_2017,minganti_spectral_2018,gelhausen_dissipative_2018,carde_nonperturbative_2026,leppenen_quantum_2024,vicentini_critical_2018,marcuzzi_universal_2014} and generally confirmed the exponential scaling with $N$, attempting to understand the value and parameter dependence of $\Delta V$ has been far less studied.  The exception here are models of a single driven-dissipative nonlinear resonator, which can be analyzed in a manner akin to a many-body system by treating photon number as an effective system size parameter $N$.  Here, the bosonic nature of the problem makes path-integral techniques especially powerful, and recent works have performed instanton calculations within a Keldysh path-integral description to calculate the potential barrier $\Delta V$ entering the slow timescale, both numerically~\cite{lee_real-time_2025} and for special cases analytically~\cite{carde_nonperturbative_2026,thompson_qubit_2022}. While extremely powerful, it remains unclear how to use similar techniques for truly many-body systems or spin systems.  

As we discuss in more detail below, our approach will be to focus on a subset of many-body open systems where an unusual version of quantum detailed balance, hidden time-reversal symmetry (hTRS), is already known to allow exact analytic descriptions.  This includes several experimentally-relevant many-body models that exhibit first-order DPTs~\cite{roberts_exact_2023,casteels_critical_2017,roberts_driven-dissipative_2020,roberts_competition_2023,lingenfelter_exact_2026}.  While hTRS does not a priori yield direct insights into dynamics, we will argue that for these systems, one can directly use properties of the exact solution to determine the potential barrier $\Delta V$ which controls the slow timescale.




\section{Ingredients for the conjecture}\label{sec:ingredients-for-conj}\label{sec:sec3}

Our approach to understanding slow relaxation near a first-order DPT in systems with hTRS has two basic ingredients:
\begin{itemize}

\item Use the analytic solution for the steady-state density matrix (more specifically a particular purification) to extract an effective one-dimensional potential function $V(x)$, where the quantity $x$ is related to an effective charge. 

\item Extract a potential barrier height $\Delta V$ from $V(x)$, and use this to predict its parameter dependence in Eq.~\eqref{eq:slowrate} exactly like one would in a classical stochastic system with detailed balance. 

\end{itemize}

We stress that the above approach for obtaining an effective potential does not require one to first derive a semiclassical description of the system, or a tractable path-integral description; instead it works directly with a purification of the steady-state density matrix. Furthermore, our approach differs from defining a potential through the logarithm of the steady-state density matrix. To set the stage, we briefly review both these topics (classical stochastic systems, detailed balance and hTRS) in this section.  


\subsection{Slow relaxation of a classical stochastic process with detailed balance}\label{subsec:classical-hopping-correspondence}\label{subsec:sec3A}
        
We first review how in classical stochastic models with detailed balance,
one can associate a potential function with the steady state, and then use this potential function to characterize slow dynamics (see e.g. Ref.~\cite{hanggi_bistable_1984}).  Understanding this intimate connection is crucial as we will eventually mimic aspects of this in our treatment of the quantum problem.

Consider a classical particle on a one-dimensional, possibly semi-infinite\footnote{We restrict ourselves to positive integers because we anticipate that the sites of the lattice correspond to particle numbers on top of a vacuum}, lattice, which can incoherently hop to adjacent sites. The probability $P_k(t)$ to find the particle on site $k$ ($k=0,1,2,...$) at time $t$ follows the classical master equation
\begin{align}
            \partial_t P_k(t) &= W^+_{k-1} P_{k-1}(t) + W^-_{k+1} P_{k+1}(t)\nonumber\\
            &- \big ( W^+_k + W^-_k\big )P_k(t),
            \label{eq:cme}
\end{align}
where $W^\pm_k$ is the rate for transitions from site $k$ to $k\pm 1$.  Note the rates can vary in space. Let $P^\text{ss}_k$ denote the steady-state probability distribution\ie $P_k(t) = P^\text{ss}_k $ is a time-independent solution of Eq.~\eqref{eq:cme}.  By construction $P^\text{ss}_k$ must satisfy the classical detailed-balance condition
        \begin{align}
             \frac{P^\text{ss}_{k+1}}{P^\text{ss}_k} = \frac{W^+_k}{W^-_{k+1}}.
            \label{eq:cme-classical-balance}
        \end{align}

We next assume that there exists a well-defined effective ``thermodynamic limit" in which the position of our particle becomes a continuous quantity. Let the parameter $N$ be an effective (dimensionless) inverse lattice constant. The continuum limit of interest then corresponds to $N \rightarrow \infty$ with $x=k/N$ playing the role of our continuum position. 
We next note that the number of discrete lattice sites between two fixed positions $x_1$ and $x_2$ will diverge in the large-$N$ limit. To ensure that dynamical timescales in Eq.~\eqref{eq:cme} have well-defined large-$N$ limits, we further scale the switching rates so that they also scale like $N$ in the large-$N$ limit.  We thus require that in the large-$N$ limits, the switching rates behave as
\begin{equation}
    \lim_{N \rightarrow \infty} \frac{W^\pm_{k=xN}}{N} = \Gamma_{\pm}(x)
\end{equation}


Given this, the steady state of the system in the large-$N$ limit can be characterized in terms of an intensive potential function $V(x)$.  We define this potential via:
\begin{equation}
    P^{\rm ss}(x) \equiv P^{\rm ss}_{k=xN}\sim  e^{-N V(x)}.
\end{equation}
Using the detailed-balance condition in Eq.~\eqref{eq:cme-classical-balance}, this potential will be directly determined by our switching rates, confirming that it is independent of $N$ in the large-$N$ limit:
        \begin{align}
            V(x) &= -\lim_{N\to\infty}\frac{\log(P^\text{ss}_k)}N = -\lim_{N\to\infty} \frac 1 N\sum_{j=0}^{k-1}\log\left [ \frac{W^+_k}{W^-_{k+1}}\right ] \nonumber\\
            &= -\int_0^x dx' \log \left [\frac{\Gamma_+(x')}{\Gamma_-(x')}\right ].
            \label{eq:noneq-potential}
        \end{align}

It immediately follows that for large $N$, the steady-state probability distribution will be concentrated on values of $x$ where the potential $V(x)$ attains its minimum.  In this context, a bistable system describes a potential $V(x)$ with two distinct, near-degenerate minima. This situation mimics a many-body system near a first-order phase transition: if the global minimum of $V(x)$ changes from one local minimum to another upon varying model parameters, this corresponds to a discontinuous change of the steady state in the thermodynamic limit. 



        
At this stage, the potential $V(x)$ is just a convenient tool for characterizing the steady-state probability distribution.  It however also has direct relevance to dynamics in the continuum limit, and in particular, will control slow dynamics in systems where one has bistability.  
In the continuum limit of interest, one can show the master equation dynamics of the system in Eq.~(\ref{eq:cme}) becomes equivalent to a one-dimensional Fokker-Planck equation for the probability density $P(x,t)$ (see, e.g.~\cite{hanggi_bistable_1984}):
        \begin{align}
            \partial_t P(x,t) = -\partial_x \big [ \mu(x) P(x,t)\big ] + \frac 1 2 \partial_x^2 \big [ D(x) P(x,t)].
            \label{eq:fokker-planck}
        \end{align}
Here the drift coefficient is $\mu(x)=\Gamma_+(x) - \Gamma_-(x)$ (describing the average velocity of the particle), and the diffusion coefficient $D(x) \sim \left( \Gamma_+(x) + \Gamma_-(x) \right)/N$.
The diffusion term is a consequence of the stochastic switching in the original model\footnote{See Ref.~\cite{hanggi_bistable_1984} for different choices for the explicit form of $D(x)$; these differences play no role to our discussion.}; its suppression in the large-$N$ limit corresponds to a weak-noise limit, and is the consequence of self-averaging in the continuum limit.  

We will now consider a bistable system such that $V(x)$ takes the form of a double-well potential with two local minima at $x_1$ and $x_2$ separated by a local maximum at $x_*$, assuming $x_1 < x_* < x_2$. Since the drift coefficient $\mu(x_i)=0$ at $x_i\in\{x_1, x_2, x_*\}$, without noise the positions would be fixed points of the dynamics (with $x_1$ and $x_2$ being stable and $x_*$ unstable).  Adding back weak noise to the system (via the diffusion term), we now have the possibility of rare stochastic switching events that take the system from one stable fixed point to the other.  Within our limiting procedure, the probability per unit time for such switching events will be exponentially small in $N$ (as the system needs to make $\mathcal O (N)$ hops in succession to transition from one attractor to another).  For such classically bistable systems, Kramers' theory allows one to quantitatively understand the slow rates corresponding to this switching  ~\cite{freidlin_random_2012,hanggi_bistable_1984,berglund_kramers_2013}. Crucially the potential $V(x)$ plays a key role in determining the slowest relaxation rate\ie the dissipative gap $\Gamma_\text{diss}$, which is the largest, nonzero eigenvalue of Eq.~\eqref{eq:fokker-planck}. The slowest relaxation rate scales as
\begin{align}
    -\lim_{N\to\infty} \frac{\log \Gamma_\text{diss}}{N} =  V(x_*) - \max_{i \in \{1, 2\} } V(x_i) \equiv \Delta V
    \label{eq:classical-dissipative-gap-scaling}
\end{align}
in the effective thermodynamic limit $N\to\infty$.  Hence, the relaxation to the steady state is exponentially slow with an exponent simply given by the product of $N$ and an effective potential barrier $\Delta V$.  

Stepping back, the essential feature of this model that we will use (by analogy) going forward is {\it a direct connection between the dissipative steady state and dynamics}.  Even without any knowledge of the specific transition rates, one can extract an effective potential from the steady state via Eq.~\eqref{eq:noneq-potential}, and then use this to predict the slow relaxation rate of the system via Eq.~\eqref{eq:classical-dissipative-gap-scaling}. The crucial ingredient here is classical detailed balance, Eq.~\eqref{eq:cme-classical-balance}. Going forward, we will suggest that a similar direct connection between non-equilibrium steady states and dynamics is possible for many-body quantum open systems that have a particular kind of quantum detailed balance known as hidden time-reversal symmetry~\cite{roberts_hidden_2021}. 


\subsection{Quantum detailed balance and hidden time-reversal symmetry}\label{subsec:htrs}\label{subsec:sec3B}

To generalize the connection in the last subsection between steady states and dynamics to quantum systems described by a Lindblad master equation 
(cf.~Eq.~\eqref{eq:lindbladian}), we will need a notion of quantum detailed balance.  We will argue that the appropriate formulation is so-called hidden time-reversal symmetry 
~\cite{roberts_hidden_2021}.  This can be viewed as the more familiar KMS detailed balance augmented by an anti-unitary time-reversal operator~\cite{carlen_gradient_2017,kossakowski_quantum_1977}, and has also been referred to as SQDB-$\theta$~\cite{fagnola_generators_2010,fagnola_generators_2012,duvenhage_quantum_2025}.  In this section, we provide a brief overview of hTRS.


    \begin{figure}
        \centering
        \includegraphics[width=\linewidth]{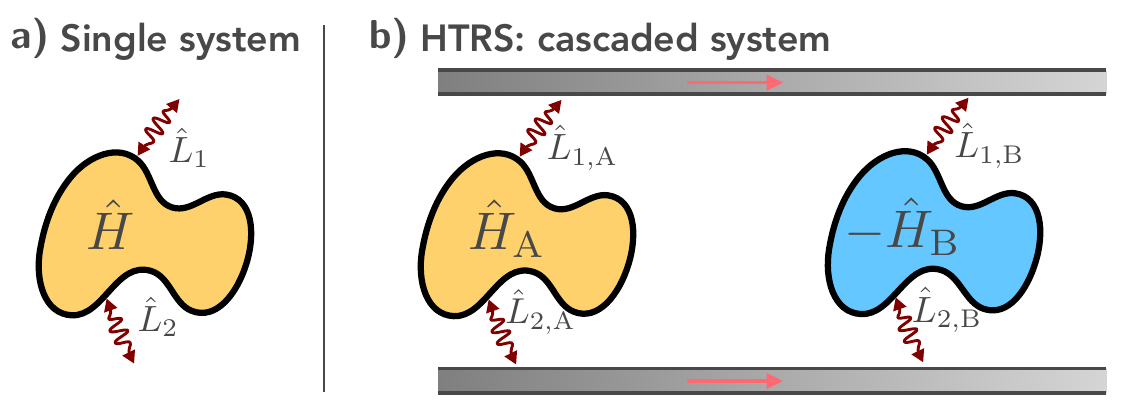}
        \caption{Construction of the doubled system in the hidden time-reversal symmetry framework. \textbf{a)} A quantum system with Hamiltonian $\hat H$ is subject to Markovian dissipation shown here using two jump operators $\hat L_1$ and $\hat L_2$. \textbf{b)} This quantum system forms system $A$ of the doubled system considered in hTRS. System $A$ and system $B$ emit photons into chiral waveguides which only propagate photons from system $A$ to system $B$, resulting in a cascaded quantum network~\cite{stannigel_driven-dissipative_2012}. Therefore system $A$ is not impacted by system $B$ while system $B$ interacts with the radiation dissipated by system $A$. If the model has hidden time-reversal symmetry, the Hamiltonian and jump operators of system $B$ are simply related to those of system $A$ and the steady state of the combined system is pure. This setup has been used to derive exact steady-state solutions, but also for quantum metrological tasks in continuous sensing~\cite{cabot_continuous_2024}.}
        \label{fig:fig3}
    \end{figure}

    We define a Lindblad master equation of the form in Eq.~(\ref{eq:lindbladian}) to have hTRS if there exists
    an invertible, antilinear operator $\hat \Psi$ ($c\hat \Psi = \hat \Psi c^*$ for any complex scalar $c$) such that
    \begin{align}
        \hat {H}_\text{eff} &= \hat \Psi \hat H_\text{eff}^\dagger \hat \Psi^{-1}\nonumber\\
        \hat {L}_k &= \hat \Psi \hat L_k^\dagger \hat \Psi^{-1} \quad \forall k.
        \label{eq:quantum-detailed-balance}
    \end{align}
    Here $\{\hat{L}_k\}$ are a set of jump operators defining the dissipative parts of the Lindbladian\footnote{Note that we are choosing a specific gauge for the jump operators in comparison with Ref.~\cite{roberts_hidden_2021}}, and $\hat H_\text{eff} = \hat H -i \sum_k \hat L_k^\dagger \hat L_k / 2$ is the standard ``no-jump" non-Hermitian Hamiltonian associated with the Lindbladian.  A direct substitution shows that if these conditions are satisfied, then the system has a full-rank steady-state density matrix given by~\cite{roberts_exact_2023}:
    \begin{align}
        \hat \rho_\text{ss} = \hat \Psi \hat \Psi^\dagger = \sum_n p_n \ket n \bra n.
    \end{align}
    Here, $|n\rangle$ are the eigenstates of $\hat \rho_\text{ss}$ with eigenvalues $p_n \geq 0$ and $\sum_n p_n=1$. Using a polar decomposition, $\hat \Psi = \sqrt{\hat\rho_{\rm ss}} \hat{\mathcal T}$, where $\hat{\mathcal T}$ is an antiunitary operator, and 
    $\sqrt{\hat \rho_{\rm ss}}$ is the conventional (positive, Hermitian) square root of the steady-state density matrix.  

While the motivation behind these definitions might seem opaque, there is some connection to classical detailed balance.  Systems with classical detailed balance have additional ``local" constraints on steady states and the rates that generate dynamics\ie~there is no probability current between any arbitrary pair of microstates (i.e.~the analogue of Eq.~\eqref{eq:cme-classical-balance}).  The hTRS constraints in Eq.~\eqref{eq:quantum-detailed-balance} are somewhat analogous: they pose independent constraints on each element (i.e.~jump operators, effective Hamiltonian) that generates dynamics.  

More relevant to our discussion is that in many cases, hTRS provides an efficient means for solving for the dissipative steady state (much as potential conditions in systems with classical detailed balance make it easier to find steady states).  While this can be done by directly working with Eqs.~\eqref{eq:quantum-detailed-balance}, there is an equivalent but in many cases more physically transparent approach.
The key idea is to introduce a second system, $B$, with an identical Hilbert space, Hamiltonian (up to a sign) and jump operators as the original system $A$.  The two systems are coupled to one another via unidirectional (chiral) waveguides, such that emitted photons can only propagate from system $A$ to system $B$, but not the other way around. Fig.~\ref{fig:fig3}b shows that system $B$ couples to these unidirectional waveguides ``downstream" and can therefore interact with photons emitted by system $A$, while photons emitted by $B$ cannot reach system $A$.  Surprisingly for systems with hTRS, this composite $A+B$ setup is guaranteed to have a pure entangled steady state.  
This is the coherent quantum absorber setup first introduced in Ref.~\cite{stannigel_driven-dissipative_2012}.


In this doubled system, we define operators acting on system $A$ as $\hat O_\text A = \hat O \otimes \hat{\mathbb 1}_\text B$ and similarly those acting on system $B$ as $\hat O_\text B = \hat{\mathbb 1}_\text A \otimes \hat O$. The dynamics of the density matrix $\hat \rho_\text{AB}$ of the combined system of $A$ and $B$ are generated by a GKSL master equation
that has the form of a cascaded quantum system~\cite{carmichael_analytical_1980,gardiner_driving_1993}:
\begin{align}
    \mathcal L_\text{AB} \hat \rho_\text{AB} \equiv \partial_t \hat \rho_\text{AB} &= -i[\hat H_\text{AB}, \hat \rho_\text{AB}]\nonumber\\ 
    &+ \sum_k \mathcal D[\hat L_{k,\text A} - \hat L_{k,\text B}]\hat \rho_\text{AB},
    \label{eq:htrs-doubled-system-dynamics}
\end{align}
with the doubled-system Hamiltonian
\begin{align}
    \hat H_\text{AB} = \hat H_\text A - \hat H_\text B - \frac i 2 \sum_k \left (L_{k,\text A}^\dagger \hat L_{k,\text B} - \hat L_{k,\text B}^\dagger \hat L_{k,\text A} \right ).
    \label{eq:doubled-system-Hamiltonian}
\end{align}
Here, the Hamiltonian of system $B$ is simply the negative of the Hamiltonian of system $A$. The cascaded form of $\mathcal{L}_{AB}$ ensures that the original system $A$ is unaffected by system $B$.

As a consequence of this cascaded form we can obtain the single-system steady state $\hat\rho_\text{ss}$ from the doubled-system steady state $\hat \rho_\text{ss,AB}$ ($\mathcal L_\text{AB}\hat \rho_\text{ss,AB}=0$) by tracing out system $B$. Surprisingly the hTRS conditions in Eqs.~\eqref{eq:quantum-detailed-balance} directly imply that $\hat \rho_\text{ss,AB}$ is pure. As we show in Appendix~\eqref{app:qdb-htrs-proof}, this follows from an equivalence between Eqs.~\eqref{eq:quantum-detailed-balance} and the following conditions
\begin{align}
    (\hat L_{k,\text A} -\hat L_{k,\text B})|\Psi_\mathcal T\rangle_\text{AB} &= 0 \quad \text{ and}\label{eq:htrs-dark-state-condition}\\
    \hat H_\text{AB}|\Psi_\mathcal T\rangle_\text{AB} &= 0.
    \label{eq:htrs-eigenstate-condition}
\end{align}
for a particular purification of $\hat \rho_\text{ss}$
\begin{align}
    |\Psi_\mathcal T\rangle_\text{AB} = \sum_n \sqrt{p_n} |n\rangle_\text{A}\otimes \hat{\mathcal T}|n\rangle_\text{B},
    \label{eq:def-purification}
\end{align}
Here, $\ket{\Psi_\mathcal T}_\text{AB}$ is expressed in terms of the eigendecomposition of $\hat\rho_\text{ss}$ and the antiunitary $\hat{\mathcal T}$ which can be shown to act simply on the eigenstates $\hat{\mathcal T}\ket n = \exp(i\phi_n)\ket n$. Importantly, Eqs.~\eqref{eq:htrs-dark-state-condition} and~\eqref{eq:htrs-eigenstate-condition} are sufficient for $|\Psi_\mathcal T\rangle_\text{AB}$ to be a pure steady state of Eq.~\eqref{eq:htrs-doubled-system-dynamics} which is the key result of hTRS~\cite{roberts_hidden_2021}. Therefore, we can solve Eqs.~\eqref{eq:htrs-dark-state-condition} and~\eqref{eq:htrs-eigenstate-condition} instead of Eqs.~\eqref{eq:quantum-detailed-balance} which often turns out to more tractable and physically transparent.
    
This framework has led to exact steady-state solutions of several driven-dissipative quantum many-body models~\cite{lingenfelter_exact_2024,yao_hidden_2024,roberts_exact_2023,roberts_competition_2023} which are generally difficult to solve using other, even numerical, methods. 
For a specific example of solving Eqs.~\eqref{eq:htrs-dark-state-condition} and~\eqref{eq:htrs-eigenstate-condition}, we refer to Appendix~\ref{app:DTFIM-purified-steady-state} where we present an exact solution of a dissipative version of the transverse-field Ising model with infinite-range interactions.

\section{A preferred basis for the steady-state purification}\label{sec:special-dark-states}\label{sec:sec4}
    We ultimately wish to use the steady-state structure of open systems with hTRS to directly characterize dynamics.  In Sec.~\ref{subsec:classical-hopping-correspondence} we saw that 
    for a simple one-dimensional classical stochastic model with detailed balance, slow relaxation is \textit{fully} determined by the steady state.  We suggest something analogous can be done in the quantum case for systems with hTRS. 
    To see why this is possible, we first discuss an additional structure in Secs.~\ref{subsec:sec4A} and~\ref{subsec:sec4B} that is inherent to all known hTRS models, and that stems from a weak $U(1)$ symmetry~\cite{albert_symmetries_2014,buca_note_2012} that exists in an appropriate zero-driving limit.  This additional structure allows a concrete mapping to a one-dimensional Hamiltonian ladder model, which in turn enables analytic solutions for the steady-state purification
    $|\Psi_\mathcal T\rangle_\text{AB}$.  We use this mapping to motivate the definition of a potential function and barrier that is determined by $|\Psi_\mathcal T\rangle_\text{AB}$ in Sec.~\ref{subsec:4C}, and then in Sec.~\ref{subsec:sec4D} conjecture that this barrier controls the dissipative gap.  In later sections, we show this conjecture is remarkably accurate in two very different dissipative models.

    
    \subsection{Mapping to a 1D chiral-symmetric two-leg  ladder}\label{subsec:ladder-model}\label{subsec:sec4A}
        As discussed in Sec.~\ref{subsec:htrs} and sketched in Fig.~\ref{fig:fig3}, Lindbladian systems with hTRS can be extended to a cascaded doubled system having a pure steady state $\ket{\Psi_\mathcal T}_\text{AB}$ which is a purification of the original system's steady state. $\ket{\Psi_\mathcal T}_\text{AB}$ must be a dark state, i.e.~it is annihilated by each correlated jump operator of the doubled system
        (cf.~Eq.~\eqref{eq:htrs-dark-state-condition}).  Further, defining 
        $\hat{\mathcal P}_\text{AB}$ to be the operator that exchanges subsystems $A$ and $B$, one can show that the purification must be even under exchange: $\hat{\mathcal P}_\text{AB} \ket{\Psi_\mathcal T}_\text{AB} = \ket{\Psi_\mathcal T}_\text{AB}$
        \footnote{This follows from Eq.~\eqref{eq:def-purification} and the fact that the anti-unitary operator $\hat {\mathcal{T}}$ is diagonal when acting on pointer states $\ket{n}$, as it leaves $\hat{\rho}_{\rm ss}$ invariant~\cite{roberts_hidden_2021}.}.
        It thus follows that $\ket{\Psi_\mathcal T}_\text{AB}$ must lie in the subspace $ \mathbf D_+$ formed by all even-parity dark states, 
        \begin{align}
            \mathbf D_+ = \Big\{\ket v_\text{AB} :\; &(\hat L_{\alpha,\text A} - \hat L_{\alpha,\text B})\ket v_\text{AB} = 0\; \forall\alpha \nonumber\\
            \text{ and } &\hat{\mathcal P}_\text{AB} \ket v_\text{AB} = \ket v_\text{AB}\Big\},
            \label{eq:even-dark-state-manifold}
        \end{align}

\begin{figure}
            \centering
            \includegraphics[width=\linewidth]{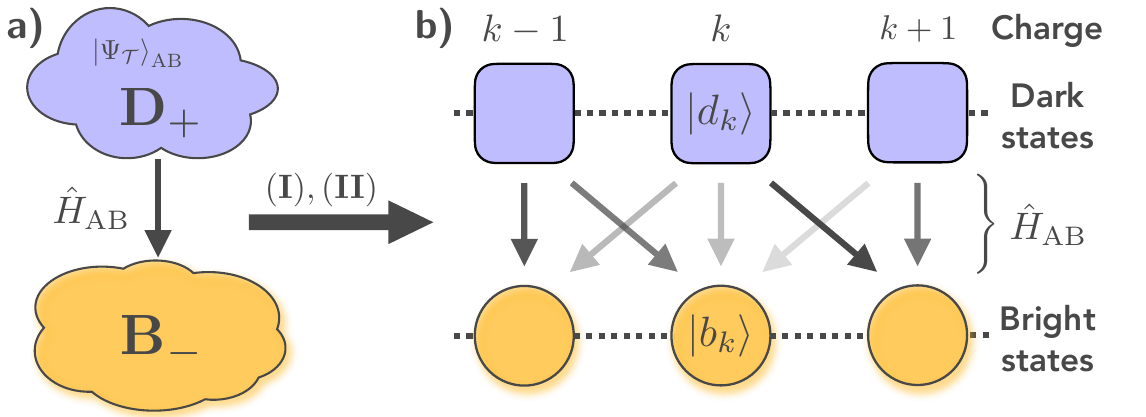}
            \caption{Pictorial illustration of the main arguments presented in Sec.~\ref{subsec:ladder-model}. \textbf{a)} The steady-state purification lives in a dark-state manifold $\mathbf D_+$ (top) which is mapped to an orthogonal manifold $\mathbf B_-$ (bottom) under $\hat H_\text{AB}$. However, these manifolds are unstructured which complicates solving Eq.~\eqref{eq:htrs-eigenstate-condition}. We argue that this problem massively simplifies assuming Eqs.~\eqref{eq:assumption-1} and~\eqref{eq:assumption-2}. \textbf{b)} Under these assumptions we can use a charge degree of freedom to arrange a special set of states in $\mathbf D_+$ (top) and $\mathbf B_-$ (bottom) as two legs of a ladder which are coupled in a local fashion by $\hat H_\text{AB}$ (solid arrows). This simplified structure allows us to obtain analytical expressions for the purified steady state $|\Psi_\mathcal T\rangle_\text{AB}$ using hTRS.
            }
            \label{fig:fig4}
        \end{figure}

        The remaining constraint determining
        $\ket{\Psi_\mathcal T}_\text{AB}$ is that it be an eigenstate of the doubled-system Hamiltonian $\hat{H}_{\rm AB}$, 
        cf.~Eqs.~(\ref{eq:doubled-system-Hamiltonian}),(\ref{eq:htrs-eigenstate-condition}).
    We now argue that with some mild additional assumptions, this Hamiltonian condition maps to a finding an eigenstate of a one-dimensional two-leg ladder Hamiltonian.  This structure will both make finding analytic solutions for the steady state tractable, and will also let us extract a meaningful one-dimensional potential $V(x)$.  The two additional assumptions required (satisfied in all known models with hTRS) are:
        \begin{enumerate}
            \item[\bf (I)] There exists a zero-driving limit where $\hat{H} \rightarrow \hat{H}_0$ in Eq.~\eqref{eq:lindbladian}, where $\hat H_0$ is the drive-free Hamiltonian.  In this limit, $\mathcal{L}$ 
            has at least one weak U(1) symmetry (possibly more), with this symmetry associated with an integer-valued charge $\hat{Q}$.

            \item[\bf (II)] Once driving terms are added to the dynamics, the weak U(1) symmetry is lost, but the driving terms are ``local" in terms of the charge $\hat{Q}$, i.e. the action of operators describing drives can only change $\hat{Q}$ by at most a finite integer $R$.

        \end{enumerate}
        Recall that for a Lindbladian $\mathcal L$ to have a weak U(1) symmetry associated with charge $\hat Q$, it must satisfy~\cite{fazio_many-body_2025}
        \begin{align}
            \mathcal L(\text{e}^{i\phi\hat Q} \hat \rho \text{e}^{-i\phi\hat Q}) = \text{e}^{i\phi\hat Q} \mathcal L(\hat \rho) \text{e}^{-i\phi \hat Q}    
            \label{eq:weak-u1-sym}
        \end{align}
        for any $\phi\in\mathbb R$ and any density matrix $\hat \rho$. We now explain how these two conditions lead to a simple ladder model as depicted in Fig.~\ref{fig:fig4}b.

        To make the general structure clear, we focus on the concrete case where each jump operator $\hat{L}_j$ in the original master equation is an operator whose action lowers the charge $\hat{Q}$ by an integer $q_j$ (i.e.~$q_j$-particle loss).  We thus have 
          \begin{align}
            \left [\hat Q, \hat H_{0}\right ] = 0 \text{ and }
            \left [\hat Q, \hat L_j \right ] = -q_j \hat L_j.
            \label{eq:assumption-1}
        \end{align}
        Note that all known many-body systems with hTRS fulfill this structure, i.e.~all dissipators are pure loss operators 
       ~\cite{roberts_competition_2023,roberts_driven-dissipative_2020,roberts_exact_2023,lingenfelter_exact_2024,yao_hidden_2024}.
        It is easy to show that Eq.~\eqref{eq:assumption-1} implies that the Lindbladian has a weak U(1) symmetry in the zero-drive limit (cf. Eq.~\eqref{eq:weak-u1-sym}). The weak symmetry and corresponding charge of $\mathcal{L}$ directly carry over to the doubled system described by $\mathcal{L}_{AB}$.  In the zero-drive limit, this Lindbladian has a weak U(1) symmetry generated by the charge 
        \begin{align}
            \hat Q_\text{tot} = \hat Q_\text{A} + \hat Q_\text B.
            \label{eq:total-charge-AB}
        \end{align}
        Further, the doubled-system lowering operator satisfies:
         \begin{align}
            [\hat Q_\text{tot}, \hat L_{j,\text{A}} - \hat L_{j,\text{B}}] = -q_j (\hat L_{j,\text{A}} - \hat L_{j,\text{B}}).
            \label{eq:charge-correlated-jump-op}
        \end{align}

        Our goal is to use this structure to simplify the Hamiltonian eigenstate condition that determines $\ket{\Psi_\mathcal T}_\text{AB}$.  First, note that Eq.~\eqref{eq:charge-correlated-jump-op} implies that $\hat Q_{\text{tot}} \mathbf D_+ \subseteq \mathbf D_+$. 
        We can thus find a basis $\ket{d_k}_\text{AB}$ for the dark-state manifold 
        $\mathbf{D}_+$ where each basis state has a definite charge: 
        $\hat Q_\text{tot} \ket{d_k}_\text{AB} = k \ket{d_k}_\text{AB}$.  We start with the simplest case where there are no other quantum numbers, and hence these basis states are unique.  
                
        Now consider the image of $\mathbf{D}_+$ under $\hat{H}_\text{AB}$.  As $\hat{H}_\text{AB}$ is parity-odd under exchange of $A$ and $B$, this image must only have odd-parity states and hence be orthogonal to $\mathbf{D}_+$.  We thus introduce a subspace of odd-parity states $\mathbf{B}_-$ 
        such that 
        \begin{align}
            \hat{H}_\text{AB} \mathbf D_+ \subseteq \mathbf{B}_-
            \label{eq:odd-bright-state-manifold}
        \end{align}
        which is seen in Fig.~\ref{fig:fig4}a.  We take $\mathbf{B}_-$ to be large enough so that $\hat Q_{\text{tot}} \mathbf B_- \subseteq \mathbf B_-$.  We can thus again find a basis for $\mathbf{B}_-$ of states $\ket{b_k}_\text{AB}$ with $\hat Q_\text{tot} \ket{b_k}_\text{AB} = k \ket{b_k}_\text{AB}$.  We again consider the simplest case where these states are unique.

        Going forward, we will view the 1D ordered basis states $\ket{d_k}_\text{AB}$, $\ket{b_k}_\text{AB}$ as defining (respectively) the top and bottom legs of a 1D ladder, cf. Fig.~\ref{fig:fig4}b. We now ask what form the relevant part of $\hat H_\text{AB}$ takes in this basis.  Parity symmetry tells that for all $k,k'$:
        \begin{equation}
            \mel{d_k}{\hat H_\text{AB}}{d_{k'}}_\text{AB} = \mel{b_k}{\hat H_\text{AB}}{b_{k'}}_\text{AB} = 0.
        \end{equation}
        In our ladder picture, this means that there are no horizontal couplings in the  model.  
        By virtue of wanting to solve Eq.~\eqref{eq:htrs-eigenstate-condition}, we only care about how $\hat{H}_\text{AB}$ acts when we start in the dark-state manifold $\mathbf D_+$ (as we are looking for an eigenstate that has no amplitudes on the bottom leg of the ladder).  We thus need to describe couplings between the two ladder legs.  Their form is constrained by our assumption (II) (i.e.~the driving can change charge by at most $R$ units): it will lead to a ladder model with $R$-local couplings.
        We have 
         \begin{align}
            \mel{b_{k'}}{\hat H_\text{AB}}{d_k}_\text{AB} = 0\text{ if }|k-k'| > R 
            \label{eq:assumption-2}
        \end{align}

        Finally, in principle $\hat{H}_\text{AB}$ acting on a state $\ket{b_k}_\text{AB}$
        could create even-parity states that are orthogonal to $\mathbf D_+$.  But given that we want a state localized on the $\ket{d_k}_\text{AB}$ leg of our ladder, we can ignore these couplings.  We can then construct a truncated (but still Hermitian) version of $\hat{H}_\text{AB}$ that only acts within the ladder subspace $\left \{ \ket{d_k}_\text{AB}, \ket{b_{k'}}_\text{AB} \right \}$.  Calling this effective Hamiltonian $\hat{H}'_\text{AB}$, it has no horizontal couplings, and the vertical couplings are given by
        \begin{align}
            \mel{b_{k'}}{\hat H'_\text{AB}}{d_k}_\text{AB} &= 
            \mel{b_{k'}}{\hat H_\text{AB}}{d_k}_\text{AB} \equiv h_{k',k}\label{eq:ladder-model-matrix-elements} \\
            \mel{d_{k}}{\hat H'_\text{AB}}{b_{k'}}_\text{AB} &= 
            \left( \mel{b_{k'}}{\hat H_\text{AB}}{d_k}_\text{AB} \right)^*.
        \end{align}

        We now have a crucial result and simplification that is general to many models:  {\it finding the purification $\ket{\Psi_\mathcal T}_\text{AB}$ reduces to finding a zero-mode of a Hermitian, $R$-local two-leg ladder model which only has terms coupling one leg to the other}.  Formally, our effective Hermitian ladder model necessarily has a chiral (or sublattice symmetry). In the examples discussed in Sec.~\ref{sec:examples-slow-relaxation} the Hermitian ladder model further simplifies to a tight-binding model with a two-site unit cell and inhomogeneous hopping amplitudes as shown in Fig.~\ref{fig:fig5}.

        The effective 1D structure we have in this ladder model now reduces the problem of finding our dark eigenstate to a multi-term recursion relation.  We first write the purification as:
        \begin{align}
             |\Psi_\mathcal T\rangle_\text{AB} = \sum_k \psi_k \ket{d_k}_\text{AB},
             \label{eq:purified-steady-state}
         \end{align}
        i.e. a generic state that only has amplitude on the upper leg of the ladder.  The eigenstate condition then reduces to insisting that for every $k$, $\langle b_k |\hat{H}_\text{AB} |\Psi_\mathcal T\rangle_\text{AB} = 0$.  This gives a constraint equation for each ``position" $k$ on our 1D lattice:
        \begin{align}
            \sum_{j=-R}^R h_{k,k+j} \psi_{k+j} = 0.
            \label{eq:htrs-recursion-relation}
        \end{align}
        In most cases of interest, the charge $\hat{Q}_{\rm tot}$ is bounded from below (i.e.~it represents a number of excitations), and hence the spatial index $k$ of our 1D lattice must satisfy $k \geq 0$, 
        i.e.~at worst we have a semi-infinite lattice.  
        As such, we can solve these equations iteratively starting with the $k=0$ equation, then increasing $k$ in steps of 1.  The equations then have the form of a multi-term recursion relation (which for small $R$ can often be analytically solved).

        The upshot of this subsection is that under relatively mild assumptions, we have found that there is a special privileged basis in which to express the purification of the steady state. Namely this is a basis of dark states $\ket{d_k}_\text{AB}$ that each have a definite charge with respect to $\hat{Q}_{\text{tot}}$.  Furthermore, this basis allows us to construct a local two-leg ladder model, facilitating exact solutions.  We will also argue that the probability amplitudes associated with this special dark-state basis, 
        $P_k = |\psi_k|^2$ (cf.~Eq.~\eqref{eq:purified-steady-state}), play a special role: they define a potential that will have direct relevance to dynamics.

    \subsection{Additional quantum numbers}\label{subsec:additional-quantum-numbers}\label{subsec:sec4B}
        Before proceeding, we discuss how to generalize the two-leg ladder construction from the previous section to the case where the even-parity dark-state manifold $\mathbf D_+$ defined in Eq.~\eqref{eq:even-dark-state-manifold} is characterized by additional quantum numbers beyond charge. This technical concern will become relevant in Sec.~\ref{subsec:sec5B} for a dissipative transverse-field Ising model. In the case of charge degeneracies, basis states can only be specified uniquely by specifying both the charge and the value of additional quantum numbers.  We associate these with an index $\alpha$, and write our basis states now as $\ket{d_k, \alpha}_\text{AB}$.  We have a similar issue with the basis for the $\mathbf B_-$ manifold, and label its fixed-charge basis states as $\ket{b_k,\beta}_\text{AB}$.

        We can still view our system as a 1D two-leg ladder as in Fig.~\ref{fig:fig4}b, but now each lattice site has an additional orbital index $\alpha,\beta$.
        This orbital index complicates solving for the steady-state purification because instead of a single matrix element $h_{k',k}$ connecting the charge-$k$ dark states to the charge-$k'$ bright states, we will have a matrix $h_{k',k}^{\alpha,\beta}$. Still for several models~\cite{roberts_exact_2023,lingenfelter_exact_2024,yao_hidden_2024,roberts_competition_2023} we can find a special orbital basis of dark states $|\tilde d_k\rangle_\text{AB}$ and bright states $|\tilde b_k\rangle_\text{AB}$ with fixed particle number $k$ such that 
        \begin{itemize}
            \item $\hat H_\text{AB}$ only couples $|\tilde d_k\rangle_\text{AB}$ to $|\tilde b_{k'}\rangle_\text{AB}$ and
            \item the steady-state purification $\ket{\Psi_\mathcal T}_\text{AB}$ can be expressed exactly in terms of Eq.~\eqref{eq:purified-steady-state} with $|\tilde d_k\rangle_\text{AB}$.
        \end{itemize}
        These special dark and bright states are given by a particular superposition over the states in $\mathbf D_+$ with fixed charge $k$
        \begin{align}
            |\tilde d_k\rangle_\text{AB} &= \sum_{\alpha} c_{k, \alpha} \ket{d_k,\alpha}_\text{AB},\label{eq:def-special-dark-states}\\
            |\tilde b_k\rangle_\text{AB} &= \sum_{\beta} \tilde{c}_{k,\beta} \ket{b_k, \beta}_\text{AB} \label{eq:def-special-bright-states}
        \end{align}
        in terms of coefficients $c_{k,\alpha}, \tilde c_{k,\beta} \in \mathbb C$. In this case, the situation reduces to the simple case without additional quantum numbers which is shown in Fig.~\ref{fig:fig4}b. While the challenge lies in finding the coefficients $c_{k,\alpha}, \tilde c_{k,\beta}$ that simplify the problem, one can often use additional symmetries and structure of $\hat H_\text{AB}$ to find these coefficients. We will see explicitly in Sec.~\ref{subsec:DTFIM} and Appendix~\ref{app:DTFIM-purified-steady-state} how we can leverage additional system symmetries to find $|\tilde d_k\rangle_\text{AB}$ and $|\tilde b_k\rangle_\text{AB}$.

    \subsection{Defining a potential from the steady-state purification}
    \label{subsec:PotentialDefinition}\label{subsec:4C}
        Using the results of the previous section we have a prescription for obtaining exact steady-state solutions of open quantum systems with hTRS. We found that solving for the steady-state purification $\ket{\Psi_\mathcal T}_\text{AB}$ corresponds to finding the zero-mode of a Hermitian, $R$-local two-leg ladder. The one-dimensional, local nature of this problem should remind us of the classical stochastic dynamics discussed in Sec.~\eqref{subsec:classical-hopping-correspondence} which established that the classical steady state can determine how quickly the system relaxes. Given these similarities, it is natural to wonder whether the correspondence between the steady state and slow relaxation in Eq.~\eqref{eq:classical-dissipative-gap-scaling} extends to the quantum dynamics.


        Since we are interested in a first-order phase transition of the steady state, we need to properly define a thermodynamic limit. Let us introduce the thermodynamic parameter $N$\eg the number of qubits or photons, such that the Lindbladian $\mathcal L_\text{AB}(N)$ explicitly depends on $N$. This $N$-dependence can enter through the number of terms, and/or through parametric dependencies\eg rescaling of interactions in nonlinear terms, as we will see in Sec.~\ref{sec:examples-slow-relaxation}. We will treat $N$ as an effective system size with corresponding extensive observables whose steady-state expectation values grow linearly with $N$. In particular, let us consider the charge picture of Sec.~\ref{subsec:ladder-model} and introduce the total charge of the steady-state purification\ie 
        \begin{align}
            \expval{\hat Q_\text{tot}}_\text{ss} = \mel{\Psi_\mathcal T}{\hat Q_\text{tot}}{\Psi_\mathcal T}_\text{AB} = \sum_k |\psi_k|^2 k.
            \label{eq:steady-state-total-charge}
        \end{align}
        We will pick the parametric $N$-dependence of $\mathcal L_\text{AB}(N)$ such that  $\langle {\hat Q_\text{tot}}\rangle_\text{ss}\sim N$ is an extensive quantity. Then the total charge density 
        \begin{align}
            n = \frac{\expval {\hat Q_\text{tot}}_\text{ss}}{N}
        \end{align}
        will be an intensive quantity in the thermodynamic limit $N\to\infty$. This definition of the thermodynamic limit then closely mirrors the continuum limit of the one-dimensional classical model discussed in Sec.~\ref{subsec:classical-hopping-correspondence}.

        We will now define a non-equilibrium potential that will be meaningful for the slow relaxation as we will explore in the next section. Let $x=k/N$ be the charge density, then we define the non-equilibrium potential $V$ with respect to the steady-state coefficients~$\psi_k$  (cf. Eq.~\eqref{eq:purified-steady-state})
        \begin{align}
            V(x) = -\lim_{N\to\infty}\frac {\log \big (|\psi_k|^2\big )} N.
            \label{eq:htrs-potential}
        \end{align}
        This potential is directly analogous to the classical stochastic model in Sec.~\ref{subsec:classical-hopping-correspondence}. If the ratio $|\psi_{k+1}/\psi_k|=\mathcal O(1)$ in the large-$N$ limit, then this non-equilibrium potential will have a well-defined thermodynamic limit without any diverging behavior (which is ideally ensured by the proper $N$-dependence of $\mathcal L_\text{AB}(N)$).

        Before we try to connect this potential to the slow relaxation, we stress that our ansatz leverages exact expressions for the steady-state coefficients $\psi_k$. Therefore, we can obtain analytical descriptions of this non-equilibrium potential for two paradigmatic models in Sec.~\ref{sec:examples-slow-relaxation}. Still, in defining the non-equilibrium potential $V(x)$ in Eq.~\eqref{eq:htrs-potential} we made a choice to use $|\psi_k|^2$. While we could have chosen a different steady-state probability distribution to define a potential, we show in Fig.~\ref{fig:fig1} and  will explore in detail in Appendix~\ref{app:dtfim-pot-barriers} that our choice correctly captures aspects of slow relaxation (a potential barrier $\Delta V$) where other ansätze for a potential fail. The success of ansatz can be understood as a consequence of a notion of locality and detailed balance in the charge-basis $|d_k\rangle_\text{AB}$.

    \subsection{Conjecture on the slow relaxation}\label{subsec:conjecture-slow-relaxation}\label{subsec:sec4D}
        Let us now study the structure of the non-equilibrium potential $V(x)$ defined by Eq.~\eqref{eq:htrs-potential} at a first-order phase transition. First of all, we quickly recapitulate our discussion in Sec.~\ref{sec:sec2}: we expect that the quantum steady state undergoes a drastic shift at the DPT which, we assume, is signified by a discontinuous jump in the steady-state charge density. From Eq.~\eqref{eq:steady-state-total-charge} we can thus infer that the probability distribution $|\psi_k|^2$ is bimodal in the large-$N$ limit and that the global maximum shifts between the bimodal peaks at the phase transition.  
        This bimodality implies that $V(x)$ has a double-well structure with minima at $x_1$ and $x_2$ separated by a local maximum at $x_*$. Therefore we can define a non-equilibrium potential barrier
        \begin{align}
            \Delta V = V(x_*) - \max_{x_i \in \{x_1, x_2\}}V(x_i)
            \label{eq:free-energy-barrier}
        \end{align}
        between the local maximum $x_*$ and the local, but not global, minimum of $V(x)$. If this were a purely classical stochastic process, then Kramers' theory would suggest that $\Delta V$ serves as a potential ansatz for the scaling of the dissipative gap $\Gamma_\text{diss}$ of the dynamics (cf. Eq.~\eqref{eq:classical-dissipative-gap-scaling}). This leads us to make the following conjecture for the dissipative gap of the quantum dynamics close to a first-order DPT.

        \begin{conjecture}
            Let $\mathcal L(\delta, N)$ describe a Lindblad master equation with a tuning parameter $\delta$ and a system size $N$. We assume that $\mathcal L(\delta, N)$ has hidden time-reversal symmetry\ie fulfills the quantum detailed-balance conditions in Eqs.~\eqref{eq:quantum-detailed-balance}. If the steady state of $\mathcal L(\delta,N)$ undergoes a first-order phase transition at $\delta=0$ in the thermodynamic limit $N\to\infty$, then the dissipative gap $\Gamma_\text{diss}(\delta, N)$\ie the slowest decay rate, of $\mathcal L(\delta,N)$ scales as
            \begin{align}
                \lim_{N\to\infty} \frac{\log \big (\Gamma_\text{diss}(\delta, N)\big )}{N} = -\Delta V(\delta)
                \label{eq:dissipative-gap-scaling-hypothesis}
            \end{align}
            close to the phase transition (small $|\delta|$). Importantly, the non-equilibrium potential barrier $\Delta V(\delta)$ (cf. Eq.~\eqref{eq:htrs-potential},~\eqref{eq:free-energy-barrier}) is \textit{fully} determined by the quantum steady state.

            \label{conj:dissipative-gap}
        \end{conjecture}

        To summarize, Conjecture~\ref{conj:dissipative-gap} proposes that knowledge of the quantum steady state can infer certain relevant properties of the quantum dynamics. In particular we hypothesize that the scaling of the slowest decay rate $\Gamma_\text{diss}$, an important figure of merit for open systems dynamics, is \emph{fully} determined by the steady state and does \emph{not} require knowledge of the microscopic dynamics. This provides physical meaning to the non-equilibrium potential $V(x)$ which is defined with respect to a particular purification of the steady state (cf. Eq.~\eqref{eq:purified-steady-state}).
        Furthermore, this intimate connection between the steady state and the dynamics highlights the similarities to the classical stochastic process discussed in Sec.~\ref{subsec:classical-hopping-correspondence} and the quantum bottleneck theorem introduced in Ref.~\cite{rakovszky_bottlenecks_2024}, but also sets our result apart as an analytical prediction for the dissipative gap. 
        

\section{Slow relaxation in driven-dissipative quantum models}\label{sec:examples-slow-relaxation}\label{sec:sec5}
    In this section we test the predictions of Conjecture~\ref{conj:dissipative-gap} for two widely-studied driven-dissipative quantum models which have hTRS. The first example is the canonical driven-dissipative Kerr oscillator: a single lossy bosonic mode with a Kerr (Hubbard) interaction is subject to a linear drive~\cite{roberts_driven-dissipative_2020,casteels_critical_2017,carde_nonperturbative_2026,lee_real-time_2025,mylnikov_qubit_2025,mylnikov_switching_2025}. The second model is a dissipative version of the transverse-field Ising model with all-to-all interactions~\cite{marcuzzi_effective_2014,roberts_exact_2023,tucker_facilitating_2020,barberena_generalized_2025,song_dissipation-induced_2025,koppenhofer_revisiting_2023,rose_metastability_2016,ostermann_breakdown_2023,paz_driven-dissipative_2021,ptaszynski_dynamical_2024,xiang_switching_2025}. 
    For both models, 
    we will follow the ``recipe" for constructing the steady-state purification $\ket{\Psi_\mathcal T}_\text{AB}$ represented in Sec.~\ref{subsec:ladder-model}.   We first identify the special dark states $\ket{d_k}_\text{AB}$ and special bright states $\ket{b_k}_\text{AB}$. 
    We then use these to construct a Hermitian ladder model, as shown in Fig.~\ref{fig:fig4}b.  The steady-state purification then corresponds to a zero-mode of this ladder model that is localized only on one leg of the ladder, and can be found as per Eq.~\eqref{eq:htrs-recursion-relation}.  Finally, we use the resulting solution to define a non-equilibrium potential according to Eq.~\eqref{eq:htrs-potential}.  Our conjecture is that the potential barrier defined by this potential (cf. Eq.~\eqref{eq:free-energy-barrier}) sets the dissipative gap, something we will confirm by comparing against explicit numerical diagonalization of the Lindbladian.  For both models, we find excellent agreement for all parameter choices tested.    
    
    
    \subsection{Driven-dissipative single-mode cavity}\label{subsec:driven-dissipative-kerr}\label{subsec:sec5A}
    
        We start with a paradigmatic interacting open quantum system: the driven-dissipative Kerr oscillator (see inset in Fig.~\ref{fig:fig6}a).  
        This model exhibits a well-defined first order DPT in an appropriate thermodynamic limit, something that was studied extensively in Ref.~\cite{casteels_critical_2017} and experimentally observed in Refs.~\cite{chen_quantum_2023,beaulieu_observation_2025}. Our goal will be to show that our conjecture lets us predict the dissipative gap near this transition. Ideally our ansatz will offer an alternative approach to more traditional instanton calculations~\cite{carde_nonperturbative_2026,mylnikov_qubit_2025,mylnikov_switching_2025} and provide insights that do not explicitly rely on the dynamics, but rather only the steady state which is a fundamental feature of classical stochastic dynamics.

        \subsubsection{Model and basic features}
            Working in a rotating frame set by the drive frequency, the system Lindbladian $\mathcal L_\text{Kerr}(N)$ is
            \begin{align}
                \mathcal L_\text{Kerr}(N) \hat \rho &= -i \left [\hat H_\text{Kerr}(N), \hat \rho \right ] + \kappa\mathcal D[\hat a]\hat \rho
                \label{eq:Kerr-qme}
            \end{align}
            where $\hat \rho$ is the density matrix of the bosonic mode $\hat a$, and
            \begin{align}
                \hat H_\text{Kerr}(N) = \Delta \hat a^\dagger \hat a -\frac {\overline U} {2N} \hat a^\dagger \hat a^\dagger \hat a \hat a + \sqrt N \,\overline F (\hat a + \hat a^\dagger).
                \label{eq:kerr-hamiltonian}
            \end{align}
            Here $\overline F$ is a linear drive with detuning $\Delta$, $\overline U$ is the Kerr interaction and $\kappa$ is the single-photon decay rate\footnote{W.l.o.g. we assume all parameters $\{\overline F,\Delta, \overline U, \kappa\}$ to be positive.}; in Appendix~\ref{app:kerr-comparison-phase-space-methods} we extend our results to additionally include two-photon driving and loss. To obtain a well-defined thermodynamic limit, we
            follow Ref.~\cite{casteels_critical_2017} and introduce a parameter $N$ in the Hamiltonian to play the role of an effective system size.  These parameter scalings ensure that for large $N$, the average photon number density  $n \equiv \expval{\hat a^\dagger \hat a}_\text{ss}/N$ scales intensively (i.e.~it is independent of $N$ in the large-$N$ limit). 

            A key requirement for our general formulation in Sec.~\ref{sec:sec4} is that our model has a weak U(1) symmetry in an appropriate undriven limit.  This is indeed the case for $\mathcal{L}_{\rm Kerr}$:  it has such a weak symmetry generated by $\hat Q = \hat a^\dagger\hat a$ in the  limit $\overline F \rightarrow 0$. 
            The non-equilibrium steady state of the model is only interesting of course when $\overline F \neq 0$.  In this case there is a competition between the driving, interactions and loss, and the steady state will generically have a non-zero photon density $n$.  At a semiclassical level, the nonlinearity $\overline U$ leads to optical bistability~\cite{lugiato_optical_1983,bonifacio_photon_1978} (see App.~\ref{app:kerr-mean-field} for a mean-field description).
    
            \begin{figure}
                \centering
                \includegraphics[width=\linewidth]{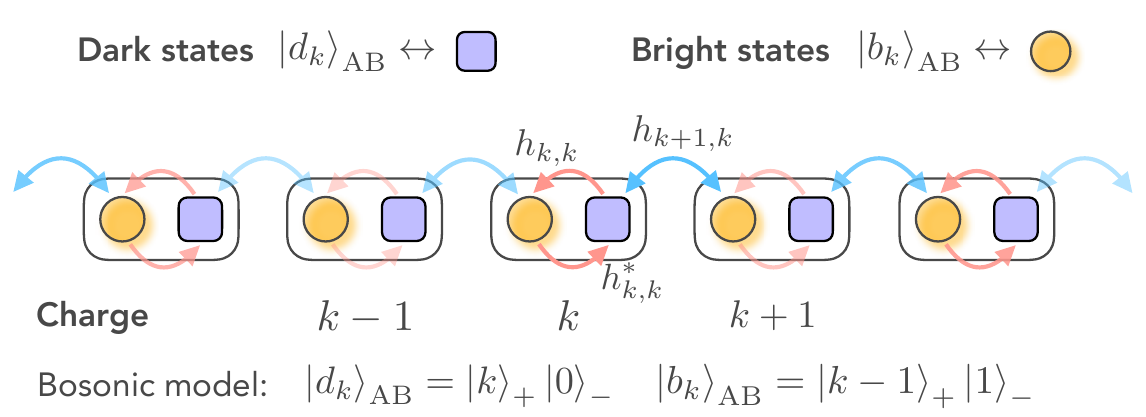}
                \caption{Mapping the Hermitian ladder model to a tight-binding chain with inhomogeneous hopping amplitudes for the nonlinear cavity model in Sec.~\ref{subsec:driven-dissipative-kerr}. Using the coupling coefficients $h_{k,k}$ and $h_{k+1,k}$ defined in Eq.~\eqref{eq:kerr-coupling-coeffs} the effective Hamiltonian $\hat H_\text{AB}'$ has the structure of a tight-binding chain with a two-site unit cell, corresponding to the dark states $\ket{d_k}_\text{AB}$ and bright states $\ket{b_k}_\text{AB}$ with charge $k$.}
                \label{fig:fig5}
            \end{figure}

            The full quantum theory does not have any steady-state bistability, and the unique steady state $\hat \rho_\text{ss}$ of Eq.~\eqref{eq:Kerr-qme} can be found analytically.  This was first derived using phase space methods~\cite{drummond_quantum_1980}, and more recently shown to be an example of hTRS~\cite{stannigel_driven-dissipative_2012,roberts_driven-dissipative_2020}.

        \subsubsection{Constructing the steady-state purification}
            
            We now embark on using the hTRS structure of our model to predict the dissipative gap.  We first need to solve for a purification of the steady state using the doubled-system formalism introduced in Sec.~\ref{subsec:htrs}. Note that our construction reproduces the result of Ref.~\cite{roberts_driven-dissipative_2020}; however, by formulating the solution using the general recipe of Sec.~\ref{subsec:ladder-model}, one gains insight into the general structure and connections to other hTRS models.
            
            In the doubled setup we have both the original system $A$ (lowering operator $\hat a$) and the mirror system $B$ (lowering operator $\hat b$). The Hamiltonian of system $B$ is the negative of the Hamiltonian in system $A$; the Hamiltonian of the doubled system is given in Eq.~\eqref{eq:doubled-system-Hamiltonian}. The two systems collectively couple to the environment via the jump operator $\hat c_- = (\hat a - \hat b)/\sqrt 2$ (up a prefactor) as we can see from the doubled-system Lindbladian defined in Eq.~\eqref{eq:htrs-doubled-system-dynamics}.
            The doubled system also has a weak U(1) symmetry in the zero-drive limit, generated by the total charge (cf. Eq.~\eqref{eq:total-charge-AB}) which is given here by
            \begin{align}
                \hat Q_\text{tot} = \hat a^\dagger \hat a+ \hat b^\dagger \hat b = \hat c_+^\dagger \hat c_+ + \hat c_-^\dagger \hat c_-.
            \end{align}
            In addition to the odd mode $\hat c_-$ we have introduced the even mode $\hat c_+ =(\hat a + \hat b) /\sqrt 2$ for convenience; here ``even" and ``odd" refer to the symmetry under exchange of $\hat a$ and $\hat b$. The Fock states $\ket k_+\ket \ell_-$ of the even and odd mode have definite charge $k+\ell$ and span the entire doubled-system Hilbert space.


            We now follow the steps in the general procedure outlined in Sec.~\ref{subsec:ladder-model} and at the start of this section.
            The first step involves constructing the even-parity, dark-state manifold $\mathbf D_+$ in Eq.~\eqref{eq:even-dark-state-manifold} which includes the special dark states $\ket{d_k}_\text{AB}\in \mathbf D_+$. It is straightforward to find that
            \begin{align}
                \mathbf D_+ = \{\ket v: (\hat a - \hat b)\ket v = \hat c_- \ket v = 0\}
            \end{align}
            is spanned by Fock states of the even mode $\hat c_+$ 
            \begin{align}
                \ket{d_k}_\text{AB} = \frac 1 {\sqrt{k!}}\left (\hat c_+^\dagger \right )^k \ket{0}_+\ket{0}_- \equiv\ket{k}_+\ket 0_-.
            \end{align}
            These states have definite total charge $k\in \mathbb N$ with respect to $\hat Q_\text{tot}$ and are symmetric under exchange of $A$ and $B$. We will use these states to express the steady-state purification $\ket{\Psi_\mathcal T}_\text{AB}$.
            
            Now, as the next step, we turn our attention to the special bright states $\ket{b_k}_\text{AB}\in \mathbf B_-$ (cf. Eq.~\eqref{eq:odd-bright-state-manifold}), as they will form the basis for our ladder model together with the special dark states. These special bright states are defined as a basis for the image of $\hat H_\text{AB}$ acting on $\mathbf D_+$\ie $\hat H_\text{AB}\mathbf D_+\subseteq \mathbf B_-$. Therefore we now analyze how the terms in $\hat H_\text{AB}$ act on the dark states $\ket{d_k}_\text{AB}$. The driving terms ($\hat a - \hat b=\hat c_-$ and $\hat a^\dagger - \hat b^\dagger=\hat c_-^\dagger$) in the doubled-system Hamiltonian $\hat H_\text{AB}$ create (or destroy) exactly one photon in the $\hat c_-$ mode. Furthermore, it is straightforward to confirm that the charge-conserving terms (e.g.~$\hat a^\dagger \hat a - \hat b^\dagger\hat b=\hat c_+^\dagger \hat c_- + \hat c_-^\dagger\hat c_+$) in $\hat H_\text{AB}$ convert exactly one photon from the even to the odd mode (or the other way around). The bright states in the manifold $\mathbf B_-$ are then identified as
            \begin{align}
                |b_k\rangle_\text{AB} = |k-1\rangle_+|1\rangle_-.
            \end{align}
            
            Equipped with these states let us explicitly construct our ladder model. The next step of the ``recipe" in Sec.~\ref{subsec:ladder-model} requires us to determine the matrix elements $h_{k',k}$ of the ladder Hamiltonian $\hat H_\text{AB}'$ which are defined in Eq.~\eqref{eq:ladder-model-matrix-elements}. A short calculation finds that the only nonzero values are
            \begin{align}
                h_{k,k} &= \mel{b_{k}}{\hat H_\text{AB}}{d_k}_\text{AB} = \left [ \Delta - \frac{\overline U}{2N}(k-1) - i\frac \kappa 2 \right ] \sqrt k, \nonumber\\
                h_{k+1,k} &= \mel{b_{k+1}}{\hat H_\text{AB}}{d_k}_\text{AB} = \sqrt{2N}\, \overline F.
                \label{eq:kerr-coupling-coeffs}
            \end{align}
            This means that the ladder in Fig.~\ref{fig:fig4}b only features connections directly vertically ($k\to k$) and diagonally to the right ($k\to k+1$) such that the interaction range on the lattice is $R=1$. In fact the model is equivalent to a one-dimensional tight-binding chain with inhomogeneous nearest-neighbor hopping as seen in Fig.~\ref{fig:fig5}.
            
            This leads us to the final step of our recipe of Sec.~\ref{subsec:ladder-model} to construct the steady-state purification $\ket{\Psi_\mathcal T}_\text{AB}$. From our general discussion it follows that $\ket{\Psi_\mathcal{T}}_\text{AB}$ is the zero-eigenvalue eigenstate of the ladder model\ie $\hat H_\text{AB}'\ket{\Psi_\mathcal T}_\text{AB}=0$. Crucially, we know that $\ket{\Psi_\mathcal T}_\text{AB}$ is only localized on the sites which correspond to the special dark states $\ket{d_k}_\text{AB}$ (cf. Eq.~\eqref{eq:purified-steady-state}). We showed in Sec.~\ref{subsec:ladder-model} that the steady-state coefficients $\psi_k$ in the dark-state basis are determined by the recursion relation in Eq.~\eqref{eq:htrs-recursion-relation}. Here we explicitly obtain
            \begin{align}
                \left [ \Delta - \frac{\overline U}{2N}(k-1) - i\frac \kappa 2 \right ] \sqrt k \psi_k + \sqrt{2N}\, \overline F \psi_{k-1} = 0
                \label{eq:kerr-recursion-relation}
            \end{align}
            which we can use to iteratively find $\psi_k$ starting from $\psi_0$. 
    
            \begin{figure}
                \centering
                \includegraphics[width=\columnwidth]{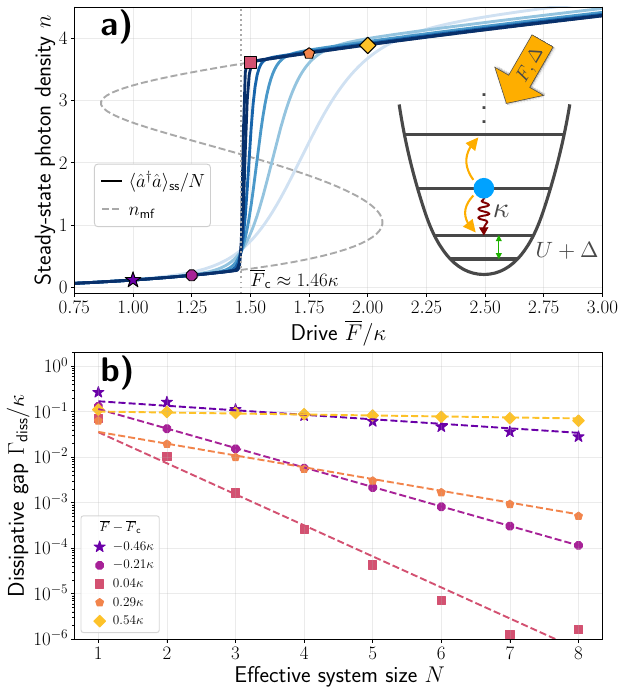}
                \caption{Optical bistability and slow relaxation in the driven-dissipative nonlinear cavity model. \textbf{a)} The steady-state photon density $n=\langle\hat a^\dagger \hat a\rangle_\text{ss}/N = \Tr(\hat a^\dagger \hat a \hat \rho_\text{ss})/N$ (solid lines) exhibits a first-order DPT in the limit $N\to\infty$, here shown for $N=1,2,4,8,16,32$ (increasing with shade). At the critical driving strength $\overline F_\text c \approx 1.46\kappa$ (dotted line) the system jumps from one bistable state $n_\text{mf}$ to another which are determined by Eq.~\eqref{eq:Kerr-mean-field} (dashed line). The enlarged markers correspond to the drives shown in panel b. The inset shows an illustration of this model. \textbf{b)} The dissipative gap $\Gamma_\text{diss}$ close to the first-order DPT obtained by numerical diagonalization (using a photon cutoff of $50$ bosons) for different drives $\overline F - \overline F_\text c$. The scaling of $\Gamma_\text{diss}$ exhibits good agreement with the prediction in Eq.~\eqref{eq:dissipative-gap-scaling-hypothesis} with the potential barrier $\Delta V$ calculated using $V_\text{Kerr}(n)$ (cf. Eq.~\eqref{eq:Kerr-free-energy}). Parameters $\overline U = \kappa, \Delta = 3\kappa$ are chosen to match Ref.~\cite{casteels_critical_2017}.}
                \label{fig:fig6}
            \end{figure}

        \subsubsection{Results for slow relaxation at first-order DPT}
            Equipped with the steady-state coefficients $\psi_k$, we now define a non-equilibrium potential according to Eq.~\eqref{eq:htrs-potential} which we will eventually use to extract a potential barrier $\Delta V$. We will then compare this potential barrier with the scaling of the dissipative gap $\Gamma_\text{diss}$ with $N$ to test Conjecture~\ref{conj:dissipative-gap}. First we plug the solution of Eq.~\eqref{eq:kerr-recursion-relation} into Eq.~\eqref{eq:htrs-potential} and arrive at one of our key results of this section: the non-equilibrium potential $V_\text{Kerr}(n)$ of the photon density $n\equiv k/N$
            \begin{align}
                V_\text{Kerr}(n) = -\int_0^n dn' \log \left [\frac{8}{n'}\frac{\overline F ^2}{(\overline U n' - 2 \Delta)^2 + \kappa^2} \right ].
                \label{eq:Kerr-free-energy}
            \end{align}
            Depending on parameter choices, this potential has either a unique minimum or a double-well structure with two minima. The double-well structure is a consequence of the nonlinearity $\overline U$ and occurs in a parameter regime where there is bistability at the mean-field level. In this regime we expect a first-order DPT corresponding to optical bistability~\cite{bonifacio_photon_1978,lugiato_optical_1983}. Therefore we can define a potential barrier $\Delta V$ close to the phase transition following Eq.~\eqref{eq:free-energy-barrier} which we can compare with numerics to test Conjecture~\ref{conj:dissipative-gap}:
            \begin{align}
                \lim_{N\to\infty} \frac{-1} N \log\big [\Gamma_\text{diss}(\overline F, \dots)\big ] = \Delta V(\overline F,\dots).
                \label{eq:kerr-potential-barrier-main}
            \end{align}
            We explicitly highlight the parameter dependence on $\overline F$ that we will subsequently study in Fig.~\ref{fig:fig6}b; we use ellipses to indicate additional parameter dependencies. Agreement between our ansatz for a potential barrier $\Delta V$ and the dissipative gap $\Gamma_\text{diss}$ would distinguish our potential from an inexhaustible number of ansätze for a potential which are not guaranteed to capture the slow relaxation properly.
    
            First, let us establish a parameter regime where a first-order DPT occurs: in Fig.~\ref{fig:fig6}a we compare the steady-state photon density $n=\expval{\hat a^\dagger \hat a}_\text{ss}/N$ with the mean-field prediction $n_\text{mf}$ (defined in Appendix~\ref{app:kerr-mean-field}) to confirm the appearance of a first-order DPT in the limit $N\to\infty$. In the mean-field analysis (dashed, gray line) there are clear signs of optical bistability in a parameter region with 3 different steady-state solutions for a fixed drive $\overline F$. Meanwhile the exact quantum steady-state solution (solid, blue lines) is always unique. As $N$ increases, one obtains a first-order phase transition between two mean-field, steady-state solutions (corresponding to either a low or high photon-density state). In the thermodynamic limit $N\to\infty$ we find the critical driving strength $\overline F_\text c \approx 1.46\kappa$ as the location of the phase transition.
            
            Now let us take a look at the dissipative gap close to the first-order DPT. In Fig.~\ref{fig:fig6}b we show numerical data for the dissipative gap of $\mathcal L_\text{Kerr}(N)$, as we tune the driving strength $\overline F$ across the bistable parameter region and therefore the phase transition; the dissipative gap is calculated using an Arnoldi method to obtain the two eigenvalues of the Lindbladian with the largest real part (with the difference in real part being $\Gamma_{\rm diss}$). Upon increasing the system size $N$ the dissipative gap is exponentially suppressed with $N$ as predicted by Eq.~\eqref{eq:dissipative-gap-scaling-hypothesis}. Crucially the slope is well-approximated by the non-equilibrium potential barrier $\Delta V$ calculated from the steady state via Eqs.~\eqref{eq:Kerr-free-energy} and~\eqref{eq:kerr-potential-barrier-main}. Note that our predictions only concern the scaling with $N$ and not the prefactor of $\Gamma_{\rm diss}$ which is fitted to best match the data to the analytical curve.  The correspondence works over a range of parameters, e.g. we see agreement over a range $\Delta V\approx 0.05\text{--}1.5$ across different values of $\overline F$ close to $\overline F_\text c$. The agreement with the predicted exponential scaling with $N$ might be a bit surprising for 
            the relatively small values of $N$ studied here; we note that this scaling behavior for small $N$ was already observed in Ref.~\cite{carde_nonperturbative_2026}. While we do observe small deviations from our prediction $\Delta V$ for the slope, these are likely numerical artifacts or finite-size effects.
            In summary the agreement in Fig.~\ref{fig:fig6}b provides good evidence for Conjecture \ref{conj:dissipative-gap} and lets us characterize the key properties of the DPT such as the location of the phase transition $\overline F_\text c$ and the potential barrier $\Delta V$ analytically without a need for finite-size scaling.

        An outstanding question remains how our ansatz for $\Gamma_\text{diss}$ compares to recent instanton calculations~\cite{carde_nonperturbative_2026,mylnikov_qubit_2025}. 
        In Appendix~\ref{app:kerr-comparison-phase-space-methods} we confirm that our potential barrier $\Delta V$ agrees with these Keldysh path-integral techniques and find excellent agreement for different parameter settings. 
        This is a priori surprising, as our ansatz is based completely on the steady state, whereas the instanton calculation is based on a model of the dynamics.  
        This nontrivial agreement validates our approach and establishes our ansatz as an alternative approach to studying the slow relaxation beyond the limit of finite-size numerics or instanton calculations. In particular our approach opens a more intriguing direction: our ansatz is readily applicable to models that exhibit hTRS but lack rigorous instanton calculations for the dissipative gap.

    \subsection{Dissipative transverse-field Ising model}\label{subsec:DTFIM}\label{subsec:sec5B}
        We now study a truly many-body model, the dissipative version of the transverse-field Ising model (DTFIM) with all-to-all interactions as illustrated in the inset of Fig.~\ref{fig:fig7}. This model has a first-order dissipative phase transition~\cite{marcuzzi_universal_2014}, and also exhibits hidden time-reversal symmetry~\cite{roberts_exact_2023};
        it is thus exactly the kind of system that is suited to our analysis.  We can again apply the machinery of Sec.~\ref{sec:sec4} to obtain the steady-state purification and make an ansatz for the scaling of the dissipative gap in terms of a potential barrier. 
        We show via comparison to numerics that the  steady-state purification can be used to predict the dissipative gap near the first-order DPT (via Conjecture~\ref{conj:dissipative-gap}).
        \subsubsection{Model and phenomenology}
            The system is composed of $N$ qubits (with Pauli matrices $\hat \sigma^\alpha_j$, $\alpha=\text{x,y,z}$) interacting via all-to-all Ising interactions and each subject to a coherent Rabi drive.
            We also include Markovian loss processes, both single-spin and collective.  Defining collective spin operators $\hat S_\alpha = \sum_j \hat \sigma^\alpha_j / 2$, and working in a rotating frame set by the common drive frequency, the dynamics are described by a Lindblad master equation (cf.~Eq.~\eqref{eq:lindbladian}) where the Lindbladian $\mathcal{L}_\text{DTFIM}(N)$ is given by
            \begin{align}
                \mathcal L_\text{DTFIM}(N) \hat \rho &= -i\left [\frac {\overline J} N\hat S_\text z^2 + \Delta \hat S_\text z + \Omega \hat S_\text x, \hat \rho\right ]\nonumber\\ 
                &+ \gamma\sum_{j=1}^N \mathcal D[\hat \sigma^-_j]\hat \rho + \frac{\overline\Gamma}N \mathcal D [\hat S_-]\hat \rho.
                \label{eq:DTFIM-qme}
            \end{align}
            Here, $\hat \sigma^-_j = (\hat \sigma^\text x_j -i \hat \sigma^\text y_j)/2 = |0_j\rangle\langle1_j|$ is the local lowering operator describing qubit decay at rate $\gamma$ and $\hat S_- = \sum_j \hat\sigma^-_j$ is the collective lowering operator which captures collective decay of the ensemble at rate $\overline \Gamma$. $\overline J$ denotes the Ising interaction strength while the detuning $\Delta$ and the driving strength $\Omega$ are parameters of the Rabi drive. The parameter dependence of $\mathcal L_\text{DTFIM}(N)$ on the number of qubits $N$ ensures a well-defined thermodynamic limit as $N\to\infty$. 
            
    
            \begin{figure}
                \centering
                \includegraphics[width=\columnwidth]{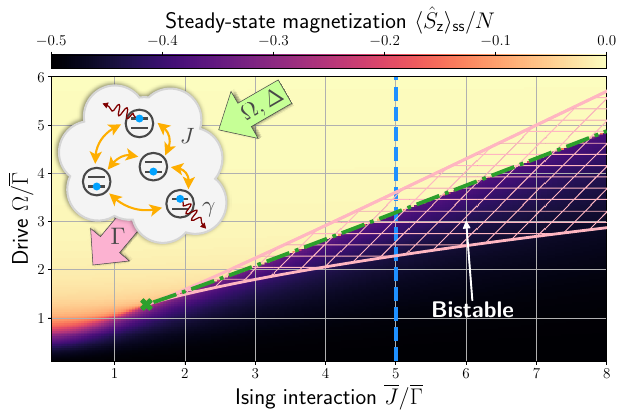}
                \caption{Steady-state phase diagram of the dissipative transverse-field Ising model (DTFIM) with all-to-all interactions. The inset shows an illustration of the model in the top left corner. The steady-state magnetization $\langle \hat S_\text z\rangle_\text{ss}=\Tr(\hat S_\text z \hat \rho_\text{ss})$ exhibits a critical point (green cross) corresponding to a second-order DPT for sufficiently strong Ising interactions. This critical point is connected to a bistable region (pink hashes) between a polarized state for small $\Omega$ and a depolarized state  for large $\Omega$. A line of first-order DPTs (green dash-dotted line) emerges between these two states in the middle of this bistable region. 
                The blue dashed line corresponds to $\overline J = 5\overline \Gamma$ which is analyzed in Fig.~\ref{fig:fig8}; the remaining parameters are $N=512, \Delta = 0, \gamma =0.5\overline \Gamma$.}
                \label{fig:fig7}
            \end{figure}
    
            The competition between driving, interactions and dissipation in this model has garnered significant theoretical and experimental interest since the late 1970s~\cite{puri_exact_1979,carmichael_analytical_1980}. The fully collective limit ($\gamma=0$) is now commonly referred to as the collective resonance fluorescence model in the literature~\cite{song_dissipation-induced_2025,agarwal_directional_2024} and can exhibit persistent oscillatory behavior in the large-$N$ limit, a phenomenon known as a boundary time crystal~\cite{iemini_boundary_2018}. Including single-qubit relaxation ($\gamma>0$) breaks the collective symmetry, resulting in an even richer phenomenology of non-equilibrium dynamics and steady-state physics~\cite{marcuzzi_universal_2014,marcuzzi_non-equilibrium_2015,paz_driven-dissipative_2021,xiang_switching_2025}. The non-equilibrium physics of these models has been explored in recent experiments with atomic ensembles~\cite{ferioli_non-equilibrium_2023,song_dissipation-induced_2025}. While there has been recent analytical progress~\cite{barberena_generalized_2025,paz_driven-dissipative_2021}, the model in Eq.~\eqref{eq:DTFIM-qme} is particularly amenable to numerical simulation up to $N\lesssim 200$ qubits due to a weak permutational symmetry~\cite{chase_collective_2008,shammah_open_2018,paz_driven-dissipative_2021}. We will leverage these efficient numerical techniques to study the scaling of the dissipative gap with system size $N$, and compare against our analytic conjecture. 

            We start by presenting the basic phase diagram of the model in Fig.~\ref{fig:fig7}.  In the limit $\Omega\to0$ the Lindbladian $\mathcal L_\text{DTFIM}(N)$ has a weak U(1) symmetry with charge $\hat Q = \hat S_\text z + N/2$, and the steady state is just vacuum. For $\Omega>0$ the coherent drive competes with the dissipation ($\gamma, \overline \Gamma$) leading to a finite-density steady state. 
             As we show in Appendix~\ref{app:dtfim-mean-field} interactions at the mean-field level lead to steady-state bistability. In contrast, the quantum model always has a unique steady state~\cite{roberts_exact_2023}, with a first-order DPT between mean-field solutions emerging in the large-$N$ limit.  
            
            In Fig.~\ref{fig:fig7} we plot the steady-state magnetization obtained from the exact solution~\cite{roberts_exact_2023},  and compare with the results of mean-field theory (cf. Appendix~\ref{app:dtfim-mean-field}). The bistable parameter region, wherein the mean-field steady-state magnetization $s_\text{z,mf}$ (cf. Eq.~\eqref{eq:DTFIM-mean-field}) is not uniquely defined, is outlined and highlighted in pink. In the middle of this region the exact steady-state solution exhibits a discontinuous jump from one mean-field solution to another, a key characteristic of a first-order DPT. Fig.~\ref{fig:fig7} shows a line of first-order phase transitions (green dash-dotted line) ending in a continuous phase transition; this is reminiscent of the liquid-vapor phase diagram and consistent with previous studies~\cite{marcuzzi_universal_2014,roberts_exact_2023}.

        \subsubsection{Constructing the steady-state purification}
            We now use the recipe of Sec.~\ref{subsec:ladder-model} to construct an exact description of the NESS via its purification 
            $\ket{\Psi_\mathcal T}_\text{AB}$ (cf.~Eq.~\eqref{eq:purified-steady-state}), showing the explicit connection to a Hermitian ladder model; further details are provided in App.~\ref{app:DTFIM-purified-steady-state}.  This connection helps elucidate the general structure and properties of the NESS.  It differs from the derivation in Ref.~\cite{roberts_exact_2023}, providing additional insights.   
            
            We first define the doubled system that will host the purification of the NESS: each of the $N$ qubits in system $A$ has a partner qubit in system $B$, yielding $N$ qubit pairs. The doubled-system Hamiltonian is given in Eq.~\eqref{eq:doubled-system-Hamiltonian} and the correlated jump operators (cf. Eq.~\eqref{eq:htrs-doubled-system-dynamics}) are
            \begin{align}
                \hat L_{0,\text{AB}} &= \sqrt{\overline \Gamma / N} (\hat S_{-,\text A} - \hat S_{-,\text B}),\nonumber\\
                \hat L_{j,\text{AB}} &= \sqrt \gamma (\hat \sigma^-_{j,\text A} - \hat \sigma^-_{j,\text B}) \quad j=1,\dots,N.
                \label{eq:dtfim-doubled-system-jump-ops}
            \end{align}
            The total charge of the doubled system (cf. Eq.~\eqref{eq:total-charge-AB}) is
            \begin{align}
                \hat Q_\text{tot} = \sum_{j=1}^N \left ( \frac{\hat \sigma^\text z_{j,\text A} + \hat \sigma^\text z_{j,\text B}}{2} + 1\right ) = \hat S_\text{z,A} + \hat S_\text{z,B} + N
                \label{eq:dtfim-total-charge}
            \end{align}
            which for $\Omega\to0$ generates a weak U(1) symmetry in the full doubled-system Lindbladian.
            
            We now follow the recipe of Sec.~\ref{subsec:ladder-model} for constructing the steady-state purification $\ket{\Psi_\mathcal T}_\text{AB}$ in the form of Eq.~\eqref{eq:purified-steady-state}. 
            The first step is to use the dark-state condition in Eq.~\eqref{eq:htrs-dark-state-condition} to find the exchange-even manifold of states $\mathbf D_+$ that are dark with respect to the dissipation (cf.~Eq.~\eqref{eq:even-dark-state-manifold}). 
            For $\gamma>0$ the dark-state condition (cf. Eq.~\eqref{eq:htrs-dark-state-condition}) constrains the local Hilbert space of each qubit pair on site $j$ to a two-dimensional subspace spanned by
            \begin{align}
                |\tilde 0_j \rangle_\text{AB} &= |0_{j}\rangle_\text{A}|0_{j}\rangle_\text{B},\nonumber\\
                |\tilde 1_j\rangle_\text{AB} &= \left(|1_{j}\rangle_\text A|0_{j}\rangle_\text B + |0_{j}\rangle_\text A|1_{j}\rangle_\text B \right) / \sqrt 2.
            \end{align}
            The span of these states are also dark with respect to the collective jump operator $\hat L_{0,\text{AB}}$.
            We thus have that $\mathbf D_+ = \text{span}\left (\left \{\ket{\tilde 0}_\text{AB}, \ket{\tilde 1}_\text{AB} \right \}^{\otimes N}\right )$.
            We see that $|\tilde 0\rangle_\text{AB}$ contributes no charge to $\hat Q_\text{tot}$ while $|\tilde 1\rangle_\text{AB}$ contributes a single charge. 

            To construct the upper rung of our ladder model, we next need to identify for each $k$ a state in $\mathbf D_+$ having a definite total charge $k$ (cf.~Eq.~\eqref{eq:purified-steady-state}). Unlike the nonlinear cavity model of Sec.~\ref{subsec:driven-dissipative-kerr}, such states are not unique (as there are many ways to pick which qubit pairs are excited).  
            However, within each subspace of charge-$k$ dark states, only one state will be relevant to us, namely the state that is permutation-invariant.  This follows from our model's weak permutation symmetry and the requirement that the NESS be permutation invariant\footnote{Since the single-system Lindbladian $\mathcal L_\text{DTFIM}(N)$ (cf. Eq.~\eqref{eq:DTFIM-qme}) is invariant under \emph{any} permutation of qubits on site $i$ and $j$ and its steady state is unique~\cite{roberts_exact_2023}, the single-system steady state $\hat \rho_\text{ss}$ is also permutationally invariant. $\hat \rho_\text{ss}$ inherits this property from the steady-state density matrix $\hat \rho_\text{ss,AB}$ of the doubled system, so it must be invariant under permuting qubit pairs on site $i$ and $j$. Importantly, the steady state $\hat \rho_\text{ss,AB}$ is pure and given by $\ket{\Psi_\mathcal T}_\text{AB} \in \mathbf D_+$.}. 
            The only permutationally invariant states $|\tilde d_k\rangle_\text{AB}$ in $\mathbf D_+$ with fixed charge $k$ (cf. Eq.~\eqref{eq:def-special-dark-states}) are generalized Dicke states~\cite{hartmann_generalized_2016,dicke_coherence_1954,raveh_dicke_2024} which are defined by
            \begin{align}
                \ket{\tilde d_k}_\text{AB} &= \frac{1}{\sqrt{\binom N k}}\sum_{\pi} \mathcal P_\pi \ket{\tilde 1}_\text{AB}^{\otimes k} \ket{\tilde 0}_\text{AB}^{\otimes N-k}.
                \label{eq:dtfim-dicke-states}
            \end{align}
            The sum is over permutations $\pi$, and $\mathcal{P}_\pi$ permutes the ordering in the product state $|\tilde 1\rangle_\text{AB}^{\otimes k} |\tilde 0\rangle_\text{AB}^{\otimes N-k}$. We  
            only sum over distinct permutations $\pi$ to avoid double counting. For example for $N=3$ and $k=2$
            \begin{align*}
                \ket{\tilde d_2}_\text{AB} = \frac{1}{\sqrt 3} \left (|\tilde 1_1,\tilde 1_2,\tilde 0_3\rangle + |\tilde 1_1,\tilde 0_2,\tilde 1_3\rangle + |\tilde 0_1,\tilde 1_2,\tilde 1_3\rangle\right ).
            \end{align*}
            These states form the upper rung of our ladder model, with the charge $k$ acting as an effective position coordinate.  
            
            We next need to construct the lower leg of the ladder model of Sec.~\ref{subsec:ladder-model}; this entails finding the special bright states $|\tilde b_k\rangle_\text{AB}\in \mathbf B_-$ in Eq.~\eqref{eq:def-special-bright-states}. These states form a basis of the image of $\hat H_\text{AB}$ acting on the span of the upper-rung states $|\tilde d_k\rangle_\text{AB}$.
            One finds that $\hat H_\text{AB}$ acting on a product-state dark state creates a single ``defect", where one qubit dimer is converted from either $\ket{\tilde 0},\ket{\tilde 1}$ to the odd-parity singlet state:
            \begin{align}
                \ket{\tilde s_j}_\text{AB} = (\ket{1_j}_\text A\ket{0_j}_\text B - \ket{0_j}_\text A\ket{1_j}_\text B)/\sqrt 2
            \end{align}
            Note that this state has a definite total charge $Q_{\rm tot} = 1$.  
            Further, since $\hat H_\text{AB}$ preserves permutational symmetry, the image of each $|\tilde d_k\rangle_\text{AB}$ (cf. Eq.~\eqref{eq:dtfim-dicke-states}) under $\hat H_\text{AB}$ is a permutationally symmetric state with exactly one ``defect" $|\tilde s\rangle_\text{AB}$. This uniquely identifies the special bright states (cf. Eq.~\eqref{eq:def-special-bright-states}) as
            \begin{align}
                \ket{\tilde b_k}_\text{AB} = \frac{1}{\sqrt{\mathcal N}}\sum_{\pi} \mathcal P_\pi \ket{{\tilde s}_j}_\text{AB} |\tilde 1\rangle_\text{AB}^{\otimes k-1} |\tilde 0\rangle_\text{AB}^{\otimes N-k}
                \label{eq:dtfim-special-bright-states}
            \end{align}
            where $\ket{\tilde b_k}_\text{AB}$ has $Q_{\rm tot} = k$, and $\mathcal N = N! / [(k-1)!(N-k)!]$. Again, we are only summing over distinct permutations $\pi$ for an arbitrary ordering of the state $\ket{{\tilde s}_j}_\text{AB} |\tilde 1\rangle_\text{AB}^{\otimes k-1} |\tilde 0\rangle_\text{AB}^{\otimes N-k}$ onto the qubit pairs. 
            One can confirm (see App.~\ref{app:DTFIM-purified-steady-state}) that $\hat H_\text{AB}$ only couples the $|\tilde d_k\rangle_\text{AB}$ to states $|\tilde b_{k'}\rangle_\text{AB}$, thus defining our ladder model  (cf. Fig.~\ref{fig:fig4}b).
            
            We can now calculate the matrix elements $h_{k',k}$ defined in Eq.~\eqref{eq:ladder-model-matrix-elements} for the ladder Hamiltonian $\hat H_\text{AB}'$ between $|\tilde d_k\rangle_\text{AB}$ and $|\tilde b_{k'}\rangle_\text{AB}$ (see Appendix~\ref{app:DTFIM-purified-steady-state} for details and Fig.~\ref{fig:fig4}b for an illustration of the ladder model):
            \begin{align}
                h_{k,k} &= \left [ \left (\Delta -i \frac{\gamma+\overline{\Gamma}/N}{2}\right ) - \frac{\overline J + i\frac{\overline\Gamma} 2}{N}(N-k) \right ]\sqrt k,\nonumber\\
                h_{k+1,k} &= \sqrt{\frac{N-k}{2}}\Omega.
                \label{eq:dtfim-matrix-elements}
            \end{align}
            All other matrix elements $h_{k,k'}$ vanish. As in the nonlinear cavity model 
            (cf.~Eq.~\eqref{eq:kerr-coupling-coeffs}) the interaction range is $R=1$ and the ladder reduces to an inhomogeneous tight-binding chain with a two-site unit cell (cf. Fig.~\ref{fig:fig5}).
            Somewhat surprisingly, we see that both the DTFIM and the driven Kerr cavity model reduce to similar-looking Hermitian tight-binding models.  The key difference is that the DTFIM model has a finite length $N$, reflecting the bounded Hilbert space (unlike the infinite-dimensional nature of the Kerr cavity model).

            The last step of our procedure is to solve for the zero-energy eigenstate of the ladder Hamiltonian $\hat H_\text{AB}'$ that is localized on the upper leg; the coefficients $\psi_k$ of this state in the $\ket{\tilde{d}_k}$ basis yield the purification of the NESS (cf. Eq.~\eqref{eq:purified-steady-state}).
            Applying the condition $\mel{\tilde b_k}{\hat H_\text{AB}'}{\Psi_\mathcal T}_\text{AB}=0$ for each $k$ (see the recursion relation in Eq.~\eqref{eq:htrs-recursion-relation}) the steady-state coefficients satisfy
            \begin{align}
                h_{k,k}\psi_k + h_{k,k-1}\psi_{k-1} = 0.
            \end{align}
        This one-term recursion relation (explicitly given in Eq.~\eqref{eq:DTFIM-recurrence-relation}) can be solved analytically in an iterative fashion, starting from $\psi_0$. This finally yields the steady-state purification 
            \begin{align}
                \ket{\Psi_\mathcal T}_\text{AB} = \sum_{k=0}^N \psi_k \ket{\tilde d_k}_\text{AB}    
                \label{eq:main-text-dtfim-purification}
            \end{align}
            in the form of Eq.~\eqref{eq:purified-steady-state}; the coefficients are given in Eq.~\eqref{eq:dtfim-steady-state-coeffs-app}. It is straightforward to show that tracing over system $B$ recovers the single-system steady state studied in Ref.~\cite{roberts_exact_2023}. We can now apply the ansatz developed in Secs~\ref{subsec:4C} and~\ref{subsec:sec4D} for determining the dissipative gap directly from the $\psi_k$.

        \subsubsection{Analytic predictions for slow relaxation near the first-order dissipative phase transition}
            We now return to our original goal: does the effective potential defined by the purification of our NESS predict slow-relaxation rates near a first-order DPT?  As per our general conjecture, we can directly use the coefficients $\psi_k$ of the purified wavefunction to define an effective non-equilibrium potential as per Eq.~\eqref{eq:htrs-potential}.  We find:
            \begin{align}
                V_\text{DTFIM}(n) &= -\lim_{N\to\infty} \frac{\log P_{k=N n}}{N} \nonumber \\
                &= - \int_0^n dn' \log\Bigg [ \frac{1-n'}{2n'}\frac{\Omega^2}{\tilde\Delta(n')^2 + \tilde\gamma(n')^2 / 4}\Bigg ]
                \label{eq:dtfim-noneq-pot}
            \end{align}
            with
            \begin{align}
                \tilde\Delta(n) & \equiv\Delta - \overline J(1-n), 
                \hspace{0.5 cm}
                \tilde\gamma(n)\equiv\gamma + \overline\Gamma (1-n)
            \end{align}
            As shown in Fig.~\ref{fig:fig1} (where $V_\text{DTFIM}$ is referred to as $V_\text{hTRS}$) this non-equilibrium potential takes the form of a double-well structure in the bistable parameter regime (pink-hashed region in Fig.~\ref{fig:fig7}). 
            
            In the bistable parameter region we can calculate the non-equilibrium potential barrier $\Delta V$ defined in Eq.~\eqref{eq:free-energy-barrier} and compare this with the scaling of the dissipative gap with system size $N$. 
            This then defines our conjecture for the dissipative gap of $\mathcal L_\text{DTFIM}(N)$ 
            \begin{align}
                \lim_{N\to\infty} \frac{\log\big [\Gamma_\text{diss}(\Omega,\dots)\big ]} N = -\Delta V(\Omega,\dots).
                \label{dtfim-potential-barrier-main}
            \end{align}
            Here, we emphasize the parameter dependence on $\Omega$ as we study this below; the ellipses leave out additional parameter dependencies of $\Delta V$.

\begin{figure}
                \centering
                \includegraphics[width=\columnwidth]{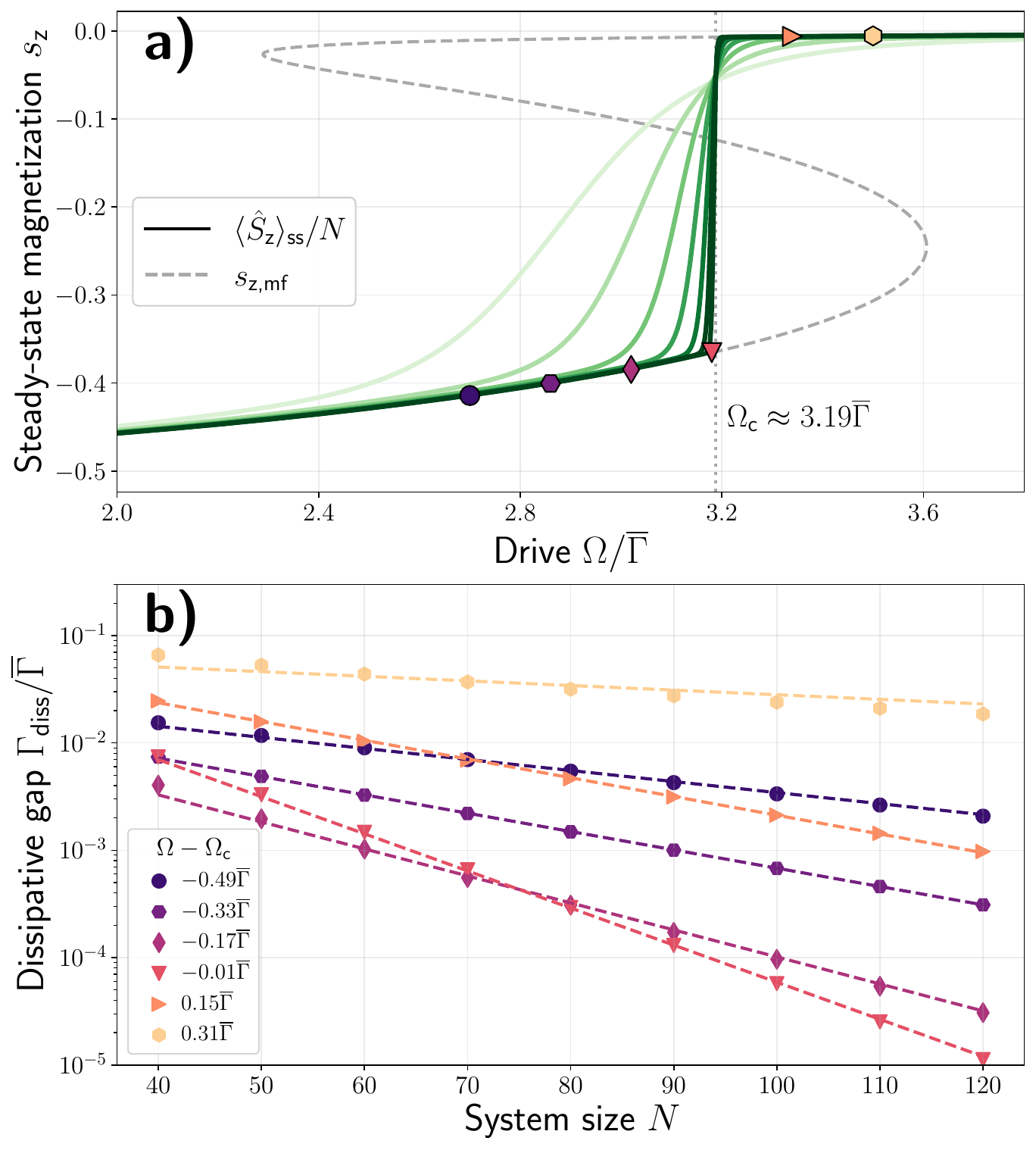}
                \caption{First-order DPT in the dissipative transverse-field Ising model (DTFIM). \textbf{a)} The steady-state magnetization $s_\text z = \langle \hat S_\text z\rangle_\text{ss} / N$
                (solid lines) exhibits a jump at $\Omega_\text c \approx 3.19\bar\Gamma$ (dotted line) between the bistable values $s_\text{z,mf}$ determined by the mean-field condition (Eq.~\eqref{eq:DTFIM-mean-field}) (dashed line) with increasing $N=16,32, 64,128, 256,512, 1024$ (increasing with shade).  The enlarged markers correspond to the drives shown in panel b. \textbf{b)} The dissipative gap $\Gamma_\text{diss}$ close to the first-order DPT determined by numerical diagonalization for small deviations from the location of the phase transition $\Omega_\text c$. The dissipative gap is in good agreement with the conjectured scaling in Eq.~\eqref{eq:dissipative-gap-scaling-hypothesis} with the potential barrier $\Delta V$ calculated using $V_\text{DTFIM}(n)$. Parameters are $\overline J = 5\overline \Gamma, \Delta = 0, \gamma =0.5\overline \Gamma$.}
                \label{fig:fig8}
            \end{figure}

            In Fig.~\ref{fig:fig8}a we plot $\langle \hat S_{\rm z} \rangle_{\rm ss}$ in the NESS for parameters chosen to give us a first-order dissipative phase transition; parameters here correspond to the blue-dashed line at $\overline J = 5\overline \Gamma$ ($\gamma=0.5\overline \Gamma$ and $\Delta=0$) in Fig.~\ref{fig:fig7}. 
            With increasing $N$ we observe a sharpening transition between two mean-field solutions; this yields a critical driving strength $\Omega_\text c \approx 3.19\overline \Gamma$. 
            We will test our conjecture for the dependence of the dissipative gap $\Gamma_\text{diss}$ with $N$ using parameters near this
            first-order DPT (namely $\Omega/\overline \Gamma\in [2.3,3.6]$). 
            
            Fig.~\ref{fig:fig8}b presents a key result of our analysis: clear evidence that our conjecture is correct, and that the dissipative steady state directly lets one predict slow metastability timescales.  We plot here the dissipative gap $\Gamma_\text{diss}$ of $\mathcal L_\text{DTFIM}(N)$ with system size $N$; the dissipative gap is extracted from the  spectrum of the Lindbladian using an Arnoldi method to obtain the two eigenvalues with the largest real part (with $\Gamma_{\rm diss}$ being the real part of their difference). 
            Our conjecture is in excellent quantitative agreement with numerics: the dissipative gap is indeed exponentially suppressed with $N$ and the corresponding slope is well approximated by the non-equilibrium potential barrier $\Delta V$.  Note that our goal is to predict the exponential scaling with $N$, and not the overall prefactor in the dissipative gap; hence, we shift each analytic curve to yield the best match to the numerical data (i.e.~an overall $N$-independent prefactor is treated as a fitting parameter).  Despite potential finite-size effects we find remarkable agreement across the entire bistable parameter regime, therefore providing further evidence for Conjecture \ref{conj:dissipative-gap}.

        \subsubsection{Analysis of the non-equilibrium effective potential and potential barrier}

\begin{figure} 
                \centering
                \includegraphics[width=\columnwidth]{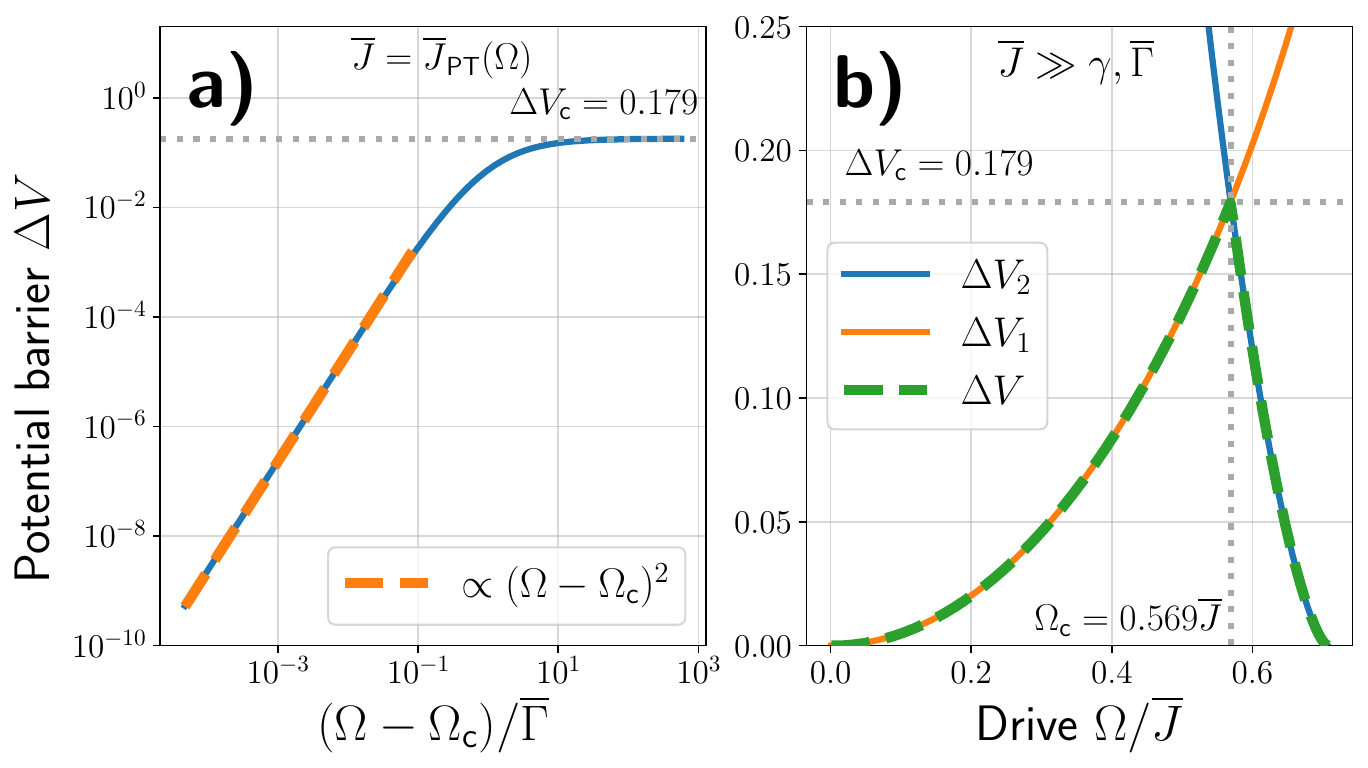}
                \caption{Behavior of the potential barrier $\Delta V$ for the DTFIM. \textbf{a)} The potential barrier evaluated along the line of first-order phase transitions in Fig.~\ref{fig:fig7}. Close to the critical point at $\Omega_\text{c}=1.28\overline \Gamma$ the barrier vanishes quadratically $\Delta V\propto (\Omega - \Omega_\text{c})^2$ before ultimately saturating for large driving strength. The parameters are $\Delta=0, \gamma=0.5\overline \Gamma$; the value for $\Omega$ then specifies the value of $\overline J_\text{PT}(\Omega)$ at which the first-order DPT occurs. \textbf{b)} The potential barrier upon varying the drive $\Omega$ in the limit of dominant Ising interactions $\overline J\gg \gamma,\overline\Gamma $ (vertical cut through Fig.~\ref{fig:fig7}). The potential barriers $\Delta V_1, \Delta V_2$ associated with the two bistable states $n_1$ and $n_2$ are minimized by $\Delta V$. Using analytical expressions for $\Delta V_1$ and $\Delta V_2$ we obtain analytical predictions for the critical drive $\Omega_\text c=0.569\overline J$ and the critical potential barrier $\Delta V_\text c = 0.179$.}
                \label{fig:fig9}
            \end{figure}

        Having established the validity of our analytic approach, we now use it to develop insights into how the effective potential barrier (and hence metastability timescales) vary as we move through our phase diagram; results are shown in  Fig.~\ref{fig:fig9} for $N \rightarrow \infty$. We stress that even with permutation symmetry, it would be challenging to obtain these results using finite-$N$ numerical diagonalization of $\mathcal L_\text{DTFIM}(N)$ and extrapolation to $N\to\infty$. 
        
        In Fig.~\ref{fig:fig9}a we show 
        how the effective potential barrier 
        varies as we move along the line of first-order DPTs depicted in Fig.~\ref{fig:fig7}.  To be explicit, for each value of $\Omega$ on the $x$-axis, we set the Ising interaction to a value that takes us to the DPT, i.e.~$\overline J=\overline J_\text{PT}(\Omega)$.
        This line of first order phase transitions ends at a critical point $\Omega \approx 1.28\overline \Gamma \equiv 
        \Omega_c'$  where the transition becomes second order (green cross in Fig.~\ref{fig:fig7}).  Near this critical point, the potential barrier vanishes quadratically\ie $\Delta V \propto (\Omega - \Omega_\text{c}')^2$ with the distance from the critical driving strength.
        This quadratic scaling of the potential barrier with $\Omega$ close to the second-order phase transition can be understood by re-writing the potential as $V_{\rm DTFIM}(n) = V_0(n) - 2 \log(\Omega) n$ where $V_0$ is independent of $\Omega$. For a double-well potential of this form, we show in Appendix~\ref{app:double-well-potential-scaling} that the quadratic scaling is a consequence of a local maximum and a local minimum merging upon tuning $\Omega\to\Omega_{\rm c}'$.

        We also see that as we move along the line of first-order transitions to larger and larger drives, the potential saturates as a function of $\Omega$ to a value we denote $\Delta V_\text c$. 
        To gain insight into this, in Fig.~\ref{fig:fig9}b we plot the potential barrier as a function of $\Omega$, but now in the limit $\overline J \gg \gamma,\Delta,\overline \Gamma$ (see App.~\ref{app:noneq-pot-dtfim} for derivation and more general parameters).  The barrier has a non-monotonic dependence with $\Omega$.  This is easily understood:  for a given choice of parameters, there are two local minima in the potential (corresponding to the two metastable states), and hence two relevant barriers $\Delta V_1$ and $\Delta V_2$.  These barriers have opposite dependencies on $\Omega$, and the effective barrier is given by the minimum of the two.  In the limit we consider, it reaches its maximal value $\Delta V_\text c \approx 0.179$ exactly at the location of the phase transition\footnote{The potential barrier is necessarily largest at $\Omega_\text c$ since the only $\Omega$ dependence is $V_\text{DTFIM}(n) \propto \log(\Omega) \times n$.} $\Omega_\text c \approx 0.569\overline J$. Both of these values are determined analytically in Appendix~\ref{app:noneq-pot-dtfim}.

            \subsubsection{Alternate methods for obtaining slow rates}

            Our results here provide (to the best of our knowledge) the first analytic predictions for the scaling of the dissipative gap of the DTFIM.  We note that analytic techniques suitable for other dissipative spin models are not immediately applicable to our model.  In Ref.~\cite{ptaszynski_quantum_2026}, metastable timescales near a first-order DPT were calculated using an instanton technique for a fully collective version of our model 
            (i.e.~$\gamma=0, \overline \Gamma > 0$). The model is of course dramatically simpler in this limit (i.e.~the Hilbert space dimension scales linearly in $N$). 
            Path-integral techniques based on a Keldysh action were used to understand a variant of the DTFIM model in Ref.~\cite{paz_driven-dissipative_2021}, though slow timescales were not calculated. In particular the all-to-all Ising were decoupled using a Hubbard-Stratonovich transformation. However, this method fails for our model when both $\overline J$ and $\overline \Gamma$ are non-zero as one cannot simultaneously decouple both the Hamiltonian and dissipative interactions.  The upshot is that the approach outlined here yields insights into our model not readily available via other techniques.  

\section{Conclusions}\label{sec:conclusions} \label{sec:sec6}

    Our work here explores a basic question in the study of open question systems: \emph{can properties of the quantum non-equilibrium steady state (NESS) of a given dissipative system provide direct insight into relaxation dynamics}? Conjecture~\ref{conj:dissipative-gap} affirms this to be the case for models with hidden time-reversal symmetry: here, the quantum steady state precisely characterizes metastable timescales at first-order dissipative phase transitions. We demonstrated how a particular, physically-relevant purification of the NESS (written in a basis of definite-charge states) could be used to directly define a potential.  The potential barrier emerging from this then successfully predicted the behavior of slow relaxation rates near first-order phase transitions in two canonical models (a driven-dissipative Kerr resonator, and a non-collective dissipative transverse-field Ising model). In both models we find excellent agreement with finite-size numerics; in the nonlinear cavity model our construction matches instanton calculations as we show in Appendix~\ref{app:kerr-comparison-phase-space-methods}.


    In future work, it would be interesting to use the approach here to study dynamics in other driven-dissipative models exhibiting hTRS (including bosonic~\cite{roberts_competition_2023} and fermionic~\cite{lingenfelter_exact_2026} lattice models and boundary driven spin chains~\cite{yao_hidden_2024}).  
    Our work also presents a new way of thinking about the structure of NESS in models with hTRS, establishing a surprising connection to zero modes of Hermitian Hamiltonian models with sublattice symmetry.  It would be interesting to see if this connection could be used to explore other aspects of dissipative systems with hTRS (e.g.~does this symmetry place other more general constraints on dynamics and the Lindbladian spectrum, can we understand the stability of such systems to perturbations that break hTRS?).

\section*{Acknowledgements}
    
    This work was supported by the Army Research Office under grant W911NF-25-1-0286, the Air Force Office of Scientific Research MURI program under Grant No. FA9550-19-1-0399, and the Simons Foundation through a Simons Investigator award (Grant No. 669487).

\bibliography{bib}

@article{roberts_exact_2023,
    title = {Exact {Solution} of the {Infinite}-{Range} {Dissipative} {Transverse}-{Field} {Ising} {Model}},
    volume = {131},
    url = {https://link.aps.org/doi/10.1103/PhysRevLett.131.190403},
    doi = {10.1103/PhysRevLett.131.190403},
    number = {19},
    urldate = {2024-12-07},
    journal = {Physical Review Letters},
    author = {Roberts, David and Clerk, A. A.},
    month = nov,
    year = {2023},
    note = {Publisher: American Physical Society},
    pages = {190403},
}

@article{ferioli_non-equilibrium_2023,
    title = {A non-equilibrium superradiant phase transition in free space},
    volume = {19},
    issn = {1745-2473, 1745-2481},
    url = {http://arxiv.org/abs/2207.10361},
    doi = {10.1038/s41567-023-02064-w},
    number = {9},
    urldate = {2024-12-07},
    journal = {Nature Physics},
    author = {Ferioli, Giovanni and Glicenstein, Antoine and Ferrier-Barbut, Igor and Browaeys, Antoine},
    month = sep,
    year = {2023},
    note = {arXiv:2207.10361 [quant-ph]},
    pages = {1345--1349},
}

@article{carmichael_analytical_1980,
    title = {Analytical and numerical results for the steady state in cooperative resonance fluorescence},
    volume = {13},
    issn = {0022-3700},
    url = {https://dx.doi.org/10.1088/0022-3700/13/18/009},
    doi = {10.1088/0022-3700/13/18/009},
    language = {en},
    number = {18},
    urldate = {2024-12-07},
    journal = {Journal of Physics B: Atomic and Molecular Physics},
    author = {Carmichael, H. J.},
    month = sep,
    year = {1980},
    pages = {3551},
}

@article{marcuzzi_universal_2014,
    title = {Universal {Nonequilibrium} {Properties} of {Dissipative} {Rydberg} {Gases}},
    volume = {113},
    url = {https://link.aps.org/doi/10.1103/PhysRevLett.113.210401},
    doi = {10.1103/PhysRevLett.113.210401},
    number = {21},
    urldate = {2023-11-22},
    journal = {Physical Review Letters},
    author = {Marcuzzi, Matteo and Levi, Emanuele and Diehl, Sebastian and Garrahan, Juan P. and Lesanovsky, Igor},
    month = nov,
    year = {2014},
    note = {Publisher: American Physical Society},
    pages = {210401},
}

@book{freidlin_random_2012,
    address = {Berlin, Heidelberg},
    series = {Grundlehren der mathematischen Wissenschaften},
    title = {Random Perturbations of Dynamical Systems},
    volume = {260},
    copyright = {https://www.springernature.com/gp/researchers/text-and-data-mining},
    isbn = {978-3-642-25846-6 978-3-642-25847-3},
    url = {https://link.springer.com/10.1007/978-3-642-25847-3},
    urldate = {2024-09-26},
    publisher = {Springer Berlin Heidelberg},
    author = {Freidlin, Mark I. and Wentzell, Alexander D.},
    year = {2012},
    doi = {10.1007/978-3-642-25847-3},
}

@article{iemini_boundary_2018,
    title = {Boundary {Time} {Crystals}},
    volume = {121},
    url = {https://link.aps.org/doi/10.1103/PhysRevLett.121.035301},
    doi = {10.1103/PhysRevLett.121.035301},
    number = {3},
    urldate = {2024-12-07},
    journal = {Physical Review Letters},
    author = {Iemini, F. and Russomanno, A. and Keeling, J. and Schirò, M. and Dalmonte, M. and Fazio, R.},
    month = jul,
    year = {2018},
    note = {Publisher: American Physical Society},
    pages = {035301},
}

@misc{xiang_switching_2025,
    title = {Switching {Dynamics} of {Metastable} {Open} {Quantum} {Systems}},
    url = {http://arxiv.org/abs/2505.05202},
    doi = {10.48550/arXiv.2505.05202},
    urldate = {2025-05-09},
    publisher = {arXiv},
    author = {Xiang, Ya-Xin and Li, Weibin and Bai, Zhengyang and Ma, Yu-Qiang},
    month = may,
    year = {2025},
    note = {arXiv:2505.05202 [quant-ph]},
}

@article{chase_collective_2008,
    title = {Collective processes of an ensemble of spin- 1 ∕ 2 particles},
    volume = {78},
    copyright = {http://link.aps.org/licenses/aps-default-license},
    issn = {1050-2947, 1094-1622},
    url = {https://link.aps.org/doi/10.1103/PhysRevA.78.052101},
    doi = {10.1103/PhysRevA.78.052101},
    language = {en},
    number = {5},
    urldate = {2024-09-27},
    journal = {Physical Review A},
    author = {Chase, Bradley A. and Geremia, J. M.},
    month = nov,
    year = {2008},
    pages = {052101},
}

@article{paz_driven-dissipative_2021,
    title = {Driven-dissipative {Ising} model: {An} exact field-theoretical analysis},
    volume = {104},
    shorttitle = {Driven-dissipative {Ising} model},
    url = {https://link.aps.org/doi/10.1103/PhysRevA.104.023713},
    doi = {10.1103/PhysRevA.104.023713},
    number = {2},
    urldate = {2024-12-07},
    journal = {Physical Review A},
    author = {Paz, Daniel A. and Maghrebi, Mohammad F.},
    month = aug,
    year = {2021},
    note = {Publisher: American Physical Society},
    pages = {023713},
}

@article{roberts_hidden_2021,
    title = {Hidden {Time}-{Reversal} {Symmetry}, {Quantum} {Detailed} {Balance} and {Exact} {Solutions} of {Driven}-{Dissipative} {Quantum} {Systems}},
    volume = {2},
    url = {https://link.aps.org/doi/10.1103/PRXQuantum.2.020336},
    doi = {10.1103/PRXQuantum.2.020336},
    number = {2},
    urldate = {2023-11-22},
    journal = {PRX Quantum},
    author = {Roberts, David and Lingenfelter, Andrew and Clerk, A.A.},
    month = jun,
    year = {2021},
    note = {Publisher: American Physical Society},
    pages = {020336},
}

@article{macieszczak_theory_2021,
    title = {Theory of classical metastability in open quantum systems},
    volume = {3},
    issn = {2643-1564},
    url = {https://link.aps.org/doi/10.1103/PhysRevResearch.3.033047},
    doi = {10.1103/PhysRevResearch.3.033047},
    language = {en},
    number = {3},
    urldate = {2024-09-19},
    journal = {Physical Review Research},
    author = {Macieszczak, Katarzyna and Rose, Dominic C. and Lesanovsky, Igor and Garrahan, Juan P.},
    month = jul,
    year = {2021},
    pages = {033047},
}

@article{hanggi_bistable_1984,
    title = {Bistable systems: {Master} equation versus {Fokker}-{Planck} modeling},
    volume = {29},
    shorttitle = {Bistable systems},
    url = {https://link.aps.org/doi/10.1103/PhysRevA.29.371},
    doi = {10.1103/PhysRevA.29.371},
    number = {1},
    urldate = {2024-12-07},
    journal = {Physical Review A},
    author = {Hanggi, Peter and Grabert, Hermann and Talkner, Peter and Thomas, Harry},
    month = jan,
    year = {1984},
    note = {Publisher: American Physical Society},
    pages = {371--378},
}

@misc{leppenen_quantum_2024,
    title = {Quantum bistability at the interplay between collective and individual decay},
    url = {http://arxiv.org/abs/2404.02134},
    doi = {10.48550/arXiv.2404.02134},
    urldate = {2024-12-08},
    publisher = {arXiv},
    author = {Leppenen, Nikita and Shahmoon, Ephraim},
    month = apr,
    year = {2024},
    note = {arXiv:2404.02134 [quant-ph]
version: 1},
}

@article{gelhausen_dissipative_2018,
    title = {Dissipative {Dicke} model with collective atomic decay: {Bistability}, noise-driven activation, and the nonthermal first-order superradiance transition},
    volume = {97},
    issn = {2469-9926, 2469-9934},
    shorttitle = {Dissipative {Dicke} model with collective atomic decay},
    url = {https://link.aps.org/doi/10.1103/PhysRevA.97.023807},
    doi = {10.1103/PhysRevA.97.023807},
    language = {en},
    number = {2},
    urldate = {2024-09-26},
    journal = {Physical Review A},
    author = {Gelhausen, Jan and Buchhold, Michael},
    month = feb,
    year = {2018},
    pages = {023807},
}

@article{minganti_spectral_2018,
    title = {Spectral theory of {Liouvillians} for dissipative phase transitions},
    volume = {98},
    issn = {2469-9926, 2469-9934},
    url = {https://link.aps.org/doi/10.1103/PhysRevA.98.042118},
    doi = {10.1103/PhysRevA.98.042118},
    language = {en},
    number = {4},
    urldate = {2024-10-21},
    journal = {Physical Review A},
    author = {Minganti, Fabrizio and Biella, Alberto and Bartolo, Nicola and Ciuti, Cristiano},
    month = oct,
    year = {2018},
    pages = {042118},
}

@article{roberts_competition_2023,
    title = {Competition between {Two}-{Photon} {Driving}, {Dissipation}, and {Interactions} in {Bosonic} {Lattice} {Models}: {An} {Exact} {Solution}},
    volume = {130},
    issn = {0031-9007, 1079-7114},
    shorttitle = {Competition between {Two}-{Photon} {Driving}, {Dissipation}, and {Interactions} in {Bosonic} {Lattice} {Models}},
    url = {https://link.aps.org/doi/10.1103/PhysRevLett.130.063601},
    doi = {10.1103/PhysRevLett.130.063601},
    language = {en},
    number = {6},
    urldate = {2024-10-30},
    journal = {Physical Review Letters},
    author = {Roberts, David and Clerk, A. A.},
    month = feb,
    year = {2023},
    pages = {063601},
}

@article{roberts_driven-dissipative_2020,
    title = {Driven-{Dissipative} {Quantum} {Kerr} {Resonators}: {New} {Exact} {Solutions}, {Photon} {Blockade} and {Quantum} {Bistability}},
    volume = {10},
    issn = {2160-3308},
    shorttitle = {Driven-{Dissipative} {Quantum} {Kerr} {Resonators}},
    url = {https://link.aps.org/doi/10.1103/PhysRevX.10.021022},
    doi = {10.1103/PhysRevX.10.021022},
    language = {en},
    number = {2},
    urldate = {2025-04-18},
    journal = {Physical Review X},
    author = {Roberts, David and Clerk, Aashish A.},
    month = apr,
    year = {2020},
    pages = {021022},
}

@article{puri_exact_1979,
    title = {Exact steady-state density operator for a collective atomic system in an external field},
    volume = {72},
    issn = {03759601},
    url = {https://linkinghub.elsevier.com/retrieve/pii/0375960179900033},
    doi = {10.1016/0375-9601(79)90003-3},
    language = {en},
    number = {3},
    urldate = {2024-03-12},
    journal = {Physics Letters A},
    author = {Puri, R.R. and Lawande, S.V.},
    month = jul,
    year = {1979},
    pages = {200--202},
}

@article{casteels_critical_2017,
    title = {Critical dynamical properties of a first-order dissipative phase transition},
    volume = {95},
    copyright = {http://link.aps.org/licenses/aps-default-license},
    issn = {2469-9926, 2469-9934},
    url = {https://link.aps.org/doi/10.1103/PhysRevA.95.012128},
    doi = {10.1103/PhysRevA.95.012128},
    language = {en},
    number = {1},
    urldate = {2024-10-03},
    journal = {Physical Review A},
    author = {Casteels, W. and Fazio, R. and Ciuti, C.},
    month = jan,
    year = {2017},
    pages = {012128},
}

@article{cabot_continuous_2024,
    title = {Continuous {Sensing} and {Parameter} {Estimation} with the {Boundary} {Time} {Crystal}},
    volume = {132},
    issn = {0031-9007, 1079-7114},
    url = {https://link.aps.org/doi/10.1103/PhysRevLett.132.050801},
    doi = {10.1103/PhysRevLett.132.050801},
    language = {en},
    number = {5},
    urldate = {2024-07-16},
    journal = {Physical Review Letters},
    author = {Cabot, Albert and Carollo, Federico and Lesanovsky, Igor},
    month = jan,
    year = {2024},
    pages = {050801},
}

@article{lingenfelter_exact_2024,
    title = {Exact {Results} for a {Boundary}-{Driven} {Double} {Spin} {Chain} and {Resource}-{Efficient} {Remote} {Entanglement} {Stabilization}},
    volume = {14},
    issn = {2160-3308},
    url = {https://link.aps.org/doi/10.1103/PhysRevX.14.021028},
    doi = {10.1103/PhysRevX.14.021028},
    language = {en},
    number = {2},
    urldate = {2025-08-19},
    journal = {Physical Review X},
    author = {Lingenfelter, Andrew and Yao, Mingxing and Pocklington, Andrew and Wang, Yu-Xin and Irfan, Abdullah and Pfaff, Wolfgang and Clerk, Aashish A.},
    month = may,
    year = {2024},
    pages = {021028},
}

@misc{yao_hidden_2024,
    title = {Hidden time-reversal in driven {XXZ} spin chains: exact solutions and new dissipative phase transitions},
    shorttitle = {Hidden time-reversal in driven {XXZ} spin chains},
    url = {http://arxiv.org/abs/2407.12750},
    urldate = {2024-07-18},
    publisher = {arXiv},
    author = {Yao, Mingxing and Lingenfelter, Andrew and Belyansky, Ron and Roberts, David and Clerk, Aashish A.},
    month = jul,
    year = {2024},
    note = {arXiv:2407.12750 [cond-mat, physics:quant-ph]},
}

@article{drummond_quantum_1980,
    title = {Quantum theory of optical bistability. {I}. {Nonlinear} polarisability model},
    volume = {13},
    issn = {0305-4470, 1361-6447},
    url = {https://iopscience.iop.org/article/10.1088/0305-4470/13/2/034},
    doi = {10.1088/0305-4470/13/2/034},
    language = {en},
    number = {2},
    urldate = {2025-09-02},
    journal = {Journal of Physics A: Mathematical and General},
    author = {Drummond, P D and Walls, D F},
    month = feb,
    year = {1980},
    pages = {725--741},
}

@article{agarwal_directional_2024,
    title = {Directional {Superradiance} in a {Driven} {Ultracold} {Atomic} {Gas} in {Free} {Space}},
    volume = {5},
    issn = {2691-3399},
    url = {https://link.aps.org/doi/10.1103/PRXQuantum.5.040335},
    doi = {10.1103/PRXQuantum.5.040335},
    language = {en},
    number = {4},
    urldate = {2025-03-06},
    journal = {PRX Quantum},
    author = {Agarwal, Sanaa and Chaparro, Edwin and Barberena, Diego and Orioli, A. Piñeiro and Ferioli, G. and Pancaldi, S. and Ferrier-Barbut, I. and Browaeys, A. and Rey, A.M.},
    month = dec,
    year = {2024},
    pages = {040335},
}

@article{shammah_open_2018,
    title = {Open quantum systems with local and collective incoherent processes: {Efficient} numerical simulations using permutational invariance},
    volume = {98},
    issn = {2469-9926, 2469-9934},
    shorttitle = {Open quantum systems with local and collective incoherent processes},
    url = {https://link.aps.org/doi/10.1103/PhysRevA.98.063815},
    doi = {10.1103/PhysRevA.98.063815},
    language = {en},
    number = {6},
    urldate = {2025-09-10},
    journal = {Physical Review A},
    author = {Shammah, Nathan and Ahmed, Shahnawaz and Lambert, Neill and De Liberato, Simone and Nori, Franco},
    month = dec,
    year = {2018},
    pages = {063815},
}

@article{marcuzzi_effective_2014,
    title = {Effective dynamics of strongly dissipative {Rydberg} gases},
    volume = {47},
    copyright = {http://iopscience.iop.org/info/page/text-and-data-mining},
    issn = {1751-8113, 1751-8121},
    url = {https://iopscience.iop.org/article/10.1088/1751-8113/47/48/482001},
    doi = {10.1088/1751-8113/47/48/482001},
    language = {en},
    number = {48},
    urldate = {2025-01-29},
    journal = {Journal of Physics A: Mathematical and Theoretical},
    author = {Marcuzzi, M and Schick, J and Olmos, B and Lesanovsky, I},
    month = dec,
    year = {2014},
    pages = {482001},
}

@article{marcuzzi_non-equilibrium_2015,
    title = {Non-equilibrium universality in the dynamics of dissipative cold atomic gases},
    volume = {17},
    issn = {1367-2630},
    url = {https://iopscience.iop.org/article/10.1088/1367-2630/17/7/072003},
    doi = {10.1088/1367-2630/17/7/072003},
    language = {en},
    number = {7},
    urldate = {2025-01-29},
    journal = {New Journal of Physics},
    author = {Marcuzzi, M and Levi, E and Li, W and Garrahan, J P and Olmos, B and Lesanovsky, I},
    month = jul,
    year = {2015},
    pages = {072003},
}

@misc{barberena_generalized_2025,
    title = {Generalized {Holstein}-{Primakoff} mapping and \$1/{N}\$ expansion of collective spin systems undergoing single particle dissipation},
    url = {http://arxiv.org/abs/2508.05751},
    doi = {10.48550/arXiv.2508.05751},
    urldate = {2025-08-11},
    publisher = {arXiv},
    author = {Barberena, Diego},
    month = aug,
    year = {2025},
    note = {arXiv:2508.05751 [quant-ph]},
}

@article{hartmann_generalized_2016,
    title = {Generalized {Dicke} states},
    volume = {16},
    issn = {15337146, 15337146},
    url = {http://www.rintonpress.com/journals/doi/QIC16.15-16-5.html},
    doi = {10.26421/QIC16.15-16-5},
    language = {en},
    number = {15\&16},
    urldate = {2025-10-03},
    journal = {Quantum Information and Computation},
    author = {Hartmann, Stephan},
    month = nov,
    year = {2016},
    pages = {1333--1348},
}

@article{stannigel_driven-dissipative_2012,
    title = {Driven-dissipative preparation of entangled states in cascaded quantum-optical networks},
    volume = {14},
    issn = {1367-2630},
    url = {https://iopscience.iop.org/article/10.1088/1367-2630/14/6/063014},
    doi = {10.1088/1367-2630/14/6/063014},
    language = {en},
    number = {6},
    urldate = {2023-12-13},
    journal = {New Journal of Physics},
    author = {Stannigel, K and Rabl, P and Zoller, P},
    month = jun,
    year = {2012},
    pages = {063014},
}

@misc{vicentini_critical_2018,
    title = {Critical slowing down in driven-dissipative {Bose}-{Hubbard} lattices},
    url = {http://arxiv.org/abs/1709.04238},
    urldate = {2024-11-07},
    publisher = {arXiv},
    author = {Vicentini, Filippo and Minganti, Fabrizio and Rota, Riccardo and Orso, Giuliano and Ciuti, Cristiano},
    month = jan,
    year = {2018},
    note = {arXiv:1709.04238},
}

@article{langer_statistical_1969,
    title = {Statistical theory of the decay of metastable states},
    volume = {54},
    copyright = {https://www.elsevier.com/tdm/userlicense/1.0/},
    issn = {00034916},
    url = {https://linkinghub.elsevier.com/retrieve/pii/0003491669901535},
    doi = {10.1016/0003-4916(69)90153-5},
    language = {en},
    number = {2},
    urldate = {2024-11-08},
    journal = {Annals of Physics},
    author = {Langer, J.S.},
    month = sep,
    year = {1969},
    pages = {258--275},
}

@article{yin_theory_2025,
    title = {Theory of {Metastable} {States} in {Many}-{Body} {Quantum} {Systems}},
    volume = {15},
    issn = {2160-3308},
    url = {https://link.aps.org/doi/10.1103/PhysRevX.15.011064},
    doi = {10.1103/PhysRevX.15.011064},
    language = {en},
    number = {1},
    urldate = {2025-10-27},
    journal = {Physical Review X},
    author = {Yin, Chao and Surace, Federica M. and Lucas, Andrew},
    month = mar,
    year = {2025},
    pages = {011064},
}

@article{macieszczak_towards_2016,
    title = {Towards a {Theory} of {Metastability} in {Open} {Quantum} {Dynamics}},
    volume = {116},
    copyright = {http://link.aps.org/licenses/aps-default-license},
    issn = {0031-9007, 1079-7114},
    url = {https://link.aps.org/doi/10.1103/PhysRevLett.116.240404},
    doi = {10.1103/PhysRevLett.116.240404},
    language = {en},
    number = {24},
    urldate = {2025-10-27},
    journal = {Physical Review Letters},
    author = {Macieszczak, Katarzyna and Guţă, Mădălin and Lesanovsky, Igor and Garrahan, Juan P.},
    month = jun,
    year = {2016},
    pages = {240404},
}

@article{rose_metastability_2016,
    title = {Metastability in an open quantum {Ising} model},
    volume = {94},
    copyright = {http://link.aps.org/licenses/aps-default-license},
    issn = {2470-0045, 2470-0053},
    url = {https://link.aps.org/doi/10.1103/PhysRevE.94.052132},
    doi = {10.1103/PhysRevE.94.052132},
    language = {en},
    number = {5},
    urldate = {2025-10-24},
    journal = {Physical Review E},
    author = {Rose, Dominic C. and Macieszczak, Katarzyna and Lesanovsky, Igor and Garrahan, Juan P.},
    month = nov,
    year = {2016},
    pages = {052132},
}

@article{lee_real-time_2025,
    title = {Real-time instanton approach to quantum activation},
    volume = {112},
    issn = {2469-9926, 2469-9934},
    url = {https://link.aps.org/doi/10.1103/5jtm-ht4n},
    doi = {10.1103/5jtm-ht4n},
    language = {en},
    number = {1},
    urldate = {2025-10-24},
    journal = {Physical Review A},
    author = {Lee, Chang-Woo and Brookes, Paul and Park, Kee-Su and Szymańska, Marzena H. and Ginossar, Eran},
    month = jul,
    year = {2025},
    pages = {012216},
}

@article{marthaler_switching_2006,
    title = {Switching via quantum activation: {A} parametrically modulated oscillator},
    volume = {73},
    copyright = {http://link.aps.org/licenses/aps-default-license},
    issn = {1050-2947, 1094-1622},
    shorttitle = {Switching via quantum activation},
    url = {https://link.aps.org/doi/10.1103/PhysRevA.73.042108},
    doi = {10.1103/PhysRevA.73.042108},
    language = {en},
    number = {4},
    urldate = {2025-10-24},
    journal = {Physical Review A},
    author = {Marthaler, M. and Dykman, M. I.},
    month = apr,
    year = {2006},
    pages = {042108},
}

@article{dykman_critical_2007,
    title = {Critical exponents in metastable decay via quantum activation},
    volume = {75},
    url = {https://link.aps.org/doi/10.1103/PhysRevE.75.011101},
    doi = {10.1103/PhysRevE.75.011101},
    number = {1},
    urldate = {2025-04-21},
    journal = {Physical Review E},
    author = {Dykman, M. I.},
    month = jan,
    year = {2007},
    note = {Publisher: American Physical Society},
    pages = {011101},
}

@article{mattes_long-range_2025,
    title = {Long-{Range} {Interacting} {Systems} {Are} {Locally} {Noninteracting}},
    volume = {134},
    url = {https://link.aps.org/doi/10.1103/PhysRevLett.134.070402},
    doi = {10.1103/PhysRevLett.134.070402},
    number = {7},
    urldate = {2025-10-31},
    journal = {Physical Review Letters},
    author = {Mattes, Robert and Lesanovsky, Igor and Carollo, Federico},
    month = feb,
    year = {2025},
    note = {Publisher: American Physical Society},
    pages = {070402},
}

@article{hanggi_reaction-rate_1990,
    title = {Reaction-rate theory: fifty years after {Kramers}},
    volume = {62},
    copyright = {http://link.aps.org/licenses/aps-default-license},
    issn = {0034-6861, 1539-0756},
    shorttitle = {Reaction-rate theory},
    url = {https://link.aps.org/doi/10.1103/RevModPhys.62.251},
    doi = {10.1103/RevModPhys.62.251},
    language = {en},
    number = {2},
    urldate = {2024-09-19},
    journal = {Reviews of Modern Physics},
    author = {Hänggi, Peter and Talkner, Peter and Borkovec, Michal},
    month = apr,
    year = {1990},
    pages = {251--341},
}

@article{ptaszynski_dynamical_2024,
    title = {Dynamical signatures of discontinuous phase transitions: {How} phase coexistence determines exponential versus power-law scaling},
    volume = {110},
    issn = {2470-0045, 2470-0053},
    shorttitle = {Dynamical signatures of discontinuous phase transitions},
    url = {https://link.aps.org/doi/10.1103/PhysRevE.110.044134},
    doi = {10.1103/PhysRevE.110.044134},
    language = {en},
    number = {4},
    urldate = {2025-11-12},
    journal = {Physical Review E},
    author = {Ptaszyński, Krzysztof and Esposito, Massimiliano},
    month = oct,
    year = {2024},
    pages = {044134},
}

@article{groszkowski_reservoir-engineered_2022,
    title = {Reservoir-{Engineered} {Spin} {Squeezing}: {Macroscopic} {Even}-{Odd} {Effects} and {Hybrid}-{Systems} {Implementations}},
    volume = {12},
    issn = {2160-3308},
    shorttitle = {Reservoir-{Engineered} {Spin} {Squeezing}},
    url = {https://link.aps.org/doi/10.1103/PhysRevX.12.011015},
    doi = {10.1103/PhysRevX.12.011015},
    language = {en},
    number = {1},
    urldate = {2024-09-12},
    journal = {Physical Review X},
    author = {Groszkowski, Peter and Koppenhöfer, Martin and Lau, Hoi-Kwan and Clerk, A. A.},
    month = jan,
    year = {2022},
    pages = {011015},
}

@article{song_dissipation-induced_2025,
    title = {A dissipation-induced superradiant transition in a strontium cavity-{QED} system},
    volume = {11},
    url = {https://www.science.org/doi/10.1126/sciadv.adu5799},
    doi = {10.1126/sciadv.adu5799},
    number = {17},
    urldate = {2025-11-12},
    journal = {Science Advances},
    author = {Song, Eric Yilun and Barberena, Diego and Young, Dylan J. and Chaparro, Edwin and Chu, Anjun and Agarwal, Sanaa and Niu, Zhijing and Young, Jeremy T. and Rey, Ana Maria and Thompson, James K.},
    month = apr,
    year = {2025},
    note = {Publisher: American Association for the Advancement of Science},
    pages = {eadu5799},
}

@misc{mylnikov_switching_2025,
    title = {Switching rates in {Kerr} resonator with two-photon dissipation and driving},
    url = {http://arxiv.org/abs/2511.13308},
    doi = {10.48550/arXiv.2511.13308},
    urldate = {2025-11-18},
    publisher = {arXiv},
    author = {Mylnikov, V. Yu and Potashin, S. O. and Ukhtary, M. S. and Sokolovskii, G. S.},
    month = nov,
    year = {2025},
    note = {arXiv:2511.13308 [quant-ph]},
}

@article{dong_engineering_2025,
    title = {Engineering an anisotropic {Dicke} model of {Rydberg} atom arrays in an optical cavity with dipole–dipole interactions},
    volume = {34},
    issn = {1674-1056},
    url = {https://doi.org/10.1088/1674-1056/ade388},
    doi = {10.1088/1674-1056/ade388},
    language = {en},
    number = {11},
    urldate = {2025-11-20},
    journal = {Chinese Physics B},
    author = {Dong, Bao-Yun and Zhou, Yanhua and Wang, Wei and Wang, Tao},
    month = nov,
    year = {2025},
    note = {Publisher: Chinese Physical Society and IOP Publishing Ltd},
    pages = {114203},
}

@article{thompson_qubit_2022,
    title = {Qubit decoherence and symmetry restoration through real-time instantons},
    volume = {4},
    issn = {2643-1564},
    url = {https://link.aps.org/doi/10.1103/PhysRevResearch.4.023020},
    doi = {10.1103/PhysRevResearch.4.023020},
    language = {en},
    number = {2},
    urldate = {2025-11-24},
    journal = {Physical Review Research},
    author = {Thompson, Foster and Kamenev, Alex},
    month = apr,
    year = {2022},
    pages = {023020},
}

@article{mori_resolving_2020,
    title = {Resolving a {Discrepancy} between {Liouvillian} {Gap} and {Relaxation} {Time} in {Boundary}-{Dissipated} {Quantum} {Many}-{Body} {Systems}},
    volume = {125},
    issn = {0031-9007, 1079-7114},
    url = {https://link.aps.org/doi/10.1103/PhysRevLett.125.230604},
    doi = {10.1103/PhysRevLett.125.230604},
    language = {en},
    number = {23},
    urldate = {2025-11-24},
    journal = {Physical Review Letters},
    author = {Mori, Takashi and Shirai, Tatsuhiko},
    month = dec,
    year = {2020},
    pages = {230604},
}

@article{wang_accelerating_2023,
    title = {Accelerating relaxation dynamics in open quantum systems with {Liouvillian} skin effect},
    volume = {108},
    issn = {2469-9950, 2469-9969},
    url = {https://link.aps.org/doi/10.1103/PhysRevB.108.054313},
    doi = {10.1103/PhysRevB.108.054313},
    language = {en},
    number = {5},
    urldate = {2025-11-24},
    journal = {Physical Review B},
    author = {Wang, Zeqing and Lu, Yao and Peng, Yi and Qi, Ran and Wang, Yucheng and Jie, Jianwen},
    month = aug,
    year = {2023},
    pages = {054313},
}

@article{lee_anomalously_2023,
    title = {Anomalously large relaxation times in dissipative lattice models beyond the non-{Hermitian} skin effect},
    volume = {108},
    issn = {2469-9950, 2469-9969},
    url = {https://link.aps.org/doi/10.1103/PhysRevB.108.064311},
    doi = {10.1103/PhysRevB.108.064311},
    language = {en},
    number = {6},
    urldate = {2025-11-24},
    journal = {Physical Review B},
    author = {Lee, Gideon and McDonald, Alexander and Clerk, Aashish},
    month = aug,
    year = {2023},
    pages = {064311},
}

@misc{berglund_kramers_2013,
    title = {Kramers' law: {Validity}, derivations and generalisations},
    shorttitle = {Kramers' law},
    url = {http://arxiv.org/abs/1106.5799},
    doi = {10.48550/arXiv.1106.5799},
    urldate = {2025-11-19},
    publisher = {arXiv},
    author = {Berglund, Nils},
    month = jan,
    year = {2013},
    note = {arXiv:1106.5799 [math]},
}

@article{fagnola_generators_2010,
    title = {Generators of {KMS} {Symmetric} {Markov} {Semigroups} on \$\$\{{\textbackslash}mathcal\{{B}\}(\{{\textbackslash}rm h\})\}\$\$ {Symmetry} and {Quantum} {Detailed} {Balance}},
    volume = {298},
    copyright = {http://www.springer.com/tdm},
    issn = {0010-3616, 1432-0916},
    url = {http://link.springer.com/10.1007/s00220-010-1011-1},
    doi = {10.1007/s00220-010-1011-1},
    language = {en},
    number = {2},
    urldate = {2025-12-02},
    journal = {Communications in Mathematical Physics},
    author = {Fagnola, Franco and Umanità, Veronica},
    month = sep,
    year = {2010},
    pages = {523--547},
}

@article{fagnola_generators_2012,
    title = {{GENERATORS} {OF} {DETAILED} {BALANCE} {QUANTUM} {MARKOV} {SEMIGROUPS}},
    url = {https://www.worldscientific.com/worldscinet/idaqp},
    doi = {10.1142/S0219025707002762},
    language = {en},
    urldate = {2025-12-02},
    journal = {Infinite Dimensional Analysis, Quantum Probability and Related Topics},
    author = {Fagnola, Franco and Umanità, Veronica},
    month = apr,
    year = {2012},
    note = {Publisher: World Scientific Publishing Company},
}

@article{duvenhage_quantum_2025,
    title = {Quantum detailed balance via elementary transitions},
    volume = {9},
    issn = {2521-327X},
    url = {http://arxiv.org/abs/2411.02339},
    doi = {10.22331/q-2025-05-15-1743},
    language = {en},
    urldate = {2025-12-03},
    journal = {Quantum},
    author = {Duvenhage, Rocco and Oerder, Kyle and Heuvel, Keagan van den},
    month = may,
    year = {2025},
    note = {arXiv:2411.02339 [quant-ph]},
    pages = {1743},
}

@article{gardiner_driving_1993,
    title = {Driving a quantum system with the output field from another driven quantum system},
    volume = {70},
    copyright = {http://link.aps.org/licenses/aps-default-license},
    issn = {0031-9007},
    url = {https://link.aps.org/doi/10.1103/PhysRevLett.70.2269},
    doi = {10.1103/PhysRevLett.70.2269},
    language = {en},
    number = {15},
    urldate = {2025-12-08},
    journal = {Physical Review Letters},
    author = {Gardiner, C. W.},
    month = apr,
    year = {1993},
    pages = {2269--2272},
}

@misc{mylnikov_qubit_2025,
    title = {Qubit decoherence in dissipative two-photon resonator: real-time instantons and {Wigner} function},
    shorttitle = {Qubit decoherence in dissipative two-photon resonator},
    url = {http://arxiv.org/abs/2512.10921},
    doi = {10.48550/arXiv.2512.10921},
    urldate = {2025-12-12},
    publisher = {arXiv},
    author = {Mylnikov, V. Yu and Potashin, S. O. and Kamenev, Alex},
    month = dec,
    year = {2025},
    note = {arXiv:2512.10921 [quant-ph]},
}

@misc{rakovszky_bottlenecks_2024,
    title = {Bottlenecks in quantum channels and finite temperature phases of matter},
    url = {http://arxiv.org/abs/2412.09598},
    doi = {10.48550/arXiv.2412.09598},
    urldate = {2025-12-17},
    publisher = {arXiv},
    author = {Rakovszky, Tibor and Placke, Benedikt and Breuckmann, Nikolas P. and Khemani, Vedika},
    month = dec,
    year = {2024},
    note = {arXiv:2412.09598 [quant-ph]},
}

@article{carlen_gradient_2017,
    title = {Gradient flow and entropy inequalities for quantum {Markov} semigroups with detailed balance},
    volume = {273},
    issn = {0022-1236},
    url = {https://www.sciencedirect.com/science/article/pii/S0022123617301878},
    doi = {10.1016/j.jfa.2017.05.003},
    number = {5},
    urldate = {2026-02-10},
    journal = {Journal of Functional Analysis},
    author = {Carlen, Eric A. and Maas, Jan},
    month = sep,
    year = {2017},
    pages = {1810--1869},
}

@article{kossakowski_quantum_1977,
    title = {Quantum detailed balance and {KMS} condition},
    volume = {57},
    issn = {1432-0916},
    url = {https://doi.org/10.1007/BF01625769},
    doi = {10.1007/BF01625769},
    language = {en},
    number = {2},
    urldate = {2026-02-10},
    journal = {Communications in Mathematical Physics},
    author = {Kossakowski, Andrzej and Frigerio, Alberto and Gorini, Vittorio and Verri, Maurizio},
    month = jun,
    year = {1977},
    pages = {97--110},
}

@article{albert_symmetries_2014,
    title = {Symmetries and conserved quantities in {Lindblad} master equations},
    volume = {89},
    url = {https://link.aps.org/doi/10.1103/PhysRevA.89.022118},
    doi = {10.1103/PhysRevA.89.022118},
    number = {2},
    urldate = {2026-02-10},
    journal = {Physical Review A},
    publisher = {American Physical Society},
    author = {Albert, Victor V. and Jiang, Liang},
    month = feb,
    year = {2014},
    pages = {022118},
}

@article{buca_note_2012,
    title = {A note on symmetry reductions of the {Lindblad} equation: transport in constrained open spin chains},
    volume = {14},
    issn = {1367-2630},
    shorttitle = {A note on symmetry reductions of the {Lindblad} equation},
    url = {https://iopscience.iop.org/article/10.1088/1367-2630/14/7/073007},
    doi = {10.1088/1367-2630/14/7/073007},
    language = {en},
    number = {7},
    urldate = {2026-02-10},
    journal = {New Journal of Physics},
    author = {Buča, Berislav and Prosen, Tomaž},
    month = jul,
    year = {2012},
    pages = {073007},
}

@article{lugiato_optical_1983,
    title = {Optical bistability},
    volume = {24},
    issn = {0010-7514, 1366-5812},
    url = {http://www.tandfonline.com/doi/abs/10.1080/00107518308210690},
    doi = {10.1080/00107518308210690},
    language = {en},
    number = {4},
    urldate = {2024-06-11},
    journal = {Contemporary Physics},
    author = {Lugiato, Luigi A.},
    month = jul,
    year = {1983},
    pages = {333--371},
}

@article{bonifacio_photon_1978,
    title = {Photon statistics of a bistable absorber},
    volume = {18},
    copyright = {http://link.aps.org/licenses/aps-default-license},
    issn = {0556-2791},
    url = {https://link.aps.org/doi/10.1103/PhysRevA.18.2266},
    doi = {10.1103/PhysRevA.18.2266},
    language = {en},
    number = {5},
    urldate = {2024-08-05},
    journal = {Physical Review A},
    author = {Bonifacio, R. and Gronchi, M. and Lugiato, L. A.},
    month = nov,
    year = {1978},
    pages = {2266--2279},
}

@misc{santis_realization_2026,
    title = {Realization of a cavity-coupled {Rydberg} array},
    url = {http://arxiv.org/abs/2602.12152},
    doi = {10.48550/arXiv.2602.12152},
    urldate = {2026-02-13},
    publisher = {arXiv},
    author = {Santis, Jacopo De and Dura-Kovács, Balázs and Öncü, Mehmet and Bouscal, Adrien and Vasileiadis, Dimitrios and Zeiher, Johannes},
    month = feb,
    year = {2026},
    note = {arXiv:2602.12152 [quant-ph]},
}

@article{fazio_many-body_2025,
    title = {Many-body open quantum systems},
    issn = {2590-1990},
    url = {https://scipost.org/SciPostPhysLectNotes.99},
    doi = {10.21468/SciPostPhysLectNotes.99},
    language = {en},
    urldate = {2026-02-20},
    journal = {SciPost Physics Lecture Notes},
    author = {Fazio, Rosario and Keeling, Jonathan and Mazza, Leonardo and Schirò, Marco},
    month = aug,
    year = {2025},
    pages = {099},
}

@article{lindblad_generators_1976,
    title = {On the generators of quantum dynamical semigroups},
    volume = {48},
    issn = {1432-0916},
    url = {https://doi.org/10.1007/BF01608499},
    doi = {10.1007/BF01608499},
    language = {en},
    number = {2},
    urldate = {2026-03-31},
    journal = {Communications in Mathematical Physics},
    author = {Lindblad, G.},
    month = jun,
    year = {1976},
    pages = {119--130},
}

@article{gorini_completely_1976,
    title = {Completely positive dynamical semigroups of {N}‐level systems},
    volume = {17},
    issn = {0022-2488},
    url = {https://doi.org/10.1063/1.522979},
    doi = {10.1063/1.522979},
    number = {5},
    urldate = {2026-03-31},
    journal = {Journal of Mathematical Physics},
    author = {Gorini, Vittorio and Kossakowski, Andrzej and Sudarshan, E. C. G.},
    month = may,
    year = {1976},
    pages = {821--825},
}

@misc{lee_timescales_2025,
    title = {Timescales, {Squeezing} and {Heisenberg} {Scalings} in {Many}-{Body} {Continuous} {Sensing}},
    url = {http://arxiv.org/abs/2505.04591},
    doi = {10.48550/arXiv.2505.04591},
    urldate = {2025-05-08},
    publisher = {arXiv},
    author = {Lee, Gideon and Belyansky, Ron and Jiang, Liang and Clerk, Aashish A.},
    month = may,
    year = {2025},
    note = {arXiv:2505.04591 [quant-ph]},
}

@article{yang_efficient_2023,
    title = {Efficient {Information} {Retrieval} for {Sensing} via {Continuous} {Measurement}},
    volume = {13},
    issn = {2160-3308},
    url = {https://link.aps.org/doi/10.1103/PhysRevX.13.031012},
    doi = {10.1103/PhysRevX.13.031012},
    language = {en},
    number = {3},
    urldate = {2026-03-31},
    journal = {Physical Review X},
    author = {Yang, Dayou and Huelga, Susana F. and Plenio, Martin B.},
    month = jul,
    year = {2023},
    pages = {031012},
}

@article{godley_adaptive_2023,
    title = {Adaptive measurement filter: efficient strategy for optimal estimation of quantum {Markov} chains},
    volume = {7},
    issn = {2521-327X},
    shorttitle = {Adaptive measurement filter},
    url = {https://quantum-journal.org/papers/q-2023-04-06-973/},
    doi = {10.22331/q-2023-04-06-973},
    language = {en},
    urldate = {2026-03-31},
    journal = {Quantum},
    author = {Godley, Alfred and Guta, Madalin},
    month = apr,
    year = {2023},
    pages = {973},
}

@article{girotti_estimating_2025,
    title = {Estimating quantum {Markov} chains using coherent absorber post-processing and pattern counting estimator},
    volume = {9},
    issn = {2521-327X},
    url = {http://arxiv.org/abs/2408.00626},
    doi = {10.22331/q-2025-08-27-1835},
    language = {en},
    urldate = {2026-03-31},
    journal = {Quantum},
    author = {Girotti, Federico and Godley, Alfred and Guţă, Mădălin},
    month = aug,
    year = {2025},
    note = {arXiv:2408.00626 [quant-ph]},
    pages = {1835},
}

@article{carde_nonperturbative_2026,
    title = {Nonperturbative {Switching} {Rates} in {Bistable} {Open} {Quantum} {Systems}: {From} {Driven} {Kerr} {Oscillators} to {Dissipative} {Cat} {Qubits}},
    volume = {136},
    issn = {0031-9007, 1079-7114},
    shorttitle = {Nonperturbative {Switching} {Rates} in {Bistable} {Open} {Quantum} {Systems}},
    url = {https://link.aps.org/doi/10.1103/q981-pd5j},
    doi = {10.1103/q981-pd5j},
    language = {en},
    number = {10},
    urldate = {2026-03-31},
    journal = {Physical Review Letters},
    author = {Carde, Léon and Gautier, Ronan and Didier, Nicolas and Petrescu, Alexandru and Cohen, Joachim and McDonald, Alexander},
    month = mar,
    year = {2026},
    pages = {100402},
}

@article{tucker_facilitating_2020,
    title = {Facilitating spin squeezing generated by collective dynamics with single-particle decoherence},
    volume = {102},
    issn = {2469-9926, 2469-9934},
    url = {https://link.aps.org/doi/10.1103/PhysRevA.102.051701},
    doi = {10.1103/PhysRevA.102.051701},
    language = {en},
    number = {5},
    urldate = {2024-03-13},
    journal = {Physical Review A},
    author = {Tucker, K. and Barberena, D. and Lewis-Swan, R. J. and Thompson, J. K. and Restrepo, J. G. and Rey, A. M.},
    month = nov,
    year = {2020},
    pages = {051701},
}

@article{raveh_dicke_2024,
    title = {Dicke states as matrix product states},
    volume = {110},
    issn = {2469-9926, 2469-9934},
    url = {https://link.aps.org/doi/10.1103/PhysRevA.110.052438},
    doi = {10.1103/PhysRevA.110.052438},
    language = {en},
    number = {5},
    urldate = {2026-04-03},
    journal = {Physical Review A},
    author = {Raveh, David and Nepomechie, Rafael I.},
    month = nov,
    year = {2024},
    pages = {052438},
}

@article{florido-llinas_product_2025,
    title = {The {Product} {Structure} of {Matrix} {Product} {States} under {Permutations}},
    volume = {6},
    url = {https://link.aps.org/doi/10.1103/8sbs-t24w},
    doi = {10.1103/8sbs-t24w},
    number = {4},
    urldate = {2026-04-03},
    journal = {PRX Quantum},
    publisher = {American Physical Society},
    author = {Florido-Llinàs, Marta and Alhambra, Alvaro M. and Trivedi, Rahul and Schuch, Norbert and Pérez-García, David and Cirac, J. Ignacio},
    month = nov,
    year = {2025},
    pages = {040338},
}

@article{koppenhofer_revisiting_2023,
    title = {Revisiting the impact of dissipation on time-reversed one-axis-twist quantum-sensing protocols},
    volume = {5},
    issn = {2643-1564},
    url = {https://link.aps.org/doi/10.1103/PhysRevResearch.5.043279},
    doi = {10.1103/PhysRevResearch.5.043279},
    language = {en},
    number = {4},
    urldate = {2024-10-30},
    journal = {Physical Review Research},
    author = {Koppenhöfer, Martin and Clerk, A. A.},
    month = dec,
    year = {2023},
    pages = {043279},
}

@misc{ostermann_breakdown_2023,
    title = {Breakdown of steady-state superradiance in extended driven atomic arrays},
    url = {http://arxiv.org/abs/2311.10824},
    urldate = {2024-02-25},
    publisher = {arXiv},
    author = {Ostermann, Stefan and Rubies-Bigorda, Oriol and Zhang, Victoria and Yelin, Susanne F.},
    month = nov,
    year = {2023},
    note = {arXiv:2311.10824 [quant-ph]},
}

@article{toth_detection_2007,
    title = {Detection of multipartite entanglement in the vicinity of symmetric {Dicke} states},
    volume = {24},
    issn = {0740-3224, 1520-8540},
    url = {http://arxiv.org/abs/quant-ph/0511237},
    doi = {10.1364/JOSAB.24.000275},
    number = {2},
    urldate = {2026-04-14},
    journal = {Journal of the Optical Society of America B},
    author = {Toth, Geza},
    month = feb,
    year = {2007},
    note = {arXiv:quant-ph/0511237},
    pages = {275},
}

@article{dicke_coherence_1954,
    title = {Coherence in {Spontaneous} {Radiation} {Processes}},
    volume = {93},
    url = {https://link.aps.org/doi/10.1103/PhysRev.93.99},
    doi = {10.1103/PhysRev.93.99},
    number = {1},
    urldate = {2026-04-14},
    journal = {Physical Review},
    publisher = {American Physical Society},
    author = {Dicke, R. H.},
    month = jan,
    year = {1954},
    pages = {99--110},
}

@misc{ptaszynski_quantum_2026,
    title = {Quantum instanton approach to metastable collective spins},
    url = {http://arxiv.org/abs/2604.15091},
    doi = {10.48550/arXiv.2604.15091},
    urldate = {2026-04-17},
    publisher = {arXiv},
    author = {Ptaszynski, Krzysztof and Chudak, Maciej and Esposito, Massimiliano},
    month = apr,
    year = {2026},
    note = {arXiv:2604.15091 [quant-ph]},
}

@misc{lingenfelter_exact_2026,
    title = {Exact steady states of interacting driven dissipative fermionic systems with hidden time-reversal symmetry},
    url = {http://arxiv.org/abs/2605.10846},
    doi = {10.48550/arXiv.2605.10846},
    urldate = {2026-05-12},
    publisher = {arXiv},
    author = {Lingenfelter, Andrew and Clerk, Aashish A.},
    month = may,
    year = {2026},
    note = {arXiv:2605.10846 [quant-ph]},
}

@article{dutta_quantum_2025,
    title = {Quantum {Origin} of {Limit} {Cycles}, {Fixed} {Points}, and {Critical} {Slowing} {Down}},
    volume = {134},
    issn = {0031-9007, 1079-7114},
    url = {https://link.aps.org/doi/10.1103/PhysRevLett.134.050407},
    doi = {10.1103/PhysRevLett.134.050407},
    language = {en},
    number = {5},
    urldate = {2025-03-27},
    journal = {Physical Review Letters},
    author = {Dutta, Shovan and Zhang, Shu and Haque, Masudul},
    month = feb,
    year = {2025},
    pages = {050407},
}

@article{chen_quantum_2023,
    title = {Quantum behavior of the {Duffing} oscillator at the dissipative phase transition},
    volume = {14},
    copyright = {2023 The Author(s)},
    issn = {2041-1723},
    url = {https://www.nature.com/articles/s41467-023-38217-x},
    doi = {10.1038/s41467-023-38217-x},
    language = {en},
    number = {1},
    urldate = {2026-06-03},
    journal = {Nature Communications},
    publisher = {Nature Publishing Group},
    author = {Chen, Qi-Ming and Fischer, Michael and Nojiri, Yuki and Renger, Michael and Xie, Edwar and Partanen, Matti and Pogorzalek, Stefan and Fedorov, Kirill G. and Marx, Achim and Deppe, Frank and Gross, Rudolf},
    month = may,
    year = {2023},
    pages = {2896},
}

@article{beaulieu_observation_2025,
    title = {Observation of first- and second-order dissipative phase transitions in a two-photon driven {Kerr} resonator},
    volume = {16},
    copyright = {2025 The Author(s)},
    issn = {2041-1723},
    url = {https://www.nature.com/articles/s41467-025-56830-w},
    doi = {10.1038/s41467-025-56830-w},
    language = {en},
    number = {1},
    urldate = {2026-06-03},
    journal = {Nature Communications},
    publisher = {Nature Publishing Group},
    author = {Beaulieu, Guillaume and Minganti, Fabrizio and Frasca, Simone and Savona, Vincenzo and Felicetti, Simone and Di Candia, Roberto and Scarlino, Pasquale},
    month = mar,
    year = {2025},
    pages = {1954},
}

@article{chamberland_building_2022,
    title = {Building a fault-tolerant quantum computer using concatenated cat codes},
    volume = {3},
    issn = {2691-3399},
    url = {http://arxiv.org/abs/2012.04108},
    doi = {10.1103/PRXQuantum.3.010329},
    number = {1},
    urldate = {2026-06-03},
    journal = {PRX Quantum},
    author = {Chamberland, Christopher and Noh, Kyungjoo and Arrangoiz-Arriola, Patricio and Campbell, Earl T. and Hann, Connor T. and Iverson, Joseph and Putterman, Harald and Bohdanowicz, Thomas C. and Flammia, Steven T. and Keller, Andrew and Refael, Gil and Preskill, John and Jiang, Liang and Safavi-Naeini, Amir H. and Painter, Oskar and Brandão, Fernando G. S. L.},
    month = feb,
    year = {2022},
    note = {arXiv:2012.04108 [quant-ph]},
    pages = {010329},
}

\onecolumngrid
\appendix

\newpage

\section*{Appendices}

    \section{Proof of hTRS conditions}\label{app:qdb-htrs-proof}
        In this section we provide a proof for Eqs.~\eqref{eq:htrs-dark-state-condition} and \eqref{eq:htrs-eigenstate-condition} using the quantum detailed-balance conditions in Eqs.~\eqref{eq:quantum-detailed-balance}. This can be seen as a generalization of Ref.~\cite{roberts_hidden_2021}.
        We start by noting that the anti-linear operator $\hat \Psi$ in Eq.~\eqref{eq:quantum-detailed-balance} can be decomposed via a polar decomposition as
        \begin{align}
            \hat \Psi = \hat \Phi \hat {\mathcal K}
        \end{align}
        where $\hat \Phi$ is a linear operator and corresponds to a Cholesky decomposition of $\hat \rho_\text{ss} = \hat \Phi \hat \Phi^\dagger$. $\hat {\mathcal K}$ is complex conjugation in some chosen basis $\{\ket k\}$ ($\hat{\mathcal K}\ket k = \ket k$) with $\hat {\mathcal K} \hat {\mathcal K} = \hat {\mathbb 1}$ and $\hat {\mathcal K} c = c^* \hat {\mathcal K}$ for any complex scalar $c$. Note that $\hat{\mathcal K}$ is related to $\hat{\mathcal T}$ (used in the main text) by a unitary operator. $\hat{\mathcal K}$ acts on linear operators $\hat A$ as
        \begin{align}
            \hat {\mathcal K} \hat A = \hat A^* \hat{\mathcal K}
        \end{align}
        where $\hat A^*$ is the complex conjugate in the basis $\{|k\rangle\}$; analogously $\hat A^T$ is the transpose of $\hat A$ in this basis. We can insert this decomposition of $\hat \Psi$ into Eqs.~\eqref{eq:quantum-detailed-balance} and obtain
        \begin{align}
            \hat X \hat \Psi = \hat \Psi \hat X^\dagger \Longleftrightarrow \hat X \hat \Phi \hat{\mathcal K} =\hat \Phi \hat X^T \hat{\mathcal K}
            \label{eq:manipulated-qdb}
        \end{align}
        where $\hat X \in \{\hat H_\text{eff}, \{\hat L_j\}\}$ is either the effective Hamiltonian $\hat H_\text{eff} = \hat H -i \sum_j \hat L_j^\dagger \hat L_j / 2$ or any jump operator $\hat L_j$. Now we will essentially vectorize Eq.~\eqref{eq:manipulated-qdb} onto a doubled Hilbert space of the original system $A$ and the ancillary system $B$--- we will need to be careful with $\hat{\mathcal K}$ which is then defined as complex conjugation of the doubled Hilbert space (instead of just the single Hilbert space). Let us define the canonical purification of the identity operator on system $A$ and $B$ with respect to $\hat{\mathcal K}$
        \begin{align}
            |\mathbb 1 \rangle_\text{AB} = \sum_k |j\rangle_\text{A}\otimes|j^*\rangle_\text{B}.
            \label{eq:purification-maximally mixed}
        \end{align}
        where $|j^*\rangle = \hat{\mathcal K}|j\rangle$ is complex conjugation\ie time reversal, on the $B$ system. This ensures that the state is invariant under gauge transformations $|j\rangle \to e^{i\phi_j}|j\rangle$. 
        This state has the useful property that the following identity holds
        \begin{align}
            (\hat O \otimes \hat{\mathbb 1})|\mathbb 1 \rangle_\text{AB} = \sum_{j,k}O_{jk} |j\rangle_{\rm A}\otimes |k^*\rangle_{\rm B} = \sum_{j,k} (O^T)_{jk}|k\rangle_{\rm A}\otimes |j^*\rangle_{\rm B} = (\hat{\mathbb 1}\otimes \hat O^T)|\mathbb 1\rangle_\text{AB}
            \label{eq:transposition-identity}
        \end{align}
        for any (linear) operator $\hat O = \sum_{j,k} O_{jk}|j\rangle\langle k|$. 
        
\begin{figure}[t]
            \centering
            \includegraphics[width=0.75\linewidth]{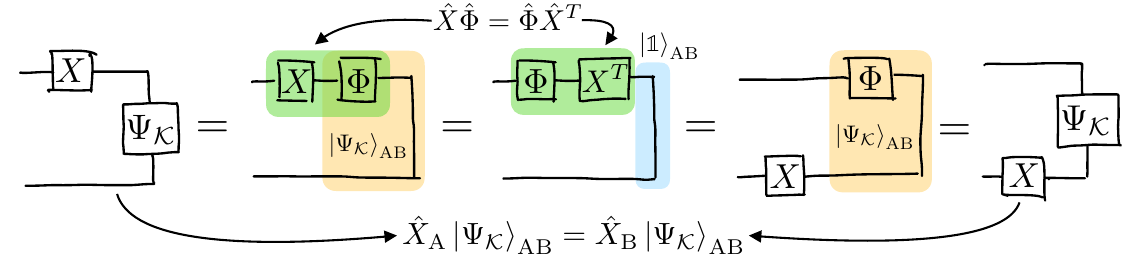}
            \caption{
            Diagrammatic proof for the equivalence of Eq.~\eqref{eq:manipulated-qdb} and Eq.~\eqref{eq:dark-state-condition},~\eqref{eq:htrs-heff-doubled-system}. On the left-hand side we identify $\hat X_\text A \ket{\Psi_{\mathcal K}}_\text{AB}$. First we use the single-system hTRS condition (cf. Eq.~\eqref{eq:manipulated-qdb}) before we use the transposition property (cf. Eq.~\eqref{eq:transposition-identity}) to arrive at $\hat X_\text B\ket{\Psi_{\mathcal K}}_\text{AB}$ which implies $(\hat X_{\rm A} - \hat X_{\rm B})|\Psi_\mathcal K\rangle_{\rm AB} =0$.}
            \label{fig:fig10}
        \end{figure}

        We can now interpret Eq.~\eqref{eq:manipulated-qdb} in terms of this doubled system. Let us first define the steady-state purification 
        \begin{align}
            |\Psi_{\mathcal K}\rangle_\text{AB} &\equiv (\hat \Phi\otimes \hat{\mathbb 1})|\mathbb 1\rangle_\text{AB}.
            \label{eq:antiunitary-purified-steady-state}
        \end{align}
        Then we can write 
        \begin{align}
            \hat X_{\rm A} \ket{\Psi_\mathcal K}_{\rm AB} \equiv (\hat X \otimes \hat{\mathbb 1})(\hat \Phi \otimes \hat{\mathbb 1}) |\mathbb 1\rangle_\text{AB} &= (\hat X\hat \Phi \otimes \hat {\mathbb 1}) |\mathbb 1\rangle_\text{AB} = (\hat \Phi \hat X^T \otimes \hat {\mathbb 1}) |\mathbb 1\rangle_\text{AB} 
            \label{eq:qbd-htrs-lhs-app}
        \end{align}
        where we used Eq.~\eqref{eq:manipulated-qdb} in the final step. Now we make use of the transposition property in Eq.~\eqref{eq:transposition-identity} to show that the RHS is equal to
        \begin{align}
            (\hat \Phi \hat X^T \otimes \hat {\mathbb 1}) |\mathbb 1\rangle_\text{AB}  = (\hat \Phi \otimes \hat X)|\mathbb 1\rangle_\text{AB}=  (\hat{\mathbb 1} \otimes \hat X)(\hat \Phi \otimes \hat{\mathbb 1})|\mathbb 1\rangle_\text{AB}\equiv \hat X_{\rm B} \ket{\Psi_\mathcal K}_{\rm AB}.
            \label{eq:qbd-htrs-rhs-app}
        \end{align}
        Here we also defined the operators $\hat X_{\rm A} \equiv \hat X \otimes \hat{\mathbb 1}$ and $\hat X_{\rm B}\equiv\hat{\mathbb 1}\otimes \hat X$ which act nontrivially only on the $A$ or $B$ subsystem.
        Combining the LHS of Eq.~\eqref{eq:qbd-htrs-lhs-app} and RHS of Eq.~\eqref{eq:qbd-htrs-rhs-app} we finally obtain
        \begin{align}
            (\hat X_{\text A} - \hat X_{\text B}) |\Psi_{\mathcal K}\rangle_\text{AB} = 0
        \end{align}
        which gives us the relations
        \begin{align}
            (\hat L_{j,\text A} - \hat L_{j,\text B}) |\Psi_\mathcal K\rangle_\text{AB} &= 0\label{eq:dark-state-condition}\\
            (\hat H_{\text{eff}, \text A} - \hat H_{\text{eff}, \text B}) |\Psi_\mathcal K\rangle_\text{AB} &= 0.
            \label{eq:htrs-heff-doubled-system}
        \end{align}
        In Fig.~\ref{fig:fig10} we provide a diagrammatic proof for the equivalence between the single-system hTRS condition in Eq.~\eqref{eq:manipulated-qdb} and these doubled-system conditions in Eqs.~\eqref{eq:dark-state-condition} and~\eqref{eq:htrs-heff-doubled-system}.

        Eq.~\eqref{eq:dark-state-condition} is equivalent to the dark-state condition in Eq.~\eqref{eq:htrs-dark-state-condition}. As for Eq.~\eqref{eq:htrs-heff-doubled-system} we can use that the doubled-system Hamiltonian $\hat H_\text{AB}$ can be written as
        \begin{align}
            \hat H_\text{AB}  &\equiv \hat H_\text{A} - \hat H_\text{B} -\frac i 2 \sum_j \Big [ \hat L_{j,\text A}^\dagger \hat L_{j, \text B} - \hat L_{j,\text B}^\dagger \hat L_{j,\text A} \Big]\\
            &= \hat H_{\text{eff}, \text A} - \hat H_{\text{eff}, \text B} - \frac i 2 \sum_j \Big [\hat L^\dagger_{j,\text A} \big (\hat L_{j,\text B} - \hat L_{j,\text A}\big ) - \hat L^\dagger_{j,\text B} \big (\hat L_{j,\text A} - \hat L_{j,\text B}\big ) \Big ]
        \end{align}
        where we used $\hat H_\text{eff} = \hat H -i \sum_j \hat L^\dagger_j \hat L_j / 2$. By applying the dark-state condition in Eq.~\eqref{eq:dark-state-condition} we finally obtain
        \begin{align}
            &(\hat H_{\text{eff}, \text A} - \hat H_{\text{eff}, \text B}) |\Psi_\mathcal K\rangle_\text{AB} = 0\Longrightarrow \hat H_\text{AB} |\Psi_\mathcal K\rangle_\text{AB} = 0. \label{eq:htrs-HAB-condition}
        \end{align}
        Eqs.~\eqref{eq:dark-state-condition} and \eqref{eq:htrs-HAB-condition} together imply that the steady state of the doubled system $|\Psi_\mathcal K\rangle_{\rm AB}$ is pure. By construction we also know that the partial trace over system $B$ recovers the single-system steady state $\hat \rho_\text{ss}$
        \begin{align}
            \Tr_\text B \Big ( |\Psi_\mathcal K\rangle\langle \Psi_\mathcal K|_\text{AB} \Big ) = \Tr_\text B \Big ( (\hat \Phi \otimes \hat{\mathbb 1})|\mathbb 1\rangle_\text{AB}\langle \mathbb 1 | (\hat \Phi^\dagger \otimes \hat{\mathbb 1}) \Big ) = \hat \Phi \hat \Phi^\dagger = \hat \rho_\text{ss}.
        \end{align}
        Here we used that the partial trace over one subsystem of $|\mathbb 1\rangle_\text{AB}$ is simply the identity matrix on the remaining system:
        \begin{align}
            \Tr_\text B \Big ( |\mathbb 1\rangle_\text{AB}\langle \mathbb 1 |  \Big ) = \hat{\mathbb 1}_\text{A}.
        \end{align}
        This proves that $|\Psi_\mathcal K\rangle_\text{AB}$ is a particular purification of $\hat \rho_\text{ss}=\sum_n p_n|n\rangle\langle n|$ ($p_n\geq0, \sum_n p_n =1$)\ie
        \begin{align}
            |\Psi_\mathcal K\rangle_\text{AB} = \sum_n \sqrt{p_n} |n\rangle_\text{A}\otimes \hat U\hat{\mathcal K}|n\rangle_\text{B}
        \end{align}
        for some unitary $\hat U$.
        
        Finally we remark that $\hat{\mathcal K}$ does play an important role if $|\Psi_\mathcal K\rangle_\text{AB}$ is expressed in any other basis than the basis of $\hat{\mathcal K}$. It leads to a time-reversal symmetry of certain two-point correlation functions as shown in Ref.~\cite{roberts_hidden_2021}. Importantly, Ref.~\cite{roberts_hidden_2021} showed that $\hat{\mathcal K}\hat \rho_\text{ss} = \hat \rho_\text{ss}\hat{\mathcal K}$\ie $\hat \rho_\text{ss}$ is diagonal in the basis $\{\ket k\}$ for models with hTRS. This implies that $|\Psi_\mathcal K\rangle_\text{AB}$ is necessarily symmetric under exchange of system $A$ and system $B$. 


    \section{Driven-dissipative Nonlinear Cavity Model}\label{app:kerr}

        \subsection{Mean-field Analysis}\label{app:kerr-mean-field}
            First of all, let us generalize the model in Eq.~\eqref{eq:Kerr-qme} to include both two-photon driving and dissipation. Then the Lindblad master equation takes the form
            \begin{align}
                \mathcal L_\text{Kerr}' \hat \rho \equiv \partial_t \hat \rho &= -i \left [\Delta \hat a^\dagger \hat a -\frac U 2 \hat a^\dagger \hat a^\dagger \hat a \hat a + F_1 (\hat a + \hat a^\dagger) + \frac{F_2}2 (\hat a^2 + \hat a^\dagger{^2}), \hat \rho\right ]+ \kappa_1\mathcal D[\hat a]\hat \rho + \kappa_2 \mathcal D[\hat a^2]\hat \rho
                \label{eq:kerr-qme-two-photon}
            \end{align}
            where $F_1, F_2$ denote the one- and two-photon driving strengths and $\kappa_1,\kappa_2$ denote the one- and two-photon relaxation rates. The corresponding mean-field equation for the coherence $\alpha\equiv \langle \hat a(t)\rangle$, which neglects quantum fluctuations, reads
            \begin{align}
                i\partial_t \alpha = F_1 + F_2 \alpha^* + \Big (\Delta -i \frac {\kappa_1} 2\Big )\alpha - (U + i\kappa_2) |\alpha|^2 \alpha
                \label{eq:kerr-general-mean-field}
            \end{align}
            where $\alpha^*$ is simply the complex conjugate of $\alpha$. Let us now focus on the mean-field limit of the model studied in the main text in Sec.~\ref{subsec:driven-dissipative-kerr}: this requires taking $F_2,\kappa_2=0$ and defining $F=F_1$ and $\kappa=\kappa_1$\footnote{We will discuss the explicit $N$ dependence of parameters below.}. 
            Phenomenologically, there is a weak U(1) symmetry for $F\to 0$ and due to the absence of driving the steady state is trivially the vacuum. For $F>0$ the drive breaks the weak U(1) symmetry and competes with the decay $\kappa>0$ leading to a non-equilibrium steady state; the Kerr nonlinearity $U$ contributes a photon-number-dependent detuning which makes the drive exceedingly off-resonant, and therefore, inefficient for large photon occupations. The mean-field equation (cf. Eq.~\eqref{eq:kerr-general-mean-field}) simplifies to
            \begin{align}
                i\partial_t \alpha = F + \Big (\Delta -i \frac \kappa 2\Big )\alpha - U |\alpha|^2 \alpha.
            \end{align}
            Despite the fact that we are dealing with a single-body problem, it will be convenient to define an effective ``thermodynamic limit". Since the Hilbert space of a bosonic mode is unbounded,  the photon population $|\alpha|^2$ can diverge and be treated as an extensive quantity\ie we introduce the parameter $N$ such that $n_\text{mf}= |\alpha|^2 / N$ is an intensive photon density. Following Ref.~\cite{casteels_critical_2017} we rescale the parameters $\overline U = N U$ and $\overline F = F / \sqrt N$ such that $\overline U \,\overline F^2 = \text{const}$ which ensures a well-defined large-$N$ limit. In the steady state of the mean-field theory we then obtain the relation
            \begin{align}
                \overline{F}^2_\text{mf} = n_\text{mf} \Big [ (\overline U n_\text{mf} -\Delta)^2 + \frac{\kappa^2} 4 \Big ]
                \label{eq:Kerr-mean-field}
            \end{align}
            which is a cubic equation for the photon density $n_\text{mf}$. Therefore there exists a bistable parameter regime with multiple real solutions for $n_\text{mf}$. For large drive $\overline F\to\infty$ the photon density $n_\text{mf} \approx (\overline F/\overline U)^{\sfrac 2 3}$ is large, while $n_\text{mf} \to 0$ is the unique solution for $\overline F\to0$, since the dissipation dominates. As the drive $\overline F$ is tuned, a bistable parameter region emerges where both solutions coexist with a third, unstable state~\cite{carde_nonperturbative_2026}. This behavior can be observed in Fig.~\ref{fig:fig4}a.

        \subsection{Steady-state purification including two-photon processes}\label{app:kerr-comparison-phase-space-methods}
            Here we will discuss our results for the driven-dissipative nonlinear cavity model presented in Sec.~\ref{subsec:driven-dissipative-kerr} in the context of several recent studies~\cite{carde_nonperturbative_2026,mylnikov_qubit_2025,mylnikov_switching_2025}. In particular we will present the exact solution for the steady-state purification in the presence of two-photon driving and dissipation which was first derived in Ref.~\cite{roberts_driven-dissipative_2020}. This enables us to compare predictions for the scaling of the dissipative gap with Refs.~\cite{carde_nonperturbative_2026,mylnikov_qubit_2025,mylnikov_switching_2025} in later section, finding numerical and analytical agreement that underscores the validity of Conjecture~\ref{conj:dissipative-gap}.
    
            Remarkably, the steady state of Eq.~\eqref{eq:kerr-qme-two-photon} can still be solved using the hTRS framework presented in Sec.~\ref{subsec:htrs}. For more details on the derivation of this steady state, we refer to Ref.~\cite{roberts_driven-dissipative_2020}. The purified steady state has the same form as in the main text
            \begin{align}
                |\Psi_\mathcal T\rangle_\text{AB} = \sum_{k=0}^\infty \psi_k |k\rangle_+|0\rangle_-
            \end{align}
            where the complex probability amplitudes $\psi_k$ are determined by a recurrence relation~\cite{roberts_driven-dissipative_2020}
            \begin{align}
                \sqrt{k+1}\big [-(U+i\kappa_2) k + 2 (\Delta - i\frac {\kappa_1} 2)\big ]\psi_{k+1} + 2 \sqrt 2 F_1 \psi_k + 2 F_2 \sqrt k \psi_{k-1} = 0.
            \end{align}
            While there is no closed-form solution of this recurrence relation to the best of our knowledge, it simplifies in the limiting cases $F_1=0$ and $F_2$ = 0. Following Conjecture~\ref{conj:dissipative-gap} we define the non-equilibrium potential for the particle density $n=k/N$
            \begin{align}
                V(n) = -\lim_{N\to\infty} \frac{\log |\psi_k|^2}{N}
            \end{align}
            in terms of the thermodynamic limit $N\to\infty$. For details on this thermodynamic limit and the effective system size $N$, we refer to our previous discussion and Sec.~\ref{subsec:driven-dissipative-kerr}. By analyzing the extrema of the potential $V(n)$ we can obtain the potential barrier height $\Delta V$ which plays a crucial role in the scaling of the dissipative gap $\Gamma_\text{diss}$ of $\mathcal L_\text{Kerr}'$ as per Conjecture~\ref{conj:dissipative-gap}. Therefore we are now equipped to make quantitative comparisons with the results of Refs.~\cite{carde_nonperturbative_2026,mylnikov_qubit_2025,mylnikov_switching_2025}.

        \subsection{Numerical comparison in the limit \texorpdfstring{$F_2\to0, \kappa_2\to0$}{}}\label{app:kerr-numerical-comp}
            We will first take the limits $F_2\to0$ and $\kappa_2\to0$ which are also presented in Sec.~\ref{subsec:driven-dissipative-kerr} of the main text. Here we will compare our results with Ref.~\cite{carde_nonperturbative_2026} which used an instanton calculation based on a Keldysh path integral to estimate the switching rate between bistable states. 
            
            On the one hand, our non-equilibrium potential $V(n)$ is given by
            \begin{align}
                V_\text{Kerr}(n) &= -\int_0^n dn' \log \left [\frac{8}{n'}\frac{\overline F_1 ^2}{(\overline U n' - 2 \Delta)^2 + \kappa_1^2} \right ]\\
                &= -3 n + 2\frac{\kappa_1}{\overline U} \arctan\left (\frac{\overline U n - 2\Delta}{\kappa_1} \right ) + n \log\left [\frac 8 n \frac{\overline F_1^2}{(\overline U n -2\Delta)^2 + \kappa_1^2} \right ] + \frac{2\Delta}{\overline U}\log\left [ \frac{(\overline U n - 2 \Delta)^2 + \kappa_1^2}{\overline F_1^2} \right] +\text{const}
            \end{align}
            in Eq.~\eqref{eq:Kerr-free-energy} of the main text. Here we explicitly integrated the potential in the second line up to an (irrelevant) constant. The locations of the extrema $n_i\in\{n_1, n_*, n_2\}$ of $V_\text{Kerr}(n)$ are determined by 
            \begin{align}
                \overline F_1^2 = \frac{n}{8}\left [ (\overline U n - 2\Delta)^2 + \kappa_1^2\right ]
            \end{align}
            which is equivalent to the mean-field equation in Eq.~\eqref{eq:Kerr-mean-field} up to rescaling $n\to n/2$. We then obtain the potential barrier
            \begin{align}
                \Delta V = V_\text{Kerr}(n_*) - \max_{i\in\{1, 2\}} V_\text{Kerr}(n_i)
                \label{eq:kerr-potential-barrier-app1}
            \end{align}
            which can easily be evaluated numerically.

\begin{figure} 
                \centering
                \includegraphics[width=0.6\columnwidth]{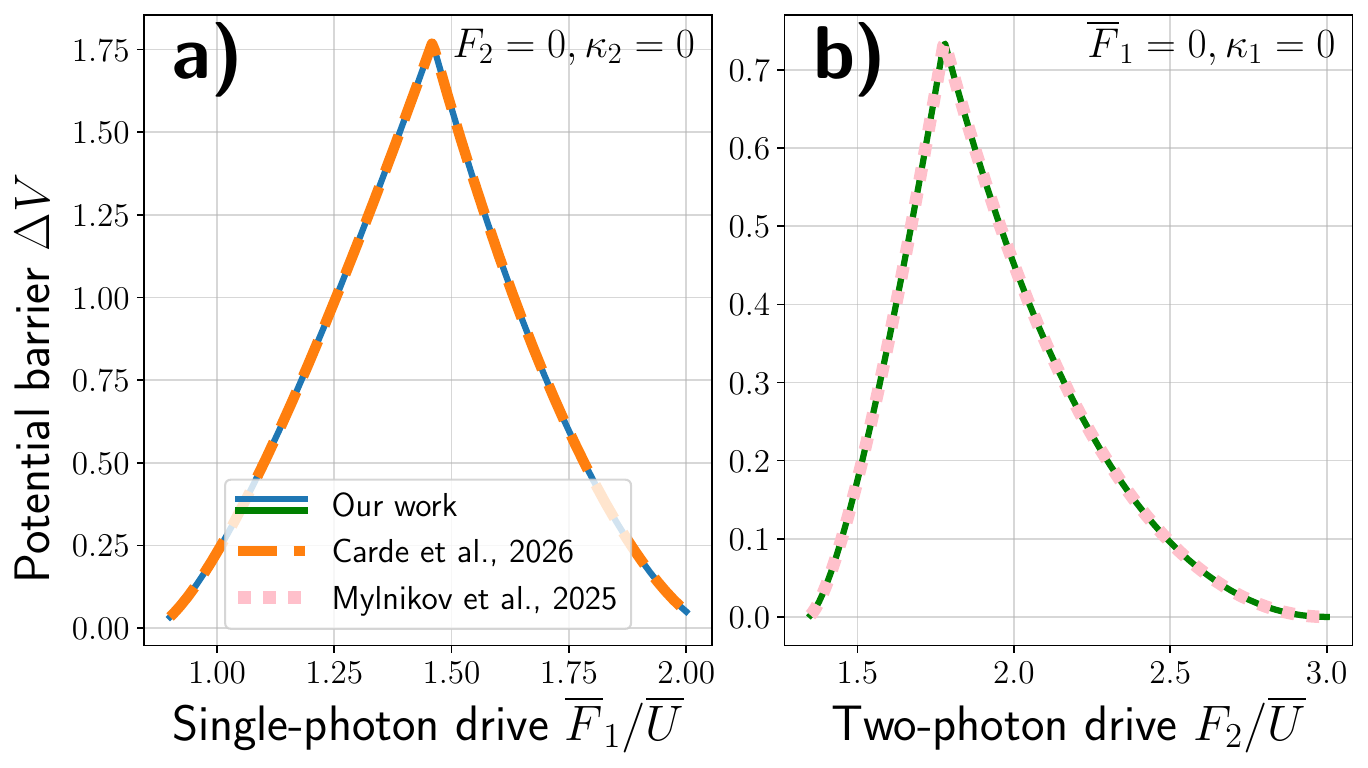}
                \caption{Potential barrier of the driven-dissipative nonlinear cavity model in different parameter regimes. We compare our result for the potential barrier with other recent predictions~\cite{carde_nonperturbative_2026,mylnikov_qubit_2025,mylnikov_switching_2025}. \textbf{a)} For the limiting case, $F_2,\overline\kappa_2=0$, considered in the main text we find excellent numerical agreement with the instanton calculation of Ref.~\cite{carde_nonperturbative_2026}. \textbf{b)} For $\overline F_1, \kappa_1=0$ and nonzero two photon driving $F_2$ and decay $\overline\kappa_2$, our prediction for the potential barrier analytically agrees with the results of Refs.~\cite{mylnikov_qubit_2025} and ~\cite{mylnikov_switching_2025}.}
                \label{fig:fig11}
            \end{figure}

            On the other hand, Ref.~\cite{carde_nonperturbative_2026} showed that the switching rate $\Gamma_{i\to j}$ between the mean-field bistable states $i,j$ is given by the Keldysh action $S_{i\to j}$ 
            \begin{align}
                \Gamma_{i\to j} \propto \exp(i S_{i\to j})
            \end{align}
            evaluated along a particular time-reversed trajectory known as the switching path. Ref.~\cite{carde_nonperturbative_2026} finds that
            \begin{align}
                i S_{i\to j} &= \int_{\alpha_i^*}^{\alpha_\text{u}^*} d\alpha^* \left [ 2\alpha - \frac{2\left (\Delta + i\frac {\kappa_1} 2 \right ) \alpha^* + 2 \overline F_1}{\overline U \alpha^*{}^2} \right ] + \int_{\alpha_i}^{\alpha_\text{u}} d\alpha \left [ 2 \alpha^* - \frac{2 \left ( \Delta -i\frac {\kappa_1} 2\right )\alpha + 2 \overline F_1}{\overline U \alpha^2}\right ]\\
                &= \left [ 2 |\alpha|^2 + 4 \text{Re}\left ( \frac{\overline F_1}{\overline U \alpha} \right ) - 4 \text{Re}\left (\frac{\Delta -i \frac {\kappa_1} 2}{\overline U}\log(\alpha) \right ) \right ]_{\alpha_i}^{\alpha_\text{u}} \equiv \Phi(\alpha)\Big|_{\alpha_i}^{\alpha_\text{u}}
                \label{eq:carde-potential}
            \end{align}
            where $\alpha_\text{u}$ is the unstable state and $\alpha_i$ is one of the bistable states. $\alpha_i$ and $\alpha_\text u$ are determined by the mean-field steady-state condition in Eq.~\eqref{eq:kerr-general-mean-field}. The potential barrier is then simply
            \begin{align}
                \Delta V' = \Phi(\alpha_\text u) - \max_i\Phi(\alpha_i)
                \label{eq:kerr-potential-barrier-app2}
            \end{align}
            such that the dissipative gap scales as
            \begin{align}
                \Gamma_\text{diss} \propto \exp(-N\Delta V')
            \end{align}
            In Fig.~\ref{fig:fig11}a we numerically compare the predictions of Eqs.~\eqref{eq:kerr-potential-barrier-app1} and~\eqref{eq:kerr-potential-barrier-app2} and find remarkable agreement up to machine precision.

        \subsection{Analytical comparison in the limit \texorpdfstring{$F_1\to0, \kappa_1\to0$}{}}\label{app:kerr-analytical-comp}
            Having established that our results match the instanton calculation presented in Ref.~\cite{carde_nonperturbative_2026} for the model discussed in the main text, we now consider the limiting case of only two-photon driving and dissipation, i.e., $F_1,\kappa_1=0$. In this limit the Lindbladian has a strong $\mathbb Z_2$ symmetry~\cite{albert_symmetries_2014,buca_note_2012} ($\hat a \to -\hat a$). While this makes parts of our analysis easier, it makes the exact steady-state solution a bit subtle as we will see below. This model was studied in detail in Ref.~\cite{mylnikov_qubit_2025,mylnikov_switching_2025} by making use of the complex-P representation of the steady state. 

            In this limit the mean-field analysis takes a particularly simple form
            \begin{align}
                i\partial_t \alpha = F_2 \alpha^* + \Delta\alpha - (U + i\kappa_2) |\alpha|^2 \alpha
            \end{align}
            such that the steady-state condition becomes
            \begin{align}
                F_2^2 |\alpha|^2 = \big [ (U|\alpha|^2 -\Delta)^2 + \kappa_2^2 |\alpha|^2\big ] |\alpha^2|.
            \end{align}
            Clearly $\alpha=0$ is a trivial steady-state solution along with the following two solutions
            \begin{align}
                n_\pm \equiv |\alpha_{\pm}|^2 = \frac{U \Delta \pm \sqrt{(U^2 +\kappa_2^2)F_2^2 - \kappa_2^2 \Delta^2}}{U^2 + \kappa_2^2}.
                \label{eq:kerr-two-photon-mf-solutions}
            \end{align}
            The useful advantage of this limit over the previous section is the simple, closed form of the mean-field solutions. 
            
            Ref.~\cite{mylnikov_qubit_2025} performs an instanton calculation for the switching rate between the metastable, mean-field solutions. The authors of Ref.~\cite{mylnikov_qubit_2025} define a complex potential in terms of the complex-P representation $P_\text{ss}(\alpha,\beta)$ 
            \begin{align}
                \Phi(\alpha,\beta) = -\log\big [ P_\text{ss}(\alpha,\beta)\big].
            \end{align}
            Using an exact expression for the complex-P representation they propose the scaling
            \begin{align}
                \Gamma_\text{diss} \propto \exp(-N\Delta \Phi)
            \end{align}
            for the instanton switching rate where the potential barrier
            \begin{align}
                \Delta \Phi &= 2\frac{\overline U \Delta + \sqrt{(\overline U^2 + \overline \kappa_2^2)F_2^2 - \overline\kappa_2^2 \Delta^2}}{\overline U^2 + \overline \kappa_2^2} -2 \frac{2\overline\kappa_2\Delta}{\overline U^2 + \overline \kappa_2^2}\left \{\arctan\left [\frac{\sqrt{(\overline U^2 + \overline \kappa_2^2)F_2^2 - \overline\kappa_2^2\Delta^2}}{\overline\kappa_2 \Delta} \right] - \arctan \left ( \frac{\overline U}{\overline \kappa_2}\right ) \right \}\nonumber\\ 
                &+ 2\frac{\overline U \Delta}{\overline U^2 + \overline\kappa_2^2}\log\left | \frac{F_2}{\Delta}\right |
                \label{eq:kerr-two-photon-barrier}
            \end{align}
            is obtained using the potential $\Phi(\alpha,\beta)$. This potential barrier exactly matches the instanton calculation of Ref.~\cite{carde_nonperturbative_2026}. For more details on this derivation we refer to Ref.~\cite{mylnikov_qubit_2025}. By numerically diagonalizing the Lindbladian and extracting the dissipative gap, Refs.~\cite{mylnikov_qubit_2025,mylnikov_switching_2025} confirmed this prediction.

            Let us now compare to our approach in Conjecture~\ref{conj:dissipative-gap} and confirm that it reproduces the instanton result in Eq.~\eqref{eq:kerr-two-photon-barrier}. The previously mentioned subtlety in the exact solution is a consequence of the strong $\mathbb Z_2$ symmetry: the Hilbert space decouples into even- and odd-parity sectors each containing (at least) one steady state. So far, we were always assuming a unique steady state. We will limit ourselves to the even-parity sector, so we are only predicting the scaling of the dissipative gap in the even-parity sector. 
            Solving for the dark-state in the doubled system, it is important that the doubled-system Lindbladian ${\mathcal L}_{\rm AB}$ (cf. Eq.~\eqref{eq:htrs-doubled-system-dynamics}) exhibits another strong $\mathbb Z_2$ symmetry in the even-parity sector corresponding to exchanging the even and odd modes $\hat c_+\leftrightarrow \hat c_-$. This leads to another steady-state degeneracy in the even-parity sector which might complicate our calculation. However, the dark states with even-parity under exchange of $A$ and $B$ are $|d_{2k}\rangle_{\rm AB} = |2k\rangle_+|0\rangle_-$ and $|d_{2k}'\rangle_{\rm AB} = |0\rangle_+|2k\rangle_-$. Since these two sets of states are simply related by exchanging $\hat c_+ \leftrightarrow \hat c_-$, so are the two degenerate steady states. Therefore, we can restrict ourselves to only working with $|d_{2k}\rangle_{\rm AB}$. If we write the steady-state purification in terms of the even-parity dark states $|d_{2k}\rangle_{\rm AB} = |2k\rangle_+|0\rangle_-$, then the other steady-state solution is simply obtained by taking $|d_{2k}\rangle_{\rm AB} \to |d_{2k}'\rangle_{\rm AB}$. Note that both of these degenerate steady states recover the same steady state after tracing over the $B$ system.
            
            Taking the appropriate parameter limits $\kappa_1\to 0, F_1 \to 0$, the recurrence relation remains valid and simplifies to a one-term recursion
            \begin{align}
                \psi_{k+1} = \frac{2F_2}{(U + i \kappa_2) k -2 \Delta}\sqrt{\frac{k}{k+1}}\psi_{k-1}
            \end{align}
            which is easily solved. The strong $\mathbb Z_2$ symmetry requires $\psi_k = 0$ for any odd integer $k=2j + 1, j\in\mathbb N$ in the even-parity sector. Therefore we define the probability distribution $p_k = |\psi_{2k}|^2$ for positive integers $k=2j\; (j\in \mathbb N)$ which obeys
            \begin{align}
                \frac{p_{k+1}}{p_k} = \left |\frac{2F_2}{(U + i\kappa_2)(2k+1)-2\Delta} \sqrt{\frac{2k+1}{2k+2}} \right |^2.
            \end{align}
            Using the rescaled parameters $\overline U = NU$ and $\overline \kappa_2 = N\kappa_2$ and taking the thermodynamic limit, we obtain the non-equilibrium potential for $n=k/N$
            \begin{align}
                V_\text{Kerr2}(n) &= -\int_0^n dn' \log\left [ \frac{F_2 }{(\overline U n' -\Delta)^2 + (\overline\kappa_2 n')^2}\right ]\nonumber\\
                &= -2n + \frac{2\overline\kappa_2\Delta}{\overline U^2 +\overline \kappa_2^2}\arctan\left [\frac{(\overline U^2 + \overline\kappa_2^2)n-\overline U \Delta}{\overline\kappa_2\Delta} \right ] + \left (\frac{\overline U\Delta}{\overline U^2 + \overline\kappa_2^2}-n \right )\log \left [ \frac{F_2^2}{(\overline U n - \Delta)^2 + (\overline \kappa_2 n)^2}\right ].
            \end{align}
            The extrema of this potential exactly coincide with the steady-state solutions at the mean-field level in Eq.~\eqref{eq:kerr-two-photon-mf-solutions}. It can be easily confirmed that 
            \begin{align}
                \Delta V = V_\text{Kerr2}(0) - V_\text{Kerr2}(n_+) = \Delta \Phi
            \end{align}
            in exact agreement with Eq.~\eqref{eq:kerr-two-photon-barrier} which was derived in Refs.~\cite{mylnikov_qubit_2025}. In Fig.~\ref{fig:fig11}b we show the potential barrier for different values of $F_2$ as a point of comparison.

            Finally, we have established that our ansatz for the potential barrier exactly reproduces recent results obtained with vastly different approaches, including a rigorous instanton calculation using a Keldysh path integral. We highlight in particular that our approach does not have any knowledge of the microscopic dynamics beyond the steady state, while an instanton calculation relies on knowledge about the dynamics; this emphasizes the surprising nature of our ansatz. Finally we have provided analytical evidence for the validity of Conjecture~\ref{conj:dissipative-gap} in a nonlinear cavity model beyond the numerics shown in the main text.

    \section{Dissipative Transverse-field Ising Model}\label{app:dtfim-details}

        \subsection{Mean-field analysis of the DTFIM}\label{app:dtfim-mean-field}
            First, let us consider a mean-field description to gain some intuition for the steady-state phase diagram of the DTFIM. Due to the all-to-all interactions, a mean-field description is generally believed to hold in the thermodynamic limit~\cite{mattes_long-range_2025}. By assuming that all correlations decouple\ie $\expval{\hat \sigma^\alpha_i \hat \sigma^\beta_j} \approx \expval{\hat \sigma^\alpha_i}\expval{\hat \sigma^\beta_j}$, we derive the following mean-field equations of motion for the coherence $s_- = \langle \hat S_-\rangle / N = s_+^*$ and the magnetization $s_\text z = \langle \hat S_\text z \rangle/N$
            \begin{align}
                \partial_t s_- &= \left [ -i(2\overline J s_\text z + \Delta) + (\overline \Gamma s_\text z -\gamma /2)\right ]s_- + i\Omega s_\text z\\
                \partial_t s_\text z &= -i \frac\Omega 2 (s_+ - s_-) -\overline \Gamma s_+ s_- -\gamma s_\text z -\frac \gamma 2.
            \end{align}
            The steady-state condition $\partial_t s_\alpha=0$ then implies
            \begin{align}
                \Omega^2 = -\frac{1+2s_\text{z,mf}}{s_\text{z,mf}}\left [ \left(2\overline J s_\text{z,mf} + \Delta\right )^2 + \left (\overline \Gamma s_\text{z,mf} -\frac \gamma 2 \right )^2\right ].
                \label{eq:DTFIM-mean-field}
            \end{align}
            The steady-state magnetization $s_\text{z,mf}$ is the solution to this equation which can easily be determined analytically or numerically. Phenomenologically, the Ising interactions add a magnetization-dependent detuning which can make the Rabi drive (off-)resonant depending on the value of $s_\text z$. Still in the weak driving regime $\Omega\to0$ the dissipation dominates leading to a polarized steady state with $s_\text z \to -1/2$, while the strong driving limit $\Omega \to\infty$ exhibits a depolarized steady state with $s_\text z\to 0$. Due to the nonlinearity of the interactions $\overline J$ and $\overline \Gamma$, a bistable parameter region emerges in-between these two trivial limits: in this bistable parameter regime a polarized and a depolarized steady state coexist with a third, unstable steady state. In Fig.~\ref{fig:fig7} we compare the magnetization of the exact steady-state solution with the mean-field result $s_\text{z,mf}$. In particular, the exact steady-state solution exhibits a first-order phase transition in the middle of this bistable parameter region. Similar to the liquid-vapor transition the resulting line of first-order phase transitions ends in a critical point (green cross) which we highlight in Fig.~\ref{fig:fig7}. For sufficiently weak Ising interactions there is no phase transition, but instead just a smooth crossover---this feature was already observed by Refs.~\cite{marcuzzi_universal_2014,roberts_exact_2023} for similar models, but without a collective decay rate $\overline \Gamma >0$. We note that the discontinuous phase transition observed here is different from the first-order phase transition at equilibrium which is a result of $\mathbb Z_2$ spontaneous symmetry breaking due to the longitudinal field $\Delta$.

        \subsection{Purified steady state of the DTFIM}\label{app:DTFIM-purified-steady-state}
            In this section we present complementary details to the derivation of the purified steady-state solution of the DTFIM in Sec.~\ref{subsec:DTFIM}; this follows our considerations in Secs.~\ref{subsec:htrs} and~\ref{sec:special-dark-states} closely. The single-system Lindbladian $\mathcal L_\text{DTFIM}$ is defined in Eq.~\eqref{eq:DTFIM-qme}. We note that our analysis is closely related and strictly equivalent to the original work in Ref.~\cite{roberts_exact_2023}, but extended to the doubled-system formalism of hTRS. Our goal is constructing a doubled system such that the combined steady state of the two systems is pure. Importantly, we here show three crucial steps of our recipe of Sec.~\ref{subsec:ladder-model}: first, we construct the special bright states $|b_k\rangle_\text{AB}$ (cf. Eq.~\eqref{eq:def-special-bright-states}) such that the image $\hat H_\text{AB}$ acting on the special dark states $|d_k\rangle_\text{AB}$ (cf. Eq.~\eqref{eq:def-special-dark-states}) is fully spanned by the set $\{|b_k\rangle_\text{AB}\}$. This task is simplified because all terms in $\hat H_{\rm AB}$ introduce exactly one ``defect'' into a product state in $\mathbf{D_+}$. Second, we will use these special bright states to obtain the matrix elements $h_{k',k} = \mel{b_{k'}}{\hat H_\text{AB}'}{d_k}_\text{AB}$ of the ladder Hamiltonian $\hat H_\text{AB}'$ (cf. Eq.~\eqref{eq:ladder-model-matrix-elements}). Finally, we solve the recursion relation in Eq.~\eqref{eq:htrs-recursion-relation} to find the steady-state coefficients $\psi_k$.

            In the main text (cf. Sec.~\ref{subsec:DTFIM}) we showed that the even-parity dark-state manifold is given by $\mathbf D_+ = \text{span}(\{\ket{\tilde 0},\ket{\tilde 1}\}^{\otimes N})$ where
            \begin{align}
                |\tilde 0_j \rangle_{\rm AB} &\equiv|00_j\rangle_\text{AB},\\
                |\tilde 1_j\rangle_{\rm AB} &\equiv \frac{|10_j\rangle_\text{AB} + |01_j\rangle_\text{AB}}{\sqrt 2}
            \end{align}
            form a two-dimensional subspace for each qubit pair on site $j$. Since there are many ways of arranging these two states on $N$ sites to obtain a state with total charge $k$, we needed to find a special basis of dark states $\{|\tilde d_k\rangle_{\rm AB}\}_{k=0,\dots,N}$ such that $|\tilde d_k\rangle_{\rm AB}$ is the only state with total charge $k$ (cf. Sec.~\ref{subsec:additional-quantum-numbers}). In Sec.~\ref{subsec:DTFIM} we argued that this degeneracy is lifted by considering the model's permutation symmetry: the steady-state purification $\ket{\Psi_\mathcal T}_{\rm AB}$ must be invariant under \emph{any} permutation of qubit pairs. Therefore the special dark states are a type of Dicke state given by
            \begin{align}
                \ket{\tilde d_k}_\text{AB} = \frac{1}{\sqrt{\binom N k}}|\tilde D^N_k\rangle_{\rm AB} = \frac{1}{\sqrt{\binom N k}}\sum_{\pi} \mathcal P_\pi |\tilde 1\rangle^{\otimes k} |\tilde 0\rangle^{\otimes N-k}
                \label{eq:app-special-dark-states-dtfim}
            \end{align}
            where the sum goes over all distinct permutations $\pi$ of labels $i=1,\dots,N$ to avoid double counting. $\mathcal P_\pi$ permutes the ordering of the product state $|\tilde 1\rangle^{\otimes k} |\tilde 0\rangle^{\otimes N-k}$.

            To find the corresponding special bright states $|\tilde b_k\rangle_{\rm AB}$, which form the basis for the bottom leg in Fig.~\ref{fig:fig4}b, we need to consider the action of $\hat H_{\rm AB}$ on the special dark states $|\tilde d_k\rangle_{\rm AB}$. We will break down $\hat H_{\rm AB}$ term-by-term and understand its action on product states in $\mathbf D_+$. We will then distinguish between non-interacting terms which act only on the qubit pair on site $i$ and interaction terms which act nontrivially on the qubit pairs on sites $i$ and $j$---surprisingly, all of these terms map to states with a single ``defect'' state placed onto a qubit pair at site $i$ of a product state. 
            Let us define the ``defect'' state as the odd-parity state 
            \begin{align}
                |\tilde s_j\rangle_{\rm AB} = \frac{| 10_j\rangle_\text{AB} - |01_j\rangle_\text{AB}}{\sqrt 2}
            \end{align}
            for a qubit pair at site $j$. Since $\hat H_{\rm AB}$ maps even- to odd-parity states and $|\tilde s_j\rangle_{\rm AB}$ is the only odd-parity state of a qubit pair, the non-interacting terms acting only on the qubit pair at site $j$ must replace $|\tilde 0_j\rangle_{\rm AB}, |\tilde 1_j\rangle_{\rm AB}$ with $|\tilde s_j\rangle_{\rm AB}$. Meanwhile it is more surprising that the interaction terms also introduce exactly one defect state into a product state of $\{|\tilde 0\rangle_{\rm AB}, |\tilde 1\rangle_{\rm AB}\}^{\otimes N}$.
            
            Before we dive into the action of $\hat H_{\rm AB}$ on $|\tilde d_k\rangle_{\rm AB}$, let us consider a property of the generalized Dicke states, that we will make use of: if we split the system into two bipartitions $S$ and its complement $\bar S$ (all sites outside of $S$), then their Schmidt decomposition is~\cite{toth_detection_2007}
            \begin{align}
                \ket{\tilde D^N_k}_{\rm AB} = \sum_{\ell=0}^{\min(k, |S|)} \ket{\tilde D^{(S)}_\ell}_{\rm AB}\ket{\tilde D^{(\bar S)}_{k-\ell}}_{\rm AB}
                \label{eq:generalized-dicke-decomposition}
            \end{align}
            where $|\tilde D^{(S)}_\ell\rangle_{\rm AB}$ is the generalized Dicke state with $\ell$ charges restricted to the subsystem $S$ of size $|S|$, while $|\tilde D^{(\bar S)}_{k-\ell}\rangle_{\rm AB}$ is the generalized Dicke state with $k-\ell$ charges on $\bar S$. For example for a single site $S=j$, we have
            \begin{align}
                \ket{\tilde D^N_k}_{\rm AB} = \ket{\tilde 0_j}_{\rm AB} \ket{\tilde D^{(\bar j)}_k}_{\rm AB} + \ket{\tilde 1_j}_{\rm AB} \ket{\tilde D^{(\bar j)}_{k-1}}_{\rm AB}.
            \end{align}

            Let us now use this property to obtain the special bright states $|\tilde b_k\rangle_\text{AB}$. First consider the non-interacting terms acting only on the qubit pair at site $j$ in Eq.~\eqref{eq:doubled-system-Hamiltonian} using Eq.~\eqref{eq:generalized-dicke-decomposition}
            \begin{alignat}{2}
                & \text{Detuning: } \quad & \sum_i (\hat \sigma^\text{z}_{\text{A},i} - \hat \sigma^\text{z}_{\text{B},i})|\tilde D^N_k\rangle_{\rm AB} &= 2\sum_i |\tilde s_i\rangle |\tilde D^{(\bar i)}_{k-1}\rangle_{\rm AB}\equiv 2|\tilde B^N_{k}\rangle_{\rm AB}\nonumber\\
                & \text{Chiral hopping: } \quad & \sum_i (\hat \sigma^+_{\text{A},i}\otimes \hat \sigma^-_{\text{B},i} - \hat \sigma^-_{\text{A},i}\otimes \hat \sigma^+_{\text{B},i})|\tilde D^N_k\rangle_{\rm AB} &= \sum_i |\tilde s_i\rangle_{\rm AB} |\tilde D^{(\bar i)}_{k-1}\rangle_{\rm AB}=|\tilde B^N_{k}\rangle_{\rm AB}\nonumber\\
                & \text{Drive: } \quad & \sum_i (\hat \sigma^\text{x}_{\text{A},i} - \hat \sigma^\text{x}_{\text{B},i})|\tilde D^N_k\rangle_{\rm AB} &=\sqrt 2 \sum_i |\tilde s_i\rangle_{\rm AB} |\tilde D^{(\bar i)}_{k}\rangle_{\rm AB}=\sqrt 2 |\tilde B^N_{k+1}\rangle_{\rm AB}.
                \label{eq:dtfim-states-local-terms}
            \end{alignat}
            In each case, the right-hand side takes the same form in terms of the unnormalized states
            \begin{align}
                \ket{\tilde B_k}_{\rm AB} \equiv \sum_i \ket{\tilde s_i}\ket{\tilde D_{k-1}^{(\bar i)}}_{\rm AB}
            \end{align}
            which crucially feature exactly one ``defect'' state $|\tilde s\rangle_{\rm AB}$ that is uniformly distributed across all $N$ qubits pairs; therefore these states are also invariant under any permutation. The fact that all of these local terms map to the same states should not be a surprise: these terms act only on the qubit pair on site $j$ and, since $\hat H_{\rm AB}$ maps the even-parity states in $\mathbf D_+$ to odd-parity states, they must map to $|\tilde s_j \rangle_{\rm AB}$ which is the only odd-parity state of the qubit pair. Recognizing that the terms in the first two lines are U(1) symmetric terms, it is clear that they can only convert one even-parity charge-1 state ($|\tilde 1 \rangle_{\rm AB}$) into an odd-parity charge-1 state ($|\tilde s \rangle_{\rm AB}$). Meanwhile the term in the third line, corresponding to the drive, breaks the U(1) symmetry and creates an additional odd-parity charge-1 state ($|\tilde s \rangle_{\rm AB}$) out of an even-parity charge-0 state ($|\tilde 0 \rangle_{\rm AB}$).
            
            It is more surprising that the interaction terms acting on qubit pairs on sites $i$ and $j$ also create exactly one defect. However, there is a simple reason for this: these terms only act nontrivially on two qubit pairs and they must map each state in $\mathbf D_+$ to a state with odd-parity under exchange of $A$ and $B$. If we use the following basis for each qubit pair $\{|\tilde 0_j\rangle_{\rm AB}, |\tilde 1_j\rangle_{\rm AB}, |\tilde s_j\rangle_{\rm AB},|11_j\rangle_{\rm AB}\}$, then it is evident that the odd-parity interaction terms must map even-parity states to states of the form
            \begin{align}
                \ket{\tilde s_i}_{\rm AB}\ket{\lambda_j}_{\rm AB} \quad \text{and} \quad \ket{\lambda_i}_{\rm AB}\ket{\tilde s_j}_{\rm AB}
                \label{eq:allowed-states-interaction-terms}
            \end{align}
            where $\ket{\lambda}_{\rm AB}\in \{|\tilde 0\rangle_{\rm AB}, |\tilde 1\rangle_{\rm AB}, |11\rangle_{\rm AB}\}$, as these are the only odd-parity states. Therefore the crucial aspect of our analysis is showing that we do not map to $|11\rangle_{\rm AB}$ since this state does not feature in the image of the non-interacting terms (cf. Eq.~\eqref{eq:dtfim-states-local-terms}). For charge-conserving interaction terms acting on two qubit pairs (as we have here), it is simple to rule this state out, since the state $|\tilde s_i\rangle_{\rm AB}|11_j\rangle_{\rm AB}$ (and $i\leftrightarrow j$) has total charge 3, while the states in $\{|\tilde 0\rangle_{\rm AB}, |\tilde 1\rangle_{\rm AB}\}^{\otimes 2}$ have at most total charge 2. 

            Using Eq.~\eqref{eq:generalized-dicke-decomposition} we can write
            \begin{align}
                \ket{\tilde D_k^N}_{\rm AB} &= \ket{\tilde 0_i}_{\rm AB}\ket{\tilde 0_j}_{\rm AB}\ket{\tilde D_k^{(\bar{ij})}}_{\rm AB} + (\ket{\tilde 1_i}_{\rm AB}\ket{\tilde 0_j}_{\rm AB} + \ket{\tilde 0_i}_{\rm AB}\ket{\tilde 1_j}_{\rm AB})\ket{\tilde D_{k-1}^{(\bar{ij})}}_{\rm AB} + \ket{\tilde 1_i}_{\rm AB}\ket{\tilde 1_j}_{\rm AB}\ket{\tilde D_{k-2}^{(\bar{ij})}}_{\rm AB}
            \end{align}
            and analyze the action of the interaction terms separately. First take the Ising interaction terms of the form
            \begin{align}
                \hat \sigma^\text{z}_{\text A,i} \hat \sigma^\text z _{\text A, j} - \hat \sigma^\text{z}_{\text B,i} \hat \sigma^\text z _{\text B, j} 
            \end{align}
            which are U(1)-symmetric and, therefore, conserve the total charge. Due to the charge conservation the $|\tilde 0_i\rangle_{\rm AB}|\tilde 0_j\rangle_{\rm AB}$ must be annihilated, since the state has zero total charge while the allowed states in Eq.~\eqref{eq:allowed-states-interaction-terms} all have nonzero total charge. We can manually confirm that $|\tilde 1_i\rangle_{\rm AB}|\tilde 1_j\rangle_{\rm AB}$ is also annihilated. This leaves $(|\tilde 1_i\rangle_{\rm AB}|\tilde 0_j\rangle_{\rm AB} + |\tilde 0_i\rangle_{\rm AB}|\tilde 1_j\rangle_{\rm AB})$ which is mapped to $(|\tilde s_i\rangle_{\rm AB}|\tilde 0_j\rangle_{\rm AB} + |\tilde 0_i\rangle_{\rm AB}|\tilde s_j\rangle_{\rm AB})$ such that we find
            \begin{align}
                \sum_{i\neq j} (\hat \sigma^\text{z}_{\text A,i} \hat \sigma^\text z _{\text A, j} - \hat \sigma^\text{z}_{\text B,i} \hat \sigma^\text z _{\text B, j} )|\tilde D^N_k\rangle &= -2\sum_{i\neq j} (|\tilde s_i\rangle |\tilde 0_j\rangle + |\tilde 0_i\rangle |\tilde s_j\rangle) |\tilde D^{(\bar{ij})}_{k-1}\rangle \nonumber\\
                &\equiv -4(N-k) |\tilde B^N_{k}\rangle
                \label{eq:dtfim-states-ising-interactions}
            \end{align}
            where we used combinatorics to arrive at the second line. The factor $N-k$ follows from the double summation in the interactions. The second interaction terms are contributions by the collective dissipation and take the form
            \begin{align}
                (\hat \sigma^+_{\text A, i} \otimes \hat \sigma^-_{\text B, j} - \hat \sigma^-_{\text A, i} \otimes \hat \sigma^+_{\text B, j}) + (\hat \sigma^+_{\text A, j} \otimes \hat \sigma^-_{\text B, i} - \hat \sigma^-_{\text A, j} \otimes \hat \sigma^+_{\text B, i})
            \end{align}
            and also possess a U(1) symmetry leading to total charge conservation. With the same charge-conservation argument as for the Ising interactions, we know that $|\tilde 0_i\rangle_{\rm AB}|\tilde 0_j\rangle_{\rm AB}$ is annihilated. We can again manually confirm that $|\tilde 1_i\rangle_{\rm AB}|\tilde 1_j\rangle_{\rm AB}$ is also annihilated. This finally leaves us with 
            \begin{align}
                &\sum_{i \neq j} \left [ (\hat \sigma^+_{\text A, i} \otimes \hat \sigma^-_{\text B, j} - \hat \sigma^-_{\text A, i} \otimes \hat \sigma^+_{\text B, j}) + (\hat \sigma^+_{\text A, j} \otimes \hat \sigma^-_{\text B, i} - \hat \sigma^-_{\text A, j} \otimes \hat \sigma^+_{\text B, i}) \right ] |\tilde D^N_k\rangle\nonumber \\ 
                &= \sum_{i \neq j} \left [ (\hat \sigma^+_{\text A, i} \otimes \hat \sigma^-_{\text B, j} - \hat \sigma^-_{\text A, i} \otimes \hat \sigma^+_{\text B, j}) + (\hat \sigma^+_{\text A, j} \otimes \hat \sigma^-_{\text B, i} - \hat \sigma^-_{\text A, j} \otimes \hat \sigma^+_{\text B, i}) \right ] (\ket{\tilde 1_i}_{\rm AB}\ket{\tilde 0_j}_{\rm AB} + \ket{\tilde 0_i}_{\rm AB}\ket{\tilde 1_j}_{\rm AB})\ket{\tilde D_{k-1}^{(\bar{ij})}}_{\rm AB}\nonumber \\
                &= \sum_{i\neq j} (|\tilde s_i\rangle |\tilde 0_j\rangle + |\tilde 0_i\rangle |\tilde s_j\rangle)\otimes |\tilde D^{(\bar{ij})}_{k-1}\rangle = 2 (N-k) |\tilde B^N_{k}\rangle.
                \label{eq:dtfim-states-collective-dissipation}
            \end{align}
            In the second line the factor $N-k$ is again a consequence of the double summation in the interactions and can be identified from combinatorics as before.

            Notice now that all special dark states $|\tilde d_k\rangle_{\rm AB}$ map exactly to $|\tilde B_k^N\rangle_{\rm AB}$ and that all $|\tilde B_k^N\rangle_{\rm AB}$ have a unique charge $k=1,\dots,N$. This is exactly what we need to form the bottom leg of the ladder model in Fig.~\ref{fig:fig4}b. Therefore we take as
            our ansatz for the special bright states $|\tilde b_k\rangle_{\rm AB}$ 
            \begin{align}
                \ket{\tilde b_k}_{\rm AB} = \frac{1}{\sqrt{\mathcal N}} \ket{\tilde B^N_k}_{\rm AB} = \frac{1}{\sqrt{\mathcal N}} \sum_{\pi} \mathcal P_\pi \ket{{\tilde s}_j}_\text{AB} |\tilde 1\rangle_\text{AB}^{\otimes k-1} |\tilde 0\rangle_\text{AB}^{\otimes N-k}
            \end{align}
            where the normalization is given by $\mathcal N = N! / [(k-1)!(N-k)!]$ and can be identified by combinatorics. Note that we are again only summing over distinct permutations $\pi$ to avoid double counting.
            
            Combining the results of Eqs.~\eqref{eq:dtfim-states-local-terms},~\eqref{eq:dtfim-states-ising-interactions} and~\eqref{eq:dtfim-states-collective-dissipation}, $\hat H_\text{AB}$ acts on $|\tilde D_k^N\rangle_{\rm AB}$ in the following fashion
            \begin{align}
                \hat H_\text{AB} |\tilde D^N_k\rangle = \frac \Omega {\sqrt{2}} |\tilde B^N_{k+1}\rangle + \left [\left (\Delta - i\frac{\gamma + \overline\Gamma / N}{2}\right ) - \left (J + i\frac {\overline \Gamma} {2} \right ) \frac{N-k}N \right ] |\tilde B^N_{k}\rangle.
            \end{align}
            Once we account for the normalization of each state we finally find the coupling coefficients
            \begin{align}
                h_{k,k} &= \mel{\tilde b_k}{\hat H_{\rm AB}}{\tilde d_k}_{\rm AB} = \left [ \left (\Delta -i \frac{\gamma+\overline{\Gamma}/N}{2}\right ) - \left (\overline J + i\frac{\overline\Gamma} 2\right )\frac{N-k}N \right ]\sqrt k\nonumber\\
                h_{k+1,k} &= \mel{\tilde b_{k+1}}{\hat H_{\rm AB}}{\tilde d_k}_{\rm AB} = \sqrt{\frac{N-k}{2}}\Omega.
                \label{eq:dtfim-coupling-coeffs-app}
            \end{align}
            Therefore, we now have the special dark and bright states as well as the coefficients of the Hamiltonian ladder model. Note that the structure of our ladder model can easily be simplified into a chain with a two-site unit cell as in Fig.~\ref{fig:fig5}; this is the same structure as for the nonlinear cavity model of Sec.~\eqref{subsec:driven-dissipative-kerr}. 
            
            In Table~\eqref{tab:ladder-coefficients-comparison-kerr-dtfim} we compare the non-zero matrix elements $h_{k',k}$ of the DTFIM and the nonlinear cavity model directly. Importantly, we find clear similarities in the parameters of the two models: the Ising interactions $\overline J$ correspond to the Kerr interactions $\overline U$, the transverse field $\Omega$ to the single-photon drive $\overline F_1$, the qubit decay rate $\gamma$ to the single-photon decay rate $\kappa$ and the detuning $\Delta$ also matches between the two models. There are two crucial differences: on the one hand, there is no (known) analogue to the two-photon drive $F_2$ in the DTFIM and on the other hand, the DTFIM has a bounded maximum number of charges. $h_{N+1,N}=0$ ensures that $\psi_k=0$ for $k>N$, while the bosonic mode has $h_{k+1,k}\neq 0$ and subsequently $\psi_k \neq 0$ for arbitrary $k\geq 0$. For $k\ll N$ the coefficients of the DTFIM and the nonlinear cavity model can be mapped onto one another which can be intuitively interpreted in terms of a Holstein-Primakoff (HP) transformation between collective spin systems and a single nonlinear cavity. This corresponds to the limit of small qubit/photon populations, where the HP transformation becomes exact. However, the conventional HP mapping cannot account for single-qubit relaxation which makes the model not fully collective. Still, Ref.~\cite{barberena_generalized_2025} recently introduced a generalized HP mapping which may be useful here; we leave a more thorough analysis to future work.
    
            \begin{table}[t]
                \centering
                \begin{tabular}{l|c|c|c}
                    Model & $h_{k+2,k}$ & $h_{k+1,k}$ & $h_{k,k}$ \\ \hline
                    Nonlinear cavity model & $2 F_2 \sqrt{k-1}$ & $2 \sqrt {2N} \overline F_1$ & $\left [ (\Delta - i \frac{\kappa_1}{2}) -\frac {\overline U + i\overline \kappa_2} {2N} (k-1)\right ]\sqrt k$ \\
                    DTFIM & 0 & $\sqrt{\frac{N-k}{2}}\Omega$ & $\left [ \left (\Delta -i \frac{\gamma+\overline{\Gamma}/N}{2}\right ) - \frac{\overline J + i\frac{\overline\Gamma} 2}{N}(N-k) \right ]\sqrt k$
                \end{tabular}
                \caption{Comparison of the non-zero matrix elements $h_{k',k}$ of the Hermitian ladder Hamiltonian $\hat H_\text{AB}'$ (cf. Eq.~\eqref{eq:ladder-model-matrix-elements}) for the nonlinear cavity model and the DTFIM.}
                \label{tab:ladder-coefficients-comparison-kerr-dtfim}
            \end{table}

            Now let us find the complex probability amplitudes $\psi_k$ of the steady-state purification $|\Psi_\mathcal T\rangle_\text{AB}$ in the special dark-state basis. As we showed in Eq.~\eqref{eq:htrs-recursion-relation} this is determined by the recursion relation $\sum_{j=-R}^{R} h_{k,k+j}\psi_{k+j}$ which simplifies to
            \begin{align}
                \frac{\Omega}{\sqrt 2 \overline J_\text{eff}} \psi_{k} = \sqrt{\frac{k+1}{N-k}} \left [ \frac{N-k-1}N - \frac{\Delta_\text{eff}}{\overline J_\text{eff}} \right ]\psi_{k+1}
                \label{eq:DTFIM-recurrence-relation}
            \end{align}
            using the coupling coefficients in Eq.~\eqref{eq:dtfim-coupling-coeffs-app}. Here we introduced the effective parameters
            \begin{align}
                \overline J_\text{eff} = \overline J + i\frac {\overline \Gamma} 2, \quad \Delta_\text{eff} = \Delta -i \frac{\gamma + \overline \Gamma/N} 2.
            \end{align}
            This finally gives us the steady-state coefficients of Eq.~\eqref{eq:purified-steady-state}
            \begin{align}
                \psi_k = \frac{1}{\sqrt \Sigma}\left ( \frac{N\Omega}{\sqrt 2 \overline J_{\rm eff}}\right )^k \sqrt{\binom{N}{k}} \frac{1}{\left (N - 1 -\frac{N\Delta_{\rm eff}}{\overline J_{\rm eff}} \right )_k}
                \label{eq:dtfim-steady-state-coeffs-app}
            \end{align}
            where $(x)_k = \prod_{j=0}^{k-1}(x-j)$ denotes the falling factorial which is closely related to a Pochhammer symbol~\cite{roberts_exact_2023}. The normalization $\Sigma$ is fixed by the condition $\sum_{k=0}^N |\psi_k|^2=1$.
            Therefore the purified steady state is given by 
            \begin{align}
                |\Psi_\mathcal T\rangle_\text{AB} = \sum_{k=0}^N \psi_k |\tilde d_k\rangle_\text{AB}.
                \label{eq:app-dtfim-purified-steady-state}
            \end{align}
            which can be interpreted as a condensate of Bell pairs uniformly distributed across the doubled system according to the non-equilibrium weights $\psi_k$. 
            
            To summarize, in this section we have provided the missing ingredients for our recipe for constructing the steady-state purification of the dissipative transverse-field Ising model discussed in the main text (cf. Sec.~\ref{subsec:DTFIM}). In particular, we have shown that the action of $\hat H_{\rm AB}$ on the special dark states simply introduces a single defect state into a product dark state, even for the interaction terms. Using this simple form we were able to derive the special bright states such that they contain the image of $\hat H_{\rm AB}$ acting on the special dark states. We then obtained the coupling coefficients for the ladder model, showing that it takes a similar structure as the nonlinear cavity model of Sec.~\ref{subsec:driven-dissipative-kerr}. Finally we used a recursion relation (cf. Eq.~\eqref{eq:htrs-recursion-relation}) to find the steady-state coefficients $\psi_k$ in Eq.~\eqref{eq:dtfim-steady-state-coeffs-app}. 
    
            We note that the Lindbladian of the DTFIM (cf. Eq.~\eqref{eq:DTFIM-qme}) can also be solved when the transverse field is not homogeneous\ie replacing
            \begin{align}
                \Omega\hat S_\text x = \frac{\Omega}{2} \sum_i \hat \sigma^\text x_i \to \sum_i \frac{\Omega_i}{2} \hat \sigma^\text x_i.
            \end{align}
            Interestingly, this only amounts to a local similarity transform of the purified steady state
            \begin{align}
                |\Psi'_\mathcal T\rangle_\text{AB} = \prod_i \hat {\tilde V}_i |\Psi_\mathcal T\rangle_\text{AB}
            \end{align}
            where $\hat {\tilde V}_i$ only acts on the qubit pair at site $i$ 
            \begin{align}
                \hat {\tilde V}_i |\tilde 0_i\rangle &= |\tilde 0_i \rangle\\
                \hat {\tilde V}_i |\tilde 1_i\rangle &= \frac{\Omega_i}{\Omega}|\tilde 1_i \rangle.
            \end{align}
            This means that every $|\tilde 1\rangle$ on site $i$ simply picks up a factor $\Omega_i/\Omega$ (up to normalization). Therefore the form of Eq.~\eqref{eq:app-dtfim-purified-steady-state} remains the same with the inhomogeneous dark states with $k$ even charges
            \begin{align}
                \ket{\tilde d_k'}_\text{AB} \propto \sum_{\substack{m_j\in\{0,1\}^N\\\sum_j m_j = k}} \prod_{j=1}^N \left ( \frac{\Omega_j}{\Omega}\right )^{m_j} \ket{\tilde m_j}.
            \end{align}
            Furthermore we mention that the homogeneous steady-state purification of Eq.~\eqref{eq:app-dtfim-purified-steady-state} has an exact and efficient encoding as a matrix-product state due to its permutational symmetry~\cite{raveh_dicke_2024,florido-llinas_product_2025}. Since the homogeneous and inhomogeneous steady-state purification are simply related through local similarity transforms, the same must also hold for the inhomogeneous steady-state purification $|\Psi'_\mathcal T\rangle_\text{AB}$. This allows for large-scale numerics of the steady state of hundreds of qubits for a disordered quantum many-body system as we show in a forthcoming work.


            \subsection{Comparison of different potentials and their potential barriers}\label{app:dtfim-pot-barriers}
                In this section we demonstrate a quantitative comparison between our ansatz for a non-equilibrium potential of the DTFIM in Eq.~\eqref{eq:dtfim-noneq-pot} and a naive potential extracted from the probability distribution of the magnetization. In particular we show numerical evidence that these two ansätze result in different predictions for the potential barrier height which provides further evidence to support our ansatz in Conjecture~\ref{conj:dissipative-gap}.
    
                First, let us define the probability distribution $P(s_\text z)$ of the magnetization $s_\text z$. Assume that $\ket {s_\text z, \alpha}$ is an eigenstate of the magnetization operator $\hat S_\text z$ with eigenvalue $s_\text z$, where $\alpha$ describes any additional degrees of freedom\eg how the magnetization is distributed across $N$ qubits. Then, we can define the probability to measure the magnetization eigenvalue $s_\text z$ with respect to a mixed quantum state $\hat \rho$ as
                \begin{align}
                    P(s_\text z) = \sum_\alpha \mel{s_\text z, \alpha}{\hat \rho}{s_\text z, \alpha}.
                \end{align}
                Therefore, we can obtain a steady-state probability distribution for the magnetization $P_\text{ss}(s_\text z)$ in terms of the steady-state density matrix $\hat \rho_\text{ss}$; note that $P_\text{ss}(s_\text z)$ was also used to analyze the thermodynamic limit of a closely related model in Ref.~\cite{leppenen_quantum_2024}. Due to the permutational symmetry of the DTFIM, we can obtain an analytical expression for $P_\text{ss}(s_\text z)$ using the exact solution in Eq.~\eqref{eq:app-dtfim-purified-steady-state}
                \begin{align}
                    P_\text{ss}(s_\text z) = \sum_{k=0}^N Q[s_\text z| k] p_k
                    \label{eq:p-sz-p-k}
                \end{align}
                where $p_k = |\psi_k|^2$ is the probability distribution associated with the steady-state purification $\ket{\Psi_\mathcal T}_\text{AB}$. W.l.o.g. we assume that the system size $N$ is even, where $Q[s_\text z, k]$ is given by
                \begin{align}
                    Q[s_\text z | k] = \frac{1}{2^{N-k}\binom{N}{k}}\sum_{\ell=|s_\text z|}^{\frac N 2} \frac{2\ell + 1}{N+1}\binom{N+1}{\frac{N}{2}-\ell}\binom{l-s_\text z}{k}\binom{\ell + k + s_\text z}{k}.
                    \label{eq:qmat-p-sz-p-k}
                \end{align}
                We can use this exact expression to convert $p_k$ into $P_\text{ss}(s_\text z)$ using simple matrix multiplication. While it is possible to use Eqs.~\eqref{eq:p-sz-p-k} and~\eqref{eq:qmat-p-sz-p-k} to obtain some insights in the thermodynamic limit, we will primarily use these expressions for a quantitative comparison---we will leave an in-depth analysis to future work. We define the potential $W$ for the magnetization probability distribution $P_\text{ss}(s_\text z)$ as
                \begin{align}
                    W(s_\text z) = -\lim_{N\to\infty} \frac{\log\left [ P_\text{ss}(s_\text z)\right ]}{N}
                    \label{eq:app-dtfim-magnetization-pot}
                \end{align}
                analogous to our non-equilibrium potential $V_\text{DTFIM}$. 

                \begin{figure}[t]
                    \centering
                    \includegraphics[width=0.6\linewidth]{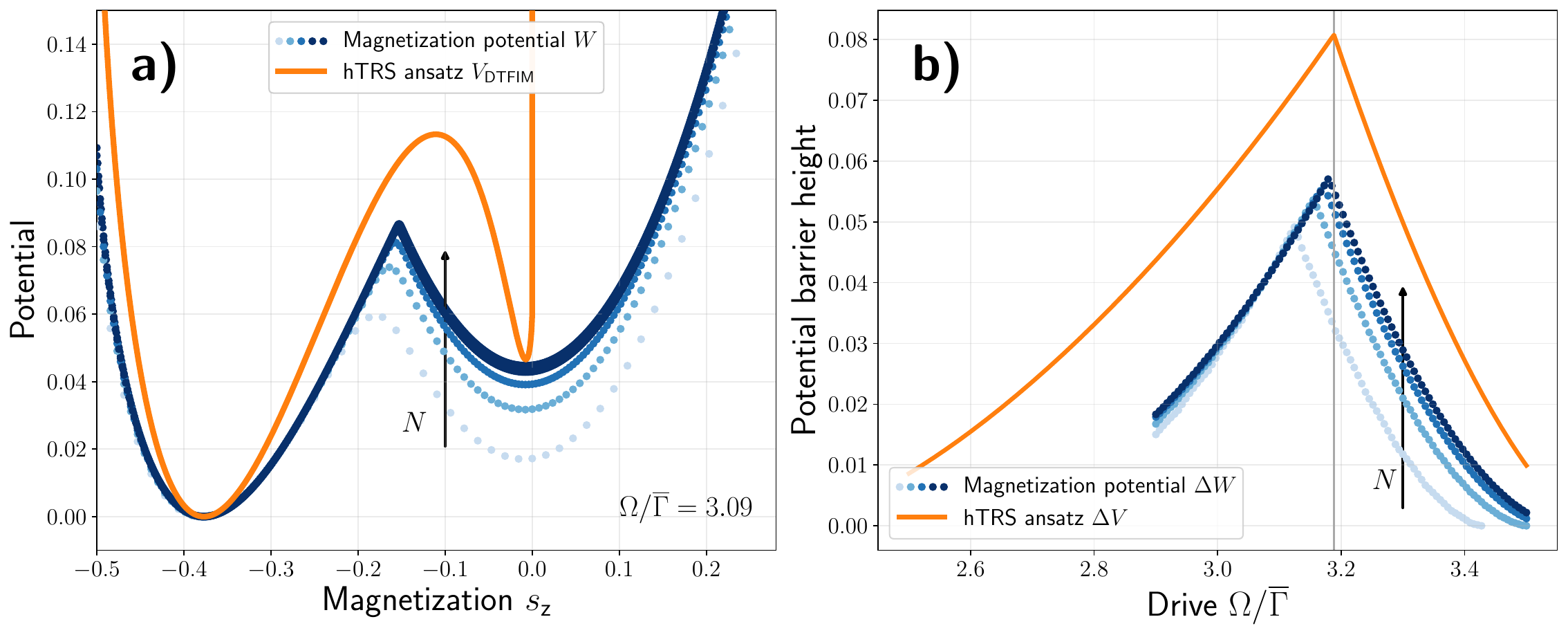}
                    \caption{Quantitative comparison of different potentials extracted from probability distributions for the magnetization in the DTFIM. \textbf{a)} We compare the magnetization potential $W_\text{ss}(s_\text z)$ (circular markers, $N=64,128,256,512,1024$ increasing with shade) with the non-equilibrium potential $V_\text{DTFIM}(s_\text z)$ (orange, solid line) defined in Eq.~\eqref{eq:dtfim-noneq-pot} for a fixed set of parameters in the thermodynamic limit $N\to\infty$. $W_\text{ss}(s_\text z)$ is defined in terms of the steady-state probability distribution $P_\text{ss}(s_\text z)$ for measuring a magnetization eigenstate with magnetization $s_\text{z}$ (cf. Eq.~\eqref{eq:app-dtfim-magnetization-pot}). Meanwhile the hTRS ansatz for a potential $V_\text{DTFIM}(s_\text z)$ is defined in terms of specific definite-charge states (see discussion in Sec.~\ref{subsec:DTFIM} and Appendix~\ref{app:DTFIM-purified-steady-state}) where the charge $n\in[0,1]$ determines the magnetization via $s_\text z=(n-1)/2$; consequently $V_\text{DTFIM}$ has no support for $s_\text z>0$ and takes the value $V_\text{DTFIM}(s_\text z > 0) =\infty$.  For easier visualization we truncate the full parameter range from $s_\text z \in [-1/2, +1/2]$ to $[-1/2,1/4]$. \textbf{b)} The potential barriers heights $\Delta W$ (circular marker, same system sizes $N$ as in a) and $\Delta V$ (solid line) plotted across the phase transition. The potential barrier $\Delta W$, numerically extracted from the magnetization potential $W_\text{ss}(s_\text z)$ systematically underestimates the non-equilibrium potential barrier $\Delta V$ which is in agreement with the dissipative gap $\Gamma_\text{diss}$ of the Lindbladian $\mathcal L_\text{DTFIM}$. Note that panel b) forms the basis for Fig.~\ref{fig:fig1}. Parameters: $\overline J = 5\overline \Gamma, \gamma =0.5\overline\Gamma, \Delta=0\overline\Gamma$.}
                    \label{fig:fig12}
                \end{figure}
    
                In Fig.~\ref{fig:fig12}a we compare the two potentials $V_\text{DTFIM}$ and $W$ for a specific set of parameters. While we lack a closed form for $W$ in the thermodynamic limit, we observe convergence as $N\to\infty$. There are two striking differences: First, the kink of $W$ at its local maximum, whereas $V_\text{DTFIM}$ is smooth across the entire domain. This kink can easily be explained in terms of minimizing two potentials $W_-(s_\text z)$ and $W_+(s_\text z)$\ie
                \begin{align}
                    W(s_\text z) = \min_{s_\text z}\left [W_-(s_\text z), W_+(s_\text z)  \right].
                \end{align}
                Second, the potential $V_{\rm DTFIM}$ is only defined for $s_{\rm z} < 0$, signified by putting $V_{\rm DTFIM}(s_{\rm z} > 0)=\infty$. From Eq.~\eqref{eq:dtfim-noneq-pot} we know that $V_{\rm DTFIM}$ is only defined for $n\in[0,1]$. Using the relation $s_{\rm z} = (n-1)/2$ it follows that $V_{\rm DTFIM}$ is only defined for $s_{\rm z}<0$.
                In Fig.~\ref{fig:fig12}a we also see that the extrema of $W$ slowly converge towards the extrema of $V_\text{DTFIM}$ which exactly coincide with the mean-field bistable solutions of Eq.~\eqref{eq:DTFIM-mean-field}. More importantly however we see in Fig.~\ref{fig:fig12}b that the potential barrier extracted from $W(s_\text z)$ converges as $N\to\infty$ and systematically underestimates our ansatz $\Delta V$, therefore underscoring the importance of working with the special basis proposed in Eq.~\eqref{eq:purified-steady-state} as opposed to other bases. 

            \subsection{Non-equilibrium potential \texorpdfstring{$V_\text{DTFIM}$}{} in the limit of dominant interactions}\label{app:noneq-pot-dtfim}
                Here we derive the non-equilibrium potential in the limit $\overline J, \overline \Gamma\gg \Delta, \gamma$ which is shown and discussed in Fig.~\ref{fig:fig9}b of the main text. This limit simplifies significantly, allowing us to derive an analytically tractable expression for the potential, the critical driving strength $\Omega_\text c$ and the potential barrier. For convenience let us define $m=1-n$ and first rewrite the non-equilibrium potential defined in Eq.~\eqref{eq:dtfim-noneq-pot}
                \begin{align}
                    V_\text{DTFIM}(m) &= - \int_0^m dm' \log\Big [ \frac{1-m'}{2m'}\frac{4(\overline J m' - \Delta)^2 + (\overline \Gamma m' +\gamma)^2}{\Omega^2}\Big ]\label{eq:dtfim-pot-app}\\
                    &= 2m + \log(1-m) + 4\frac{\overline J \gamma + \overline \Gamma\Delta}{4\overline J^2 + \overline \Gamma^2} \arctan\left [ \frac{(4\overline J ^2 + \overline \Gamma^2)m - 4\overline J \Delta + \overline \Gamma \gamma}{2(\overline J \gamma + \overline \Gamma \Delta)}\right ]\nonumber\\ 
                    &+ \frac{4\overline J \Delta - \overline \Gamma \gamma}{4\overline J ^2 + \overline \Gamma^2}\log\left [ \frac{4(\overline J m - \Delta)^2 + (\overline \Gamma m + \gamma)^2}{\Omega^2}\right ] - m\log \left [ \frac{1-m}{2m}\frac{4(\overline J m - \Delta)^2 + (\overline \Gamma m + \gamma)^2}{\Omega^2}\right ]
                \end{align}
                where we explicitly solve the integral in the second line. The extrema of Eq.~\eqref{eq:dtfim-pot-app} are determined by
                \begin{align}
                    (1-m)\left [ 4(\overline J m - \Delta)^2 + (\overline \Gamma m + \gamma)^2\right ] = 2\Omega^2 m.
                \end{align} 
                
                Upon taking the limit $\Delta /\overline J\to 0$ and $\gamma/\overline \Gamma\to 0$ for any $m>0$ we find that the extrema are subject to
                \begin{align}
                    (1-m) (4\overline J^2 + \overline\Gamma^2) m^2 = 2\Omega^2 m
                \end{align}
                which has the trivial solution $m=0$ and
                \begin{align}
                    m_\pm = \frac{1 \pm \sqrt{1 - \frac{8\Omega^2}{4\overline J^2 + \overline \Gamma^2}}}{2}
                \end{align}
                which has real solutions for $\Omega < \sqrt{(4\overline J^2 + \overline \Gamma^2) / 8}$. Notice that these states are ordered such that $0 < m_- < m_+$, allowing us to immediately identify $m_1 = 0, m_*=m_-$ and $m_2 = m_+$. The potential simplifies to
                \begin{align}
                    V_\text{DTFIM}(m) = 2m + \log(1-m) - m \log[m (1-m)] + m\log\left (\frac{2\Omega^2}{4\overline J^2 + \overline\Gamma^2} \right )
                \end{align}
                which gives
                \begin{align}
                    V_\text{DTFIM}(0) &= 0 \;\text{ and }\\
                    V_\text{DTFIM}(m_\pm) &= \log\left ( \frac{1\mp \sqrt{1 - \frac{8\Omega^2}{4\overline J^2 + \overline \Gamma^2}}}2\right ) \pm \sqrt{1 - \frac{8\Omega^2}{4\overline J^2 + \overline\Gamma^2}} + 1
                \end{align}
                after some straightforward algebra. Therefore we finally obtain the potential barrier\footnote{Note that $\log$ refers to the natural logarithm with base $e$ here.}
                \begin{align}
                    \Delta V =\begin{cases}
                        \log\left ( \frac{1 + \sqrt{1 - \frac{8\Omega^2}{4\overline J^2 + \overline \Gamma^2}}} 2\right ) + 1 - \sqrt{1 - \frac{8\Omega^2}{4\overline J^2 + \overline\Gamma^2}}, \quad \Omega < \Omega_\text c\\
                        \log\left ( \frac{1 + \sqrt{1 - \frac{8\Omega^2}{4\overline J^2 + \overline\Gamma^2}}}{1 - \sqrt{1 - \frac{8\Omega^2}{4\overline J^2 + \overline\Gamma^2}}}\right ) - 2 \sqrt{1 - \frac{8\Omega^2}{4\overline J^2 +\overline\Gamma^2}},\quad \Omega > \Omega_\text c
                    \end{cases}
                    \label{eq:dtfim-barrier-limit}
                \end{align}
                which form the two branches (solid lines) shown in Fig.~\ref{fig:fig9}b. Here the critical drive is determined by
                \begin{align}
                    \log\left ( \frac{1-\sqrt{1 - \frac{8\Omega_\text c ^2}{4\overline J^2 + \overline\Gamma^2}}} 2\right ) = -1 -\sqrt{1-\frac{8\Omega_\text c ^2}{4\overline J^2 + \overline \Gamma^2}}
                \end{align}
                which can be expressed in terms of the Lambert $W$ function
                \begin{align}
                    \Omega_\text c = \sqrt{\frac{1-\left [1+W\left(\frac{-2}{e^2}\right ) \right ]}{8}}\sqrt{4\overline J^2 + \overline\Gamma^2}\approx 0.2845 \sqrt{4\overline J^2 + \overline \Gamma^2}.
                \end{align}
    
                Plugging this into Eq.~\eqref{eq:dtfim-barrier-limit} we obtain the maximum potential barrier $\Delta V_\text c$ at $\Omega_\text c$, i.e., the location of the phase transition,
                \begin{align}
                    \Delta V_\text c = \log\left [ \frac{2+W\left(\frac{-2}{e^2}\right )}{2}\right] - W\left(\frac{-2}{e^2}\right ) \approx 0.179.
                \end{align}
    
                Recently this limit of the DTFIM with $\overline J=0$ has attracted significant attention including experimental investigations~\cite{song_dissipation-induced_2025,ferioli_non-equilibrium_2023}. These studies are particularly focused on the late-time physics to observe the emergence of the phase transition. In this context, it is important to take into account the slow relaxation timescales which lead to hysteresis in the thermodynamic limit. Therefore we believe that it would be useful to first probe the mesoscopic regime with ten to a hundred qubits, where it might be feasible to observe the emergence of slow relaxation. Given recent experimental advancements~\cite{song_dissipation-induced_2025}, we expect that this can be engineered using ultracold atoms trapped in tweezer arrays and strongly coupled to an optical cavity~\cite{dong_engineering_2025,santis_realization_2026}.

    \section{Quadratic scaling of double-well potential close to second-order phase transition}\label{app:double-well-potential-scaling}
        In this section we explain the quadratic scaling of the potential barrier $\Delta V$ upon approaching the second-order phase transition which was observed in Fig.~\ref{fig:fig9}a. It will turn out that this scaling is completely generic for a double-well potential at a parameter point, where a local maximum and a local minimum merge.

        Let us consider a potential $U(x, \alpha)$ of the form
        \begin{align}
            U(x,\alpha) = \alpha x + U_0(x)
        \end{align}
        where $U_0(x)$ is $\alpha$-independent. Furthermore, we assume that $U(x,\alpha)$ takes the form of a double-well potential in the parameter $x$ for a properly chosen value of $\alpha$. This means that
        \begin{align}
            \partial_x U(x,\alpha) = \alpha + \partial_x U_0(x) = 0
            \label{eq:generic-double-well-potential-extrema-appendix}
        \end{align}
        reduces to a cubic equation with solutions $x_1 < x_* <x_2$; we assume that $x_1, x_2$ correspond to local minima and $x_*$ to a local maximum. We define the potential barrier
        \begin{align}
            \Delta U = U(x_*) - \max_{i\in\{1,2\}} U(x_i)
        \end{align}
        which will be our quantity of interest.
        
        Let there be a parameter value $\alpha_\text c$ such that a local minimum, say $x_1$, merges with the local maximum $x_*$\ie $x_1 \to x_*$ as $\alpha\to\alpha_\text c$. Then we have a saddle-point at $x_*$ with
        \begin{align}
            \partial_x U(x, \alpha_\text c)\Big |_{x=x_*} = 0 \quad\&\quad \partial_x^2 U(x, \alpha_\text c)\Big |_{x=x_*} = 0.
        \end{align}
        Now let us take $\delta \alpha \ll 1$ and perform a Taylor expansion of $U(x,\alpha_\text c + \delta \alpha)$ with $\epsilon = x - x_*\ll 1$:
        \begin{align}
            U(x_* + \epsilon, \alpha_\text c + \delta \alpha) = U(x_* + \epsilon, \alpha_\text c) + \delta \alpha \epsilon +\mathcal O(\delta \alpha^2) = U(x_*, \alpha_\text c) + \frac 1 6 \partial_x^3 U_0(x)\Big |_{x=x_*} \epsilon^3 + \delta \alpha \epsilon + \mathcal{O}(\delta\alpha^2, \epsilon^2). 
        \end{align}
        Since we are still in an extremum of $U$ for $\epsilon=\epsilon_*$, we know that
        \begin{align}
            \partial_x U(x,\alpha_\text c + \delta \alpha)\Big |_{x=x_* + \epsilon_*} = \delta \alpha + \frac{1}{2}\partial_x^3 U_0(x)\Big |_{x=x_*} \epsilon_*^2 = 0.
        \end{align}
        We can then simplify
        \begin{align}
            U(x_*+\epsilon_*,\alpha_\text c + \delta \alpha) = U(x_*,\alpha_\text c) + \frac{2}{3}\delta \alpha \epsilon_*.
        \end{align}
        Doing the same but with $x_1$ instead of $x_*$ also gives the same scaling such that the potential barrier $\Delta U$ scales as
        \begin{align}
            \Delta U = U(x_*) - U(x_1) \sim \delta \alpha \epsilon_*. 
        \end{align}
        For a fixed $\delta\alpha$ the locations $x_1$ and $x_*$ are determined by Eq.~\eqref{eq:generic-double-well-potential-extrema-appendix}. Since $\epsilon_* = x_1 - x_*\to 0$ as $\delta\alpha\to 0$, we can make the generic ansatz that $\epsilon_*\sim \delta \alpha$ which recovers the quadratic scaling
        \begin{align}
            \Delta U \sim \delta \alpha \epsilon_* \sim \delta \alpha^2.
        \end{align}
        This explains that the quadratic scaling of the potential barrier in Fig.~\ref{fig:fig9} is a generic feature of double-well potentials close to a second-order phase transition. In the case of the potential $V_{\rm DTFIM}$ in Eq.~\eqref{eq:dtfim-noneq-pot} we find that $\alpha = -2 \log(\Omega)$.
                    

\end{document}